\documentclass[pteplogo]{ptephy_v2}

\preprintnumber{JLAB-THY-26-4819} 
\usepackage[colorlinks=true,citecolor=blue,linkcolor=blue,urlcolor=blue]{hyperref}

\usepackage{amsmath,amssymb,amsfonts,mathrsfs,bbold}
\usepackage{yfonts}
\usepackage{graphicx}
\usepackage{color}
\usepackage{lipsum}
\usepackage{slashed}
\usepackage[normalem]{ulem}
\usepackage{url}
\usepackage{siunitx}
\usepackage{amsthm}
\usepackage{physics}
\usepackage{bm}
\usepackage{upgreek}
\newcommand{\ben}{\begin{eqnarray*}}
\newcommand{\een}{\end{eqnarray*}}

\newcommand{\non}{\nonumber\\}

\newcommand{\pT}{\ensuremath{P_{T}}}

\begin{document}
\allowdisplaybreaks

\title{Single inclusive hadron and jet production in lepton-hadron scattering}


\author[1,2]{Jian-Wei Qiu}
\author[3,4]{Kazuhiro Watanabe}
\affil[1]{Theory Center, Jefferson Lab,
Newport News, Virginia 23606, USA}
\affil[2]{Department of Physics, William \& Mary,
Williamsburg, Virginia 23187, USA
\email{jqiu@jlab.org}}

\affil[3]{Department of Physics, 
Graduate School of Science, 
Tohoku University, Sendai 980-8578, Japan}
\affil[4]{Faculty of Science and Technology, 
Seikei University, Musashino, 
Tokyo 180-8633, Japan
\email{kazuhiro.watanabe.b8@tohoku.ac.jp}}

\begin{abstract}
We present the first calculation of single inclusive hadron and jet production at large transverse momentum in lepton-hadron scattering in a joint QCD+QED factorization approach. The scattering cross section is factorized into a convolution of infrared-safe hard coefficient functions with universal lepton distribution functions (LDFs) and parton distribution functions (PDFs) of the colliding lepton and hadron, respectively, together with fragmentation functions (FFs) of the observed hadron (or jet). With joint QCD+QED factorization, the Dokshitzer-Gribov-Lipatov-Altarelli-Parisi (DGLAP) type evolution equations for LDFs, PDFs, and FFs necessarily have evolution kernels calculated in both QCD and QED. We derive a default set of LDFs for our calculations and discuss a strategy to extract universal, non-perturbative LDFs from future data. We present our calculations for single inclusive hadron and/or jet production at the energies of Jefferson Lab and the future Electron-Ion Collider.
\end{abstract}

\subjectindex{B61, B62, B63, B65}

\maketitle

\tableofcontents

\section{Introduction}
\label{sec1:intro}

The lepton-hadron Deep Inelastic Scattering (DIS), with a large momentum transfer, has played a unique role in the discovery of the proton's partonic structure~\cite{Bloom:1969kc} and the development of Quantum Chromodynamics (QCD) as a fundamental theory of strong interactions (for reviews, see~\cite{Brambilla:2014jmp,Gross:2022hyw}). With the high luminosity lepton-hadron fixed target facility at the Jefferson Lab (JLab), the future Electron-Ion Collider (EIC) under construction at the Brookhaven National Lab, and facilities around the world, lepton-hadron scattering with large momentum transfer will continue to play an important and critical role in the study of exploring the three-dimensional internal structure of hadrons~\cite{Accardi:2012qut,Proceedings:2020eah,AbdulKhalek:2021gbh,Boer:2024ylx,Proceedings:2026xrb}.

By detecting a scattered lepton of momentum $\ell'$ in the collision between a lepton of momentum $\ell$ and a hadron of momentum $p$, the momentum transfer between the colliding lepton and the hadron, $q=\ell - \ell'$, can be carried by a vector boson that acts as a short-distance hard probe when its virtuality $Q^2\equiv -q^2\gg 1/R_h^2\sim \Lambda_\mathrm{QCD}^2$ with hadron radius $R_h$. By varying $Q^2$ and Bjorken variable $x_B=Q^2/2p\cdot q$, this vector boson has been considered as a good and controllable hard probe for exploring QCD dynamics and partonic structure inside the colliding hadron~\cite{Roberts:1990ww,Tung:2001cv}.

However, the large momentum transfer between the colliding lepton and hadron necessarily generates both QCD and QED collision-induced radiation, which affects the measured cross section and the distributions of produced particles in the final states. A recent development is the identification of a new pinch-singular region in the DIS cross section, in which the exchanged virtual photon momentum $q$ could go on-shell ($Q^2\to 0$)~\cite{Cammarota:2025jyr}. 
The emergence of the pinch singularity
leads to requirements for experimental cuts on the final-state kinematics, along with parameter(s) defining the cuts, to ensure the deeply inelastic nature of the underlying parton-level process~\cite{Charchula:1994kf,Aschenauer:2013iia}. That is, such collision-induced radiation, if not taken into account and precisely quantified, could obscure the true partonic structure of the colliding hadron relative to what we can measure in experiments, due to uncertainty in determining the hard probe's momentum. It is therefore critically important to account for such collision-induced effects in the true partonic-structure information that we aim to probe.

In Refs.~\cite{Liu:2021jfp,Liu:2020rvc}, it was argued that collision-induced QED radiation in lepton-hadron collisions can and should be treated on an equal footing with collision-induced QCD radiation within a joint QCD and QED (hereafter QCD+QED) factorization approach. By systematically factorizing all collinear-sensitive, non-perturbative, and process-independent contributions from collision-induced QED radiation into universal lepton distribution functions (LDFs) of colliding lepton and lepton fragmentation functions (LFFs) if the scattered lepton is observed, this joint QCD+QED factorization approach ensures that the process-dependent contributions are either perturbatively calculable in both QCD and QED or suppressed by inverse powers of the large momentum transfer of the lepton-hadron collisions and can be approximately neglected. For inclusive DIS, it was explicitly demonstrated in Refs.~\cite{Cammarota:2024vxj,Cammarota:2025jyr} that next-to-leading order perturbative corrections to the factorized short-distance hard parts are infrared safe in both QCD and QED in this joint factorization approach, eliminating the need to introduce unknown parameter(s) other than the standard factorization scale, in contrast to the traditional radiative correction (RC) approach to treat collision-induced QED radiation~\cite{Mo:1968cg,Bardin:1989vz,Badelek:1994uq,Kripfganz:1990vm,Spiesberger:1994dm,Blumlein:2002fy}. 
Although the new LDFs and LFFs are non-perturbative and unknown, like the parton distribution functions (PDFs) of the colliding hadron, they are universal and process-independent, so that predictions of this joint QCD+QED factorization approach can be tested and verified when different observables, which are factorized into the same set of universal functions, are compared.

In this paper, we present the first calculation of inclusive production of a single hadron $H$ (or a jet) of momentum $P$ in lepton-hadron scattering, $e(\ell) + h(p)\to H(P)+X$ with $h$ the nucleon and $H=\pi,K, ...$, in the joint QCD+QED factorization approach. Without measuring the scattered lepton of momentum $\ell'$, as was done for the traditional inclusive lepton-hadron DIS 
and Semi-Inclusive DIS (SIDIS) in which a hadron is measured in coincidence with the scattered lepton of $\ell'$,
the transverse momentum of the observed hadron, \pT, defines the hard scale of this scattering process~\cite{Kang:2011jw,Hinderer:2015hra,Abelof:2016pby,Qiu:2020xum}. Kinematically, this single inclusive hadron production at high-\pT\ is similar and complementary to the inclusive DIS by measuring a produced hadron of momentum $P$ instead of the scattered lepton of momentum $\ell'$. Like the inclusive DIS, the inclusive production of a single hadron of mass $m_H$ with $P_{T}^2 \gg m_H^2$ is a scattering process with a single hard scale $P_{T}$. It could be studied employing the joint QCD and QED collinear factorization approach. Understanding how this high-\pT\ hadron is produced in lepton-hadron scattering is not only providing an independent observable to test the joint factorization approach and to extract the LDFs, but also introducing a prerequisite for studying lepton-hadron SIDIS at JLab and the future EIC~\cite{Liu:2020rvc,Boussarie:2023izj}.

In Sec.~\ref{sec:factorization}, we introduce the joint collinear factorization formalism of QCD+QED for the inclusive production of a single hadron at high-\pT. The renormalization-group improvement of this joint factorization formalism naturally yields Dokshitzer-Gribov-Lipatov-Altarelli-Parisi (DGLAP) type evolution equations for LDFs, PDFs, and FFs. Those are necessary to have evolution kernels calculated in both QCD and QED, which require the factorization scale $\mu$ to be sufficiently large so that QCD perturbation theory can be used to evaluate the evolution kernels.
In Sec.~\ref{sec:ldfs}, we discuss the strategy to solve evolution equations for LDFs, as PDFs and FFs, with evolution kernels calculated in both QCD and QED. To ensure the reliability of calculated evolution kernels, we choose to solve the evolution equations with an input factorization scale $\mu_0 = m_c$, the charm quark's mass, since the evolution kernels for a photon to split into a light quark-antiquark pair could be non-perturbative in QCD if the scale is less than $m_c$, even if they might be perturbative in QED.

Although LDFs at the input scale $\mu_0=m_c$ can be extracted from fitting experimental data, almost all published lepton-hadron scattering data have already been processed using traditional radiative correction (RC) frameworks~\cite{Mo:1968cg,Bardin:1989vz,Badelek:1994uq,Kripfganz:1990vm,Spiesberger:1994dm,Blumlein:2002fy}. Consequently, these existing data cannot be directly compared with our calculations to extract LDFs without RCs being removed from these data, as recently addressed in the analysis by the ZEUS Collaboration~\cite{ZEUS:2023zie}.

For the numerical calculations in this paper and helping analysis of future data in this joint QCD+QED factorization approach, in Sec.~\ref{sec:ldfs}, we derive a set of model or default LDFs by introducing a set of LDFs at the input scale $\mu_0=m_c$ and evolving them to a larger factorization scale $\mu \geq \mu_0$. We also discuss and test the modification to evolution equations for PDFs and FFs within this joint QCD+QED factorization approach. We find that the numerical impact of QED corrections to the evolution of hadron PDFs and FFs is much smaller than that of QCD corrections to the evolution of LDFs. As an approximation, by neglecting the small QED corrections to hadron PDFs and FFs~\cite{Cridge:2021pxm,Xie:2021equ,NNPDF:2024djq}, we used available JAM20 hadron PDFs~\cite{Moffat:2021dji} and various available FFs for our numerical calculations in this paper, since hadron PDFs are better determined and consistent between different sets.
In contrast, hadron FFs have much larger uncertainties.

For our numerical calculations in Sec.~\ref{sec:single-hadron}, we used various sets of hadron FFs, including JAM20 FFs~\cite{Moffat:2021dji} and MAP1.0 FFs~\cite{Khalek:2021gxf}, to test the sensitivity of our proposed observables to the extraction of hadron FFs. With our evolved default LDFs, together with the hadron PDFs and FFs, we calculate and make our predictions for single inclusive hadron production at high-\pT\ in lepton-hadron scattering at JLab and future EIC energies. Our results can be used for the planning of future experimental measurements. By comparing our predictions with future data, we will be able to test the joint factorization approach and extract the true LDFs at the input scale, 
thereby obtaining the universal LDFs through joint QCD+QED evolution. Universal LDFs can be used in future studies of high-energy lepton-hadron and lepton-lepton scatterings. If the lepton-hadron data are precise enough, we could also extract lepton distributions of hadrons that are very important for precision studies at the LHC and future high-energy hadron-hadron facilities~\cite{NNPDF:2024djq,Manohar:2024ndc}.

Furthermore, in Sec.~\ref{sec:single-hadron}, we study single high-\pT\ hadron production in lepton-nucleus scattering within the joint factorization approach, assuming that the hard scale $P_T$ is sufficiently large and higher-twist corrections are negligible. We employ collinear nuclear PDFs for the colliding nucleus. 
The collinear PDFs of a free nucleon can be modified inside a nucleus by the cold nuclear medium, giving rise to effective nuclear PDFs and distinct phenomena such as shadowing, anti-shadowing, and the EMC effect, etc.~\cite{Eskola:1993mb,Eskola:2002yc,Eskola:2021nhw,Kusina:2020lyz,AbdulKhalek:2022fyi,Helenius:2021tof}, which subsequently modify the hadron production yield in lepton-nucleus scattering relative to the lepton-nucleon scattering.
We claim that high-$P_T$ single hadron production in lepton-nucleus collisions is an important probe for constraining nuclear PDFs, especially in the EMC region at high-$x$.

Probing hadron structure by measuring various jet-relevant observables has also been an active area of study.
In Sec.~\ref{sec:single-jet}, we present our numerical calculations for inclusive production of a single high-\pT\ jet in lepton-hadron collisions at EIC energies. 
We demonstrate inclusive jet production by employing jet functions, instead of hadron FFs used in high-$P_T$ single hadron production. High-$P_T$ jet production will also serve as an important tool for probing cold nuclear effects, and related discussions will be presented.

Finally, we provide our summary and outlook in Sec.~\ref{sec:summary}.

\section{Factorization Formalism}
\label{sec:factorization}

In this section, we introduce the joint QCD+QED collinear factorization formalism for the single inclusive high-\pT\ hadron production in lepton-hadron scattering, which depends on universal LDFs of the colliding lepton in addition to PDFs of the colliding hadron and FFs of the observed hadron. Like PDFs and FFs, LDFs satisfy DGLAP-type evolution equations, with kernels calculated perturbatively in both QCD and QED within this joint factorization approach. Below, we discuss our strategy for solving these universal functions. 

QCD factorization for single inclusive hadron production at high-\pT\ in hadronic collisions, $A(p_A)+B(p_B) \to H(P)+X$, was discussed in Ref.~\cite{Nayak:2005rt} and further justified in Ref.~\cite{QSY:2026} with the following factorization formalism: 
\begin{eqnarray}
\label{eq:hh-fac}
E\frac{d\sigma_{AB\to H(P)X}}{d^3P}
&=& \sum_{a,b,c} \,
\int \frac{dz}{z^2}\, D_{c\to H}(z,\mu^2)
\int dx_a\, f_{a/A}(x_a,\mu^2)
\int dx_b\, f_{b/B}(x_b,\mu^2)
\nonumber \\
& \ & \times\,
E_c\frac{d \hat{\sigma}_{ab \to c(p_c)X}}{d^3p_c}
\left(x_a,x_b,z, p_c=\frac{P}{z},\mu^2\right)
+ {\cal O}\left(\frac{1}{P_{T}^2}\right)
\end{eqnarray}
where $a$, $b$, and $c$ run through all QCD parton flavors. $D_{c\to H}(z,\mu^2)$, the FF at the factorization scale $\mu\sim P_{T}$, represents the probability distribution for the observed hadron $H$ of momentum $P$ to be emerged from the produced parton $c$ of momentum $p_c$, carrying a momentum fraction $z=P/p_c$. $f_{a/A}(x_a,\mu^2)$ and $f_{b/B}(x_b,\mu^2)$ are the PDFs to find a parton of flavor $a$ or $b$ inside the colliding hadron $A$ or $B$ with the hadron's longitudinal momentum fraction $x_a$ or $x_b$.  In Eq.~\eqref{eq:hh-fac}, $d\hat{\sigma}_{ab \to c(p_c)X}$ represents partonic scattering cross sections, where all perturbative collinear divergences along the observed hadron momenta, $p_A, p_B, P$, are removed and absorbed into corresponding universal distributions, $f_{a/A}$, $f_{b/B}$, and $D_{c\to H}$, respectively. The partonic cross sections are then infrared safe and can be systematically calculated in QCD perturbation theory in powers of $\alpha_s$~\cite{Nayak:2005rt,QSY:2026}.

Since the photon, as a gauge boson of QED, does not carry electric charge and commutes with gluon fields of QCD, it is straightforward to verify that the same sequence of arguments, which is used for justifying the QCD factorization in Eq.~\eqref{eq:hh-fac} in Refs.~\cite{Nayak:2005rt,QSY:2026}, can be carried through to justify the factorization for the production of single inclusive hadron at high-\pT\ in lepton-hadron collisions, $e(\ell) + h(p)\to H(P)+X$, with the following formalism:
\begin{eqnarray}
\label{eq:lh-fac}
E\frac{d\sigma_{eh\to H(P)X}}{d^3P}
&=& \frac{1}{2S} \sum_{i,b,c} \,
\int_{z_\mathrm{min}}^1 \frac{dz}{z^2}\, D_{c\to H}(z,\mu^2)
\int_{\xi_\mathrm{min}}^1 \frac{d\xi}{\xi}\, f_{i/e}(\xi,\mu^2)
\int _{x_\mathrm{min}}^1 \frac{dx}{x}\, f_{b/h}(x,\mu^2)
\nonumber \\
& \ & \times\,
\widehat{H}_{ib\to c(p_c)X}\left(\xi,x,z, p_c=\frac{P}{z},\mu^2\right)
+ {\cal O}\left(\frac{1}{P_{T}^2}\right)
\\
&\equiv& \frac{1}{2S}\sum_{i,b,c}
D_{c\to H}(z)\otimes_{z} 
f_{i/e}(\xi) \otimes_{\xi}
f_{b/h}(x) \otimes_x\,
\widehat{H}_{ib\to cX}(\xi,x,z, p_c) \, ,
\nonumber
\end{eqnarray}
where $S=(\ell+p)^2$, and 
$\xi$ and $x$ represent the longitudinal momentum fraction of the beam lepton and that of the target hadron, respectively; $\otimes_j$ with $j=\xi,x,z$ denote the convolution of momentum fraction ``$j$'' as defined in Eq.~\eqref{eq:lh-fac}. $i$, $b$ and $c$ formally run through all QED lepton flavors and photon as well as QCD quark flavors and gluon for high-energy lepton-hadron scatterings. 
However, contributions to the cross section are dominated by a combination of $i=(e,\gamma,\bar{e})$ and $b,c$ equal to $(q,g,\bar{q})$ for the collision energies at JLab and the future EIC. In Eq.~\eqref{eq:lh-fac}, $f_{i/e}(\xi,\mu^2)$ represents the new non-perturbative LDFs, $D_{c\to H}(z,\mu^2)$ and $f_{b/B}(x_b,\mu^2)$ represent the hadron FFs and PDFs, respectively, with their operator definitions extended to include QED components~\cite{Cammarota:2025jyr}. The integration limits in Eq.~\eqref{eq:lh-fac} are given by
\begin{align}
\begin{split}
z_\mathrm{min} & = - \frac{T+U}{S}\, , \\
\xi_\mathrm{min} & = - \frac{T}{zS+U} \, , \\
x_\mathrm{min} & = - \frac{\xi U}{\xi z S +T} \, ,
\end{split}
\label{eq:limits}
\end{align}
with $T=(p - P)^2=-\sqrt{S}P_T e^{+y}$ and $U=(\ell-P)^2=-\sqrt{S}P_T e^{-y}$.
In Eq.~(\ref{eq:lh-fac}), $\widehat{H}_{ib\to c(p_c)X}$ are short-distance hard parts, defined as:
\begin{equation}
\widehat{H}_{ib\to c(p_c)X}\left(\xi,x,z, p_c=\frac{P}{z},\mu^2\right)
\equiv (2\hat{s})\, E_c\frac{d \hat{\sigma}_{ib \to c(p_c)X}}{d^3p_c}
\left(\xi,x,z, p_c=\frac{P}{z},\mu^2\right)
\label{eq:H}
\end{equation}
where $\hat{s}=\xi x S$ and $\hat{\sigma}_{ib \to c(p_c)X}$ are partonic scattering cross sections, where all perturbative collinear divergences along the observed external momenta, $\ell, p, P$, are absorbed into corresponding universal distributions, $f_{i/e}$, $f_{b/h}$ and $D_{c\to H}$, respectively. These short-distance hard parts are infrared safe and can be perturbatively calculated in QCD and QED in powers of ($\alpha_{em}^m \alpha_s^n$) as $\widehat{H}^{(m,n)}_{ib\to c(p_c)X}$. By comparing the factorization formalisms in Eqs.~\eqref{eq:hh-fac} and \eqref{eq:lh-fac}, it is clear that the single inclusive high-\pT\ hadron production in lepton-hadron collisions is a suitable observable for extracting and/or testing the universal LDFs.

\begin{figure}[t]
\begin{center}
\includegraphics[width=0.4\textwidth]{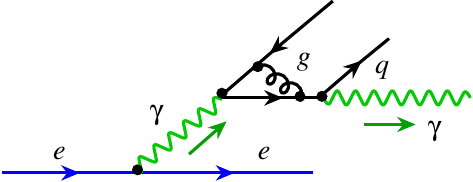}
\caption{A sample of hadronic correction to the photon LDF in an unpolarized electron. The photon evolution to a light quark-antiquark pair and subsequent evolution in QCD could be a non-perturbative process before a charged quark or antiquark evolves into a photon. }
\label{fig:LDFs}
\end{center}
\end{figure}

The physical lepton-hadron scattering cross section on the left-hand side of Eq.~\eqref{eq:lh-fac} does not depend on how we factorize the cross section on the right-hand side. The factorization scales $\mu$ to separate the collinear sensitivity along the observed hadron $H$, colliding hadron $h$, and colliding lepton $e$ do not have to be the same. At the same time, they are of the same order of magnitude to minimize the size of logarithms of ratios of these scales in the hard parts. The full derivative of the physical cross section with respect to any choice of the factorization scale should vanish perturbatively to all orders,
\begin{equation}
\frac{d}{d\ln\mu^2_f}\left[E\frac{d\sigma_{eh\to H(P)X}}{d^3P}\right] = 0
\label{eq:renorm}
\end{equation}
where $\mu_f$ with $f=e,h,H$ is the factorization scale to separate collinear sensitive contributions along the momentum of the observed colliding lepton $e$ and hadron $h$, and produced hadron $H$, respectively. This renormalization group improvement to the factorization formalism in Eq.~\eqref{eq:lh-fac} naturally leads to DGLAP-type evolution equations for PDFs, FFs, and LDFs~\cite{Dokshitzer:1977sg,Gribov:1972ri,Lipatov:1974qm,Altarelli:1977zs}; in particular, the evolution equation for LDFs reads
\begin{equation}
\mu_e^2 \frac{d}{d\mu_e^2} f_{i/e}(\xi,\mu_e^2) 
= \sum_{j} P_{ij}\left(\frac{\xi}{\xi'},\alpha_s(\mu_e),\alpha_{em}(\mu_e)\right) 
\otimes_{\xi'} f_{j/e}(\xi',\mu_e^2)
\label{eq:DGLAP_LDFs}
\end{equation}
with the evolution kernels $P_{ij}$ derived from the $\mu_e$ dependence of the perturbatively calculated hard parts $\widehat{H}^{(m,n)}_{ib\to c(p_c)X}$ or directly from the scale dependence of the operators defining the distributions. Similarly, Eq.~\eqref{eq:renorm} with the derivative of the factorization scale $\mu_h$ and $\mu_H$ leads to the evolution equations for PDFs and FF, respectively. With the joint QCD+QED factorization, PDFs and FFs are no longer purely QCD objects since electromagnetically charged quarks can radiate a photon that can decay into a lepton pair. 
Meanwhile, the LDFs are not purely QED objects since a lepton can radiate a photon that can decay into a quark-antiquark pair, as shown in Fig.~\ref{fig:LDFs}.

Consequently, like PDFs and FFs, the LDFs are also non-perturbative quantities in this joint QCD+QED factorization approach since a photon can decay into a pair of light quark-antiquark at a non-perturbative scale in QCD. Their evolution kernels can include terms calculated in QCD, QED, or both at higher orders, as further discussed in the next section.

\section{Lepton Distribution Functions}
\label{sec:ldfs}

In this section, we discuss our strategy for deriving universal, non-perturbative LDFs within the joint QCD+QED factorization approach. We present our approach to solve the coupled QCD and QED evolution equations for LDFs and introduce a model for LDFs at the input scale $\mu_0=m_c$. A similar approach is also used to evolve non-perturbative PDFs and FFs, which are no longer purely QCD correlation functions.

\subsection{Evolution equations}
\label{subsec:evolution}

With the joint QCD+QED factorization, the LDFs are necessarily gauge invariant in both QCD and QED~\cite{Cammarota:2025jyr}. DGLAP-type evolution equations for LDFs in Eq.~\eqref{eq:DGLAP_LDFs} necessarily have evolution kernels that are perturbatively calculated in both QCD and QED,
\begingroup
\renewcommand*{\arraystretch}{1.2}
\begin{align}
	\frac{\partial}{\partial \ln \mu^2}
	\left(
	\begin{array}{c}
		f_{e/e}\\
		f_{\bar{e}/e}\\
		f_{\gamma/e}\\
		f_{q/e}\\
		f_{\bar{q}/e}\\
		f_{g/e}\\
	\end{array}
	\right)
	&=
	\left(
	\begin{array}{cccccc}
		P^{(1,0)}_{ee} & P^{(2,0)}_{e\bar{e}} & P^{(1,0)}_{e\gamma} & P^{(2,0)}_{eq} & P^{(2,0)}_{e\bar{q}} & P^{(2,1)}_{eg} \\
		P^{(2,0)}_{\bar{e}e} & P^{(1,0)}_{\bar{e}\bar{e}} & P^{(1,0)}_{\bar{e}\gamma} & P^{(2,0)}_{\bar{e}q} & P^{(2,0)}_{\bar{e}\bar{q}} & P^{(2,1)}_{\bar{e}g} \\
		P^{(1,0)}_{\gamma e} & P^{(1,0)}_{\gamma \bar{e}} & P^{(1,0)}_{\gamma\gamma} & P^{(1,0)}_{\gamma q} & P^{(1,0)}_{\gamma \bar{q}} & P^{(1,1)}_{\gamma g} \\
		P^{(2,0)}_{qe} & P^{(2,0)}_{q\bar{e}} & P^{(1,0)}_{q\gamma} & P^{(0,1)}_{qq} & P^{(0,2)}_{q\bar{q}} & P^{(0,1)}_{qg} \\
		P^{(2,0)}_{\bar{q}e} & P^{(2,0)}_{\bar{q}\bar{e}} & P^{(1,0)}_{\bar{q}\gamma} & P^{(0,2)}_{\bar{q}q} & P^{(0,1)}_{\bar{q}\bar{q}} & P^{(0,1)}_{\bar{q}g} \\
		P^{(2,1)}_{ge} & P^{(2,1)}_{g\bar{e}} & P^{(1,1)}_{g\gamma} & P^{(0,1)}_{gq} & P^{(0,1)}_{g\bar{q}} & P^{(0,1)}_{gg} \\
	\end{array}
	\right)
	\otimes_{\xi}
	\left(
	\begin{array}{c}
    	f_{e/e}\\
		f_{\bar{e}/e}\\
		f_{\gamma/e}\\
		f_{q/e}\\
		f_{\bar{q}/e}\\
		f_{g/e}\\
	\end{array}
	\right),
	\label{eq:evo_LDFs}
\end{align}
\endgroup
where $q$ and $\bar{q}$ run over all active parton flavors, we neglected other lepton flavors in this paper, and we have used the following shorthand notation for QCD and QED splitting functions:
\begin{align}
P_{ij}(\xi,\mu^2)=\sum_{m,n=0}^{\infty}\left(\frac{\alpha_{em}(\mu^2)}{2\pi}\right)^{m}\left(\frac{\alpha_s(\mu^2)}{2\pi}\right)^{n}\hat{P}_{ij}^{(m,n)}(\xi)
\equiv \sum_{m,n=0}^{\infty}P_{ij}^{(m,n)}(\xi,\mu^2)
\label{eq:kernel_def}
\end{align}
with $P_{ij}^{(0,0)}=0$. In Eq.~\eqref{eq:evo_LDFs}, we have expressed the evolution equations for LDFs in a matrix form and listed the first non-vanishing splitting functions of evolution kernels. For the evolution of LDFs, the splitting functions of the top-left quadrant of the matrix correspond to pure QED evolution, and those of the lower-right quadrant of the matrix correspond to pure QCD evolution. Meanwhile, the splitting functions in the other two quadrants describe the mixing evolution from QCD to QED, and vice versa. We also have similar structures for the evolution of PDFs and FFs in the joint factorization.

In the rest of this paper, we restrict our discussion to the lowest order (LO) in powers of $\alpha_s$ and $\alpha_{em}$ and neglect $P^{(2,0)}$, $P^{(0,2)}$, $P^{(1,1)}$, and $P^{(2,1)}$ in our analysis. The corresponding LO QCD and QED splitting functions are summarized in Appendix~\ref{app:kernels}. In this paper, we consider only a single lepton family and have $n_l=1$ in $P^{(1,0)}_{\gamma\gamma}$ in Eq.~\eqref{eq:evo_gg}.

As in solving DGLAP evolution equations for PDFs in QCD, by organizing the set of evolution equations into a single coupled quark-gluon evolution equation, along with a number of non-singlet or valence-quark evolution equations, we can try to solve the joint QCD+QED evolution equations as follows. By introducing the shorthand notation, $F_{j^\pm}=f_{j/e}\pm f_{\bar{j}/e}$ for lepton and quark flavor $j$, the flavor composition for the LDFs is given by
\begin{align}
F_{\pm 3}&=F_{u^\pm}-F_{d^\pm},
\label{eq:pm3}\\
F_{\pm 8}&=F_{u^\pm}+F_{d^\pm}-2F_{s^\pm},\\
F_{\pm 15}&=F_{u^\pm}+F_{d^\pm}+F_{s^\pm}-3F_{c^\pm},\\
F_{\pm 24}&=F_{u^\pm}+F_{d^\pm}+F_{s^\pm}+F_{c^\pm}-4F_{b^\pm},\\
F_{\pm}&=F_{\pm 35}=F_{u^\pm}+F_{d^\pm}+F_{s^\pm}+F_{c^\pm}+F_{b^\pm}.
\end{align}
It is very convenient for solving the evolution equations to write up the evolution equations in terms of the combinations $F_{j^\pm}$, so that Eq.~\eqref{eq:evo_LDFs} can be cast into
\begin{align}
\frac{\partial \mathbf{F}}{\partial \ln \mu^2}=\mathbf{P}\otimes\mathbf{F},
\label{eq:evolution_Fs} \\
\frac{\partial F_{e^-}}{\partial \ln \mu^2}=P_{ee}^{(1,0)}F_{e^-},
\label{eq:evolution_Fe}\\
\frac{\partial F_{-}}{\partial \ln \mu^2}=P_{qq}^{(0,1)}F_{-},
\label{eq:evolution_F-}\\
\frac{\partial F_{-i}}{\partial \ln \mu^2}=P_{qq}^{(0,1)}F_{-i},
\label{eq:evolution_Fi}
\end{align}
where $i=3,8,15,24,35$, the $\mathbf{F}=\left(F_{e^+},f_{\gamma},F_{+},F_{+24},F_{+15},F_{+8},F_{+3},f_{g}\right)^\top$ consistent with $n_l=1$ in this paper, and the $(8\times8)$ matrix $\mathbf{P}$ can have the following block structure:
\begingroup
\renewcommand*{\arraystretch}{1.2}
\begin{align}
\mathbf{P}
&=
\left(
\begin{array}{cc}
\mathbf{P}_\mathrm{EE} & \mathbf{P}_\mathrm{EC} \\
\mathbf{P}_\mathrm{CE} & \mathbf{P}_\mathrm{CC}
\end{array}
\right)
\label{eq:splitting_function_matrixform}
\end{align}
\endgroup
where the LO matrices in both QCD and QED are respectively given by
\begingroup
\renewcommand*{\arraystretch}{1.2}
\begin{align}
\mathbf{P}_\mathrm{EE}
&=
\left(
\begin{array}{cc}
P^{(1,0)}_{ee} & 2P^{(1,0)}_{e\gamma} \\
P^{(1,0)}_{\gamma e} & P^{(1,0)}_{\gamma\gamma} 
\end{array}
\right)\, ,
\label{eq:splitting_function_ee} \\
\mathbf{P}_\mathrm{EC}
&=
\left(
\begin{array}{cccccc}
0 & 0 & 0 & 0  & 0 & 0 \\
\dfrac{e_+^2}{5}P^{(1,0)}_{\gamma e} & \dfrac{e_{+24}^2}{20}P^{(1,0)}_{\gamma e} & \dfrac{e_{+15}^2}{12}P^{(1,0)}_{\gamma e} & \dfrac{e_{+8}^2}{6}P^{(1,0)}_{\gamma e} & \dfrac{e_{+3}^2}{2}P^{(1,0)}_{\gamma e} & 0
\end{array}
\right)\, ,
\label{eq:splitting_function_ec} \\
\mathbf{P}_\mathrm{CE}
&=
\left(
\begin{array}{cc}
0 & 2e_{+}^2 N_c P^{(1,0)}_{e\gamma}  \\
0 & 2e_{+24}^2 N_c P^{(1,0)}_{e\gamma}   \\ 
0 & 2e_{+15}^2 N_c P^{(1,0)}_{e\gamma} \\
0 & 2e_{+8}^2 N_c P^{(1,0)}_{e\gamma} \\
0 & 2e_{+3}^2 N_c P^{(1,0)}_{e\gamma} \\
0 & 0
\end{array}
\right)\, ,
\label{eq:splitting_function_ce} \\
\mathbf{P}_\mathrm{CC}
&=
\left(
\begin{array}{cccccc}
P_{qq}^{(0,1)} & 0 & 0 & 0  & 0 & 2n_{f}P_{qg}^{(0,1)} \\
0 & P_{qq}^{(0,1)} & 0 & 0  & 0 & 0 \\
0 & 0 & P_{qq}^{(0,1)} & 0  & 0 & 0 \\
0 & 0 & 0 & P_{qq}^{(0,1)}  & 0 & 0 \\
0 & 0 & 0 & 0  & P_{qq}^{(0,1)} & 0 \\
P_{gq}^{(0,1)} & 0 & 0 & 0  & 0 & P_{gg}^{(0,1)} 
\end{array}
\right)\, ,
\label{eq:splitting_function_cc} 
\end{align}
\endgroup
with
\begin{align}
e_+^2&=e_u^2+e_d^2+e_s^2+e_c^2+e_b^2,\quad
e_{+24}^2=e_u^2+e_d^2+e_s^2+e_c^2-4e_b^2,\non
e_{+15}^2&=e_u^2+e_d^2+e_s^2-3e_c^2,\quad
e_{+8}^2=e_u^2+e_d^2-2e_s^2,\quad
e_{+3}^2=e_u^2-e_d^2.
\end{align}

Given a set of LDFs at the input scale $\mu_0=m_c$, we can derive the LDFs at a different scale $\mu$ by solving the single coupled (or singlet) evolution equation in Eq.~\eqref{eq:evolution_Fs} and three non-singlet or so-called valence-type evolution equations in Eqs.~\eqref{eq:evolution_Fe}-\eqref{eq:evolution_Fi}, instead of the fully coupled evolution equation set in Eq.~\eqref{eq:evo_LDFs}.

\subsection{Input lepton distribution functions}
\label{subsec:ldfs}

Historically, LDFs were obtained perturbatively in QED by setting $\mu_0=m_e$, the electron mass.  For example, $f_{e/e}(\xi,\mu^2)$, the electron distribution of an electron beam at a factorization scale $\mu$, is given by~\cite{Liu:2021jfp}
\begin{equation}
f^\mathrm{NLO}_{e/e}(\xi,\mu^2) =
 \delta(1-\xi) 
 + \frac{ \alpha_{em}}{2\pi} \left[\frac{1+\xi^2}{1-\xi}\ln\frac{\mu^2}{(1-\xi)^2m_e^2}\right]_+
\label{eq:LDFs_ee} 
\end{equation}
where superscript ``NLO'' indicates the perturbative accuracy at the next-to-leading order (NLO) in $\alpha_{em}$ and the ``+'' prescription is defined in Appendix~\ref{app:kernels}. The higher-order corrections in QED are suppressed by high powers of $\alpha_{em}$. Similarly, the photon distribution of an electron, $f_{\gamma/e}(\xi,\mu^2)$, is given by the so-called Weizsa\"{c}ker-William (WW) distribution at the LO in $\alpha_{em}$
with $\overline{\mathrm{MS}}$-scheme~\cite{Hinderer:2015hra}:
\begin{align}
	f^\mathrm{WW1}_{\gamma/e}(\xi,\mu^2)=\frac{\alpha_{em}(\mu^2)}{2\pi}P_{\gamma e}(\xi)\left[\ln\left(\frac{\mu^2}{\xi^2m_e^2}\right)-1\right].
	\label{eq:ww-distribution}
\end{align}
We note that, like PDFs, perturbatively calculated LDFs are not unique and depend on the choice of factorization scheme. Indeed, another set of WW distribution was given in Ref.~\cite{Kniehl:1996we}
\begin{align}
	f^\mathrm{WW2}_{\gamma/e}(\xi,\mu^2)=\frac{\alpha_{em}(\mu^2)}{2\pi}P_{\gamma e}(\xi)\left[\ln\left(\frac{\mu^2}{\mu_{\min}^2}\right)-2m_e^2 \xi \left(\frac{1}{\mu^2}-\frac{1}{\mu_{\min}^2}\right)\right]
	\label{eq:ww-distribution2}
\end{align}
with $\mu_{\min}^2=(m_e^2\xi^2)/(1-\xi)$.

However, within the joint QCD+QED factorization approach, the evolution of LDFs is only perturbative when the factorization scale is sufficiently large, so that the QCD splitting functions are perturbatively reliable. Therefore, we cannot solve for LDFs by evolving the evolution equations in Eqs.~\eqref{eq:evolution_Fs}-\eqref{eq:evolution_Fi} from $\mu_0=m_e$. As we normally do when solving for PDFs and FFs, we set $\mu_0=m_c$ in this paper.

The LDFs at the input scale are non-perturbative and should be extracted from experimental data by fitting. However, almost all published data from lepton-hadron scattering have some kind of QED RC implemented. They cannot be used for our purposes without reanalyzing the data, as in the recent effort by the ZEUS collaboration~\cite{ZEUS:2023zie}. For the numerical calculations in this paper, as an approximation, we introduce the following simple model for our default LDFs at $\mu_0 = m_c$, which can be easily improved once we obtain sufficient data for comparison.  Like the input distributions for PDFs, we choose the following generic functional form for the LDFs at the input scale,
\begin{equation}
f_{i/e}(\xi,\mu_0^2) \equiv
N_i\, \frac{\xi^{\alpha_i} (1-\xi)^{\beta_i}}{B[1+\alpha_i, 1+\beta_i]}
\label{eq:input_LDFs}
\end{equation}
where $B$ represents the Euler-Beta function, and the flavor index runs over $i=(e,\gamma,\bar{e}=e^+, q,\bar{q},g)$. Each flavor is characterized by three parameters ($N_i,\alpha_i,\beta_i$), where $N_i$ determines the normalization, while $\alpha_i$ and $\beta_i$ control the shape of the $\xi$-dependence.
This generic functional form is useful for solving the evolution equations in the Mellin moment space of LDFs. It can be further improved by multiplying a polynomial in $\xi$ once more precise experimental data are available for comparison.

Unlike PDFs, which are large in the region of small parton momentum fraction $x$ and vanish very quickly when $x\to 1$, we expect the LDFs to have a significantly different behavior as functions of momentum fraction $\xi$ due to the elementary nature of the electron. For the electron distribution function of an electron, $f_{e/e}(\xi,\mu_0^2)$, we expect $\alpha_e \gg \beta_e$ in contrast to the behavior of a typical quark distribution, which should have $\alpha_q \ll \beta_q$. Although $f_{e/e}(\xi,\mu_0^2)$ should be much larger than typical PDFs when $\xi\to 1$, $f_{e/e}(\xi,\mu_0^2)$ should vanish at $\xi=1$, which is very different from the fixed order perturbatively calculated LDFs in Eq.~\eqref{eq:LDFs_ee}. Such a unique feature of the electron LDF reflects non-perturbatively that the probability for finding an electron to carry 100\% of its parent electron's momentum should be vanishingly small, once we allow the electron to radiate. 

For our default LDFs, as an approximation, we assume that all quark, antiquark, and gluon contents of LDFs at scale $\mu > \mu_0$ are perturbatively generated by the evolution equations in Eqs.~\eqref{eq:evolution_Fs}-\eqref{eq:evolution_Fi}; hence, we shall set $f_{q/e}(\xi,\mu_0^2) = 0$, $f_{\bar{q}/e}(\xi,\mu_0^2) = 0$, $f_{g/e}(\xi,\mu_0^2) =0$. Consequently, all perturbatively generated quark, antiquark, and gluon LDFs at factorization scale $\mu\geq \mu_0$ effectively represent the lower limits of these LDFs. 

For determining the lepton flavor LDFs at the input scale $\mu_0=m_c$, we adopt the following approximation and procedure.

\medskip
\noindent{\it 1) Valence electron distribution of an electron:}
\medskip

With the limited phase space for the evolution and the small value of $\alpha_{em}$, we approximate the valence electron distribution at the input scale $\mu_0$ as
\begin{equation}
f_{e/e}^\mathrm{V}(\xi,\mu_0^2) \,
\equiv f_{e/e}(\xi,\mu_0^2)
-f_{\bar{e}/e}(\xi,\mu_0^2) 
\approx f_{e/e}^\mathrm{NLO}(\xi,\mu_0^2).
\label{eq:valence}
\end{equation}
The right side above 
$f_{e/e}^\mathrm{NLO}(\xi,\mu_0^2)$
is given in Eq.~\eqref{eq:LDFs_ee}, and satisfies $\int_0^1 d\xi\, f_{e/e}^\mathrm{V}(\xi,\mu_0^2) = 1$. 
The definition of valence lepton distribution in Eq.~\eqref{eq:valence} is also consistent with our introduction of $F_{j^{\pm}}= f_{j/e}\pm f_{\bar{j}/e}$, just above Eq.~(\ref{eq:pm3}), for effectively solving the set of evolution equations in Eq.~\eqref{eq:evo_LDFs} when applying $j=e$.
Since the $\delta(1-\xi)$ term and the ``+'' prescription term in $f_{e/e}^\mathrm{NLO}(\xi,\mu_0^2)$ are meaningful only under the integration over $\xi$, we parameterize $f_{e/e}^\mathrm{V}(\xi,\mu_0^2)$ using the generic form in Eq.~\eqref{eq:input_LDFs} with two parameters ($\alpha_V,\beta_V$). We determine these two parameters by fitting the Mellin moments of the ansatz,
\begin{equation}
    \widetilde{f}_{e/e}^V(n,\alpha_V,\beta_V) = \int_0^1 d\xi\, \xi^{n-1}
    \frac{\xi^{\alpha_V} (1-\xi)^{\beta_V}}{B[\alpha_V+1,\beta_V+1]}
    = \frac{B[\alpha_V+n,\beta_V+1]}{B[\alpha_V+1,\beta_V+1]} \, ,
\label{eq:valence-moments}
\end{equation}
with the Mellin moments of perturbative $f_{e/e}^\mathrm{NLO}(\xi,\mu_0^2)$,
\begin{align}
    \widetilde{f}_{e/e}^\mathrm{NLO}(n,\mu^2)=\,&\int_0^1d\xi\,\xi^{n-1}\,f_{e/e}^\mathrm{NLO}(\xi,\mu^2)\non
=\,&1+\frac{\alpha_{em}(\mu^2)}{2\pi}\Bigg[
\ln\left(\frac{\mu^2}{m_e^2}\right) \left(-\frac{1}{n}-\frac{1}{n+1}-2 (\psi^{(0)}(n)+\gamma_E )+\frac{3}{2}\right)\non
&+2 \left(-\frac{\gamma_E  n (n+1) (n (\gamma_E 
   n+\gamma_E +2)+1)+1}{n^2 (n+1)^2}-\frac{3}{n^2+n}-\psi^{(0)}(n)^2\right.\non
&\left.-\frac{(2 n (\gamma_E  n+\gamma_E +1)+1) \psi^{(0)}(n)}{n
   (n+1)}+\psi^{(1)}(n)-\frac{\pi^2}{6}+\frac{7}{4}\right)\Bigg], 
\label{eq:perturbative-LDF-Mellin}
\end{align}
where $\gamma_E$ represents the Euler's constant and $\psi^{(k)}$ represents the $k$-th derivative of the digamma function. $\alpha_V$ and $\beta_V$ are obtained by minimizing the following cost function:
\begin{align}
J(\alpha_V,\beta_V,m,\mu^2)=\sum_{n=1}^m\left[1-\frac{\tilde{f}^V_{e/e}(n,\alpha_V,\beta_V)}{\widetilde{f}_{e/e}^\mathrm{NLO}(n,\mu^2)}\right]^2\, ,
\label{eq:cost}
\end{align}
where $m$ specifies the number of Mellin moments included in the fit.

Evidently, parameter regions with large $\alpha_V$ and small $\beta_V$ provide better approximations to the perturbative LDF, Eq.~\eqref{eq:perturbative-LDF-Mellin}. At the input scale $\mu=m_c$, we examined representative parameter sets $(\alpha_V,\beta_V)=(50,0.125)$ and $(60,0.1)$. Both sets lie in the low $J$ region and reproduce the perturbative Mellin moments reasonably well up to $n\approx 4$; therefore, we should truncate the sum on the right side of Eq.~\eqref{eq:perturbative-LDF-Mellin} at $m=5$. Since the two sets yield very similar behavior, either set could, in principle, be adopted as the  valence-electron LDF at the input scale. In this work, we choose $(\alpha_V,\beta_V)=(60,0.1)$ as the default parameter set because it is located further inside the preferred large-$\alpha_V$, small-$\beta_V$ region and yields a slightly sharper shape, consistent with the perturbative distribution. 
Fitting data can further improve the input parameters in the future.

\medskip
\noindent{\it 2) Photon, sea-electron and positron distributions:}
\medskip

We approximate the photon and sea-electron distributions at $\mu_0=m_c$ to be perturbatively generated by the evolution equations in Eqs.~\eqref{eq:evolution_Fs}-\eqref{eq:evolution_Fi}
from $\mu=m_e$ to $\mu=m_c$ with only the QED evolution kernels and the following initial condition, 
\begin{align}
\begin{split}
f_{e/e}(\xi,m_e^2)  &= \frac{\xi^{\alpha_0}(1-\xi)^{\beta_0}}{B[\alpha_0+1,\beta_0+1]} , 
\\
f_{\gamma/e}(\xi,m_e^2)  &= 0\, , 
\\
f_{\bar{e}/e}(\xi,m_e^2)  &= 0\, .
\end{split}
\label{eq:QED_input}
\end{align}
We can determine the reasonable parameter set $(\alpha_0,\beta_0)$ by requiring that the perturbatively evolved valence electron distribution at $\mu=m_c$ closely reproduces the valence electron distribution, $f_{e/e}^\mathrm{V}(\xi,m_c^2)$ in Eq.~\eqref{eq:valence}, which should be independently determined in the previous step. We found that $(\alpha_0,\beta_0)\approx (70,0.1)$ effectively reproduces the perturbatively generated photon, sea-electron, and positron LDFs at $\mu=m_c$.

\medskip
\noindent{\it 3) Parameterization of LDFs at input scale $\mu_0=m_c$:}
\medskip

\begin{figure}[t]
\begin{center}
    \includegraphics[width=0.5\textwidth]{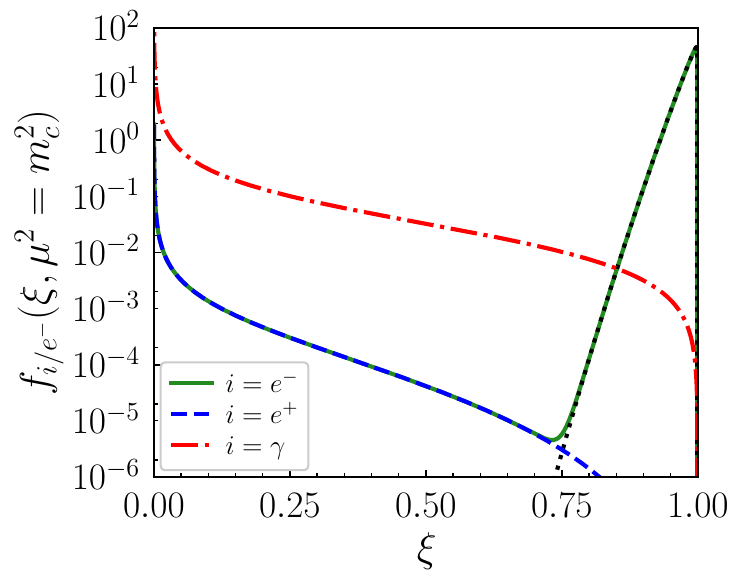}
    \caption{Our default input LDFs at $\mu_0=m_c$ with parameters in Table~\ref{table:parameters}. Solid green line: total electron (valence + sea). Dotted black line: valence electron. Dashed blue line: positron. Dash-dotted red line: photon.}
    \label{fig:input_LDFs}
\end{center}
\end{figure}

We shall use the following generic functional form for each lepton flavor $i=(e,\gamma,\bar{e})$,
\begin{align}
f_{i/e}(\xi,m_c^2)=N_i\, \xi^{\alpha_i}(1-\xi)^{\beta_i} \, ,
\label{eq:parameterization}
\end{align}
which represents our model or default set of LDFs at the input scale $\mu_0 = m_c$.  We determine the parameters $(N_i,\alpha_i,\beta_i)$ by fitting our perturbatively generated LDFs with the initial conditions in Eq.~\eqref{eq:QED_input}, along with the following sum-rule constraints, 
\footnote{For our numerical check, we introduce a lower cutoff in the integration with respect to $\xi$: $\xi_{\min}=10^{-8}$.}
\begin{align}
&\int_0^1 d\xi \, \left[
f_{e/e}(\xi,\mu_0^2) - f_{\bar{e}/e}(\xi,\mu_0^2) \right] = 1\, ,
\label{eq:number} \\
& \int_0^1 d\xi \, \xi \left[
f_{e/e}(\xi,\mu_0^2) + f_{\gamma/e}(\xi,\mu_0^2) + f_{\bar{e}/e}(\xi,\mu_0^2) \right] = 1 \, ,
\label{eq:momentum}
\end{align}
which correspond to the electron number conservation and the momentum conservation, respectively. Table~\ref{table:parameters} presents the resulted parameters $(N_i,\alpha_i,\beta_i)$, which specifies our default LDFs.  We plot the corresponding LDFs in Fig.~\ref{fig:input_LDFs}. 

\begin{table}[t]
\caption{Parameters for our default input LDFs with the generic functional form in Eq.~\eqref{eq:parameterization}. Note that the parameters for the valence electron distribution are shown.}
\centering
\begin{tabular}{|c|c|c|c|}
\hline
  {\hskip 0.5in}     &  {\hskip 0.3in}  $N_i$  {\hskip 0.3in} & {\hskip 0.3in}  $\alpha_i$ {\hskip 0.3in}  &  {\hskip 0.3in} $\beta_i$  {\hskip 0.3in} \\ 
\hline
 $f_{e/e}^\mathrm{V}$           &  $\num[round-mode=places,round-precision=3]{96.8082656415002}$  &  60  &  0.1  \\
 \hline
 $f_{\gamma/e}$ &  $\num[round-mode=places,round-precision=3]{0.03458101}$ & $\num[round-mode=places,round-precision=3]{-0.99521272}$ &  $\num[round-mode=places,round-precision=3]{1.08208418}$ \\
 \hline
 $f_{\bar{e}/e}$   &  $\num[round-mode=places,round-precision=3]{1.89447717e-04}$ & $\num[round-mode=places,round-precision=3]{-1.00053266e+00}$  &  $\num[round-mode=places,round-precision=3]{3.18235287e+00}$  \\
\hline
\end{tabular}
\label{table:parameters}
\end{table}

\begin{figure}[t]
\begin{center}
    \includegraphics[width=\textwidth]{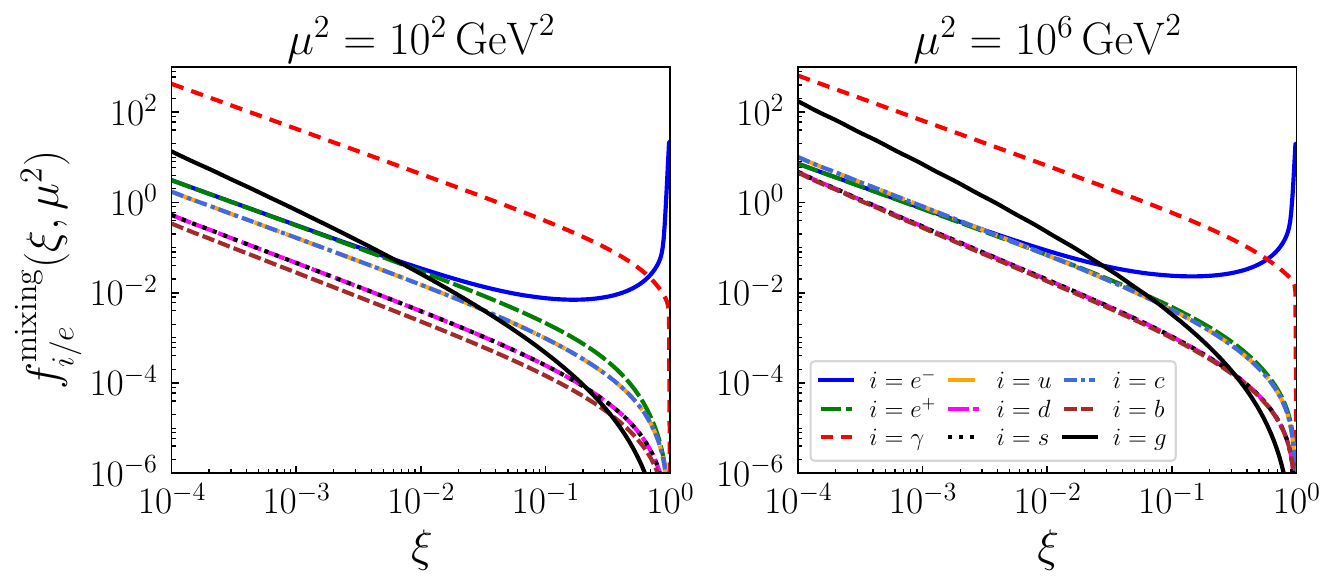}
    \caption{Evolved default LDFs at different factorization scales $\mu^2$. }
    \label{fig:evolve_mixing}
\end{center}
\end{figure}

Using the input LDFs shown in Fig.~\ref{fig:input_LDFs}, we derive the default LDFs at any higher scale $\mu > \mu_0$ by solving the evolution equations given in Eqs.~\eqref{eq:evolution_Fs}-\eqref{eq:evolution_Fi}. The resulting numerical grid points are structured to be compatible with the LHAPDF6 data format~\cite{Buckley:2014ana}, enabling smooth interpolation to arbitrary points in the $(\xi, \mu)$ space.
Figure~\ref{fig:evolve_mixing} shows our default LDFs at $\mu=10\,\mathrm{GeV}$ (left panel) and $\mu=10^3\,\mathrm{GeV}$ (right panel), respectively.  The superscript ``mixing" indicates that we have used the evolution equations with the block diagonal QED and QCD evolution kernels $\mathbf{P}_\mathrm{EE}$ and $\mathbf{P}_\mathrm{CC}$, respectively, as well as the mixing kernels $\mathbf{P}_\mathrm{EC}$ and $\mathbf{P}_\mathrm{CE}$.  Consequently, we generated quark and gluon LDFs dynamically only by the perturbative evolution, which represents a lower limit of quark and gluon LDFs since we did not model to include any quark and gluon distributions at the input scale $\mu_0$.  

Like the DGLAP evolution of PDFs, LDFs are enhanced at lower values of the momentum fraction $\xi$ while they are reduced at higher values of $\xi$. As expected, away from $\xi\sim 1$, the photon distribution of an electron dominates, while the gluon distribution catches up quickly at small $\xi$ because $\alpha_s\gg \alpha_{em}$. 

\begin{figure}[t]
\begin{center}
\includegraphics[width=\textwidth]{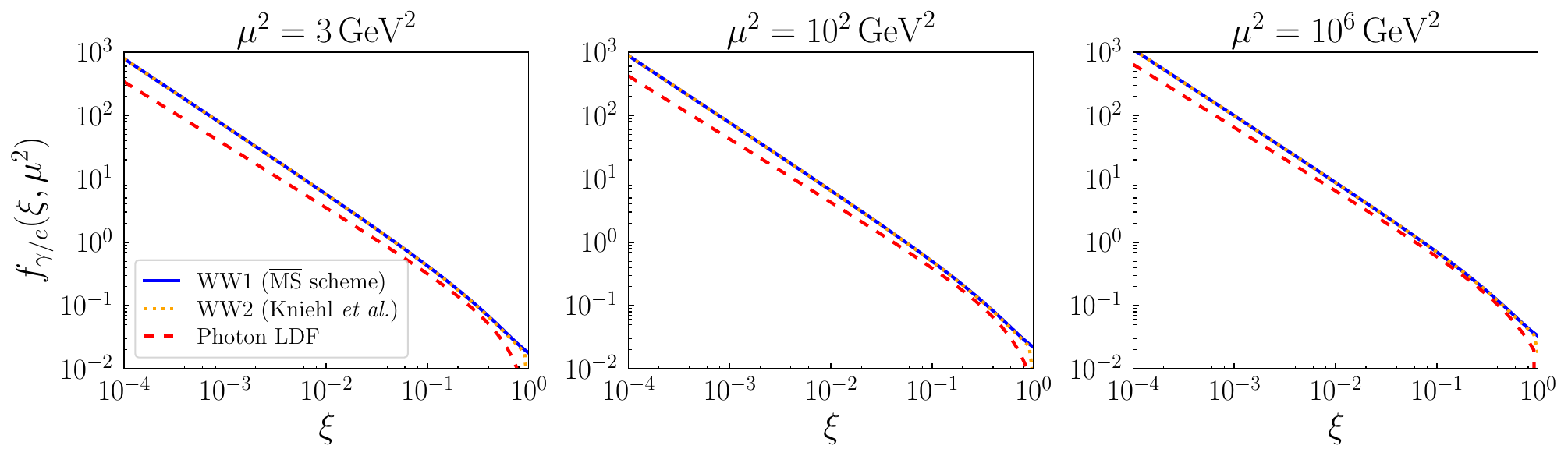}
\includegraphics[width=\textwidth]{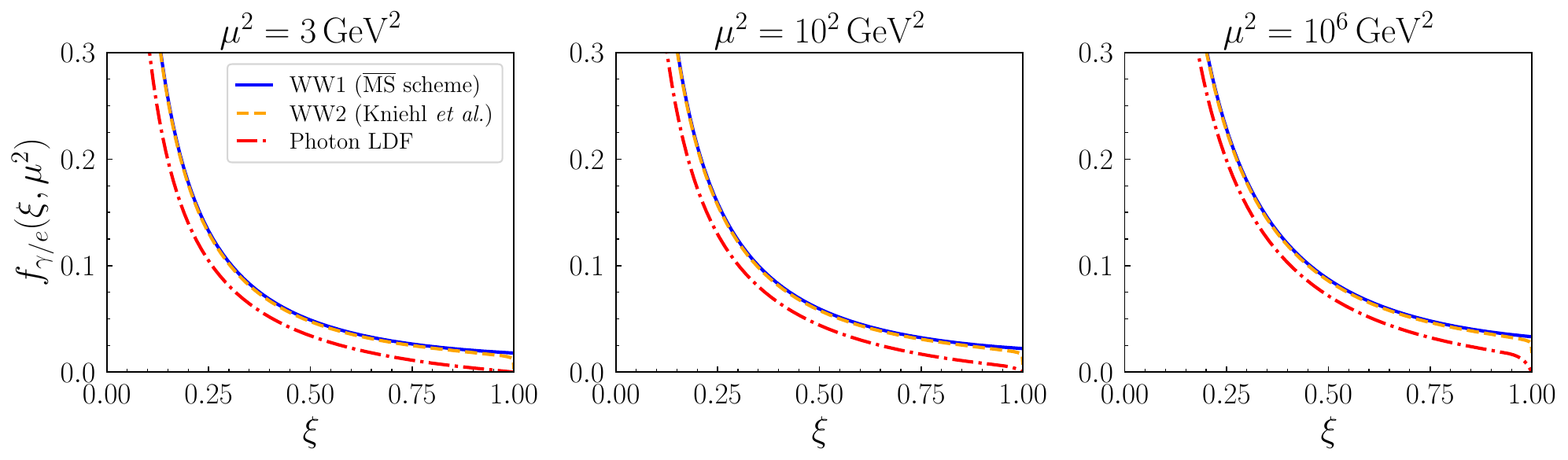}
\caption{Numerical comparison between the WW distributions in Eqs.~\eqref{eq:ww-distribution} and \eqref{eq:ww-distribution2} and the photon LDF including quantum evolution effects in $\xi$-space for given values of $\mu^2$. The electron LDF is also shown for comparison. \textbf{Top:} Log-log scale. \textbf{Bottom:} Linear scale.}
\label{fig:WWmodel}
\end{center}
\end{figure}

With the mixed evolution between QED and QCD, we expect that the photon distribution of our default LDF set differs from the perturbatively calculated WW photon distribution in Eq.~\eqref{eq:ww-distribution} or in Eq.~\eqref{eq:ww-distribution2}, which have been commonly used for calculations of photoproduction at HERA. For a quantitative comparison, Fig.~\ref{fig:WWmodel} shows both WW photon distributions, along with the photon distribution of our default LDFs.   

As expected, our default photon distribution obtained by solving the evolution equations in both QCD and QED is significantly smaller than the perturbatively calculated WW photon distribution in QED due to the generation of quark and gluon distributions of the LDFs. The difference is enhanced at lower values of $\xi$, as evident in Fig.~\ref{fig:WWmodel} (top). In addition, the perturbatively calculated WW photon distribution does not vanish as $\xi\to 1$, while the photon distribution of our LDFs vanishes by default, as shown in Fig.~\ref{fig:WWmodel} (bottom). This visible difference in the photon distributions could lead to discrepancies in numerical calculations or predictions of single hadron photoproduction in lepton-hadron scattering.

\section{Single inclusive hadron production}
\label{sec:single-hadron}

For numerical evaluations of single inclusive light hadron production in lepton-hadron collisions, we need universal hadron PDFs and FFs in addition to the LDFs discussed in the last section. In the joint QCD+QED factorization approach, similar to the evolution of LDFs, the evolution of PDFs and FFs also has evolution kernels calculated in both QCD and QED, as well as kernels for mixed evolution between partons and leptons, which naturally, for example, lead to the photon distribution of the hadron~\cite{NNPDF:2024djq,Manohar:2024ndc}.  

As pointed out in Ref.~\cite{Cammarota:2025jyr}, having the photon distribution of a colliding hadron is necessary and critical to ensure the infrared safety of the short-distance hard coefficient functions of factorized lepton-hadron scatterings beyond the LO in QED in this joint QCD+QED factorization approach.  The cancellation of infrared divergence associated with collision-induced photon radiation requires us to include QED radiation from all involved particles with electric charges, such as quarks and electrons, and their antiparticles, at any given order of perturbative calculations. In addition, it has been recognized that the photon distribution of a colliding hadron at LHC energies could introduce percent-level numerical uncertainties to the precision tests and the search for new physics beyond the Standard Model (BSM)~\cite{NNPDF:2024djq,Manohar:2024ndc,Martin:2004dh,Roth:2004ti,Manohar:2016nzj}.  

However, for the energies of lepton-hadron scatterings to be discussed in this paper, the numerical impact of the photon distribution of the colliding hadron should be much smaller than that at LHC energies. 
As a reasonable approximation, we will neglect such small uncertainties from the photon distribution in hadron PDFs and FFs in our numerical calculations below.
In addition, we found that the numerical impact on the evolution of LDFs from QCD splitting functions is much stronger than the influence of QED splitting functions on the evolution of PDFs and FFs due to the size of the respective coupling constants.  Therefore, we will use the LDFs with full QCD and QED evolution at the first non-trivial order for our numerical calculations in this Section. 
Regarding proton PDFs and hadron FFs, we will employ the available JAM20 set~\cite{Moffat:2021dji}.
For the comparison, we will also use 
CT18 set~\cite{Hou:2019efy} for proton PDFs and 
MAP1.0 set~\cite{Khalek:2021gxf} for hadron FFs.  

\subsection{Photoproduction in lepton-hadron collisions}
\label{subsec:photoproduction}

For the production of single inclusive hadron at high-$P_T$ in lepton-hadron collisions, $e(\ell)+h(p)\to H(P)+X$, the perturbatively calculable hard parts $\widehat{H}_{ib\to cX}(\xi,x,z, p_c)$ in Eq.~\eqref{eq:lh-fac} obtain the leading order contribution from subprocesses, $e+q\to e+q$ with different quark flavors $q$, where the produced quark fragments to the observed hadron $H$ and the scattered lepton has a large transverse momentum to balance the transverse momentum of the observed hadron. The hard parts also obtain the leading order contribution from subprocesses, $\gamma+q\to q+g$, where $\gamma$ is treated as a real photon and the produced quark or gluon fragments to the observed hadron via $D_{q\to H}$ and $D_{g\to H}$, respectively. We refer to all contributions from subprocesses involving a real photon from the colliding lepton as contributions from photoproduction in lepton-hadron collisions.  

With the scattered lepton measured in lepton-hadron collisions, the exchange photon of momentum $q^\mu=(\ell - \ell')^\mu$ between the colliding lepton and hadron is virtual, and its virtuality $Q^2=-q^2$ depends on the transverse component of the scattered lepton momentum $\ell'_T$. 
The larger the transverse momentum of the scattered lepton $\ell'_T$ is, the more off-shell the exchanged virtual photon is, and the more suppressed the measured lepton-hadron cross section is, since it is inversely proportional to the virtuality of the exchanged photon $Q^2$.
Therefore, the cross section for single inclusive high-$P_T$ hadron production in lepton-hadron collisions could be enhanced if the exchanged photon is nearly on-shell ($Q^2 \to 0$). The photoproduction of high-$P_T$ jet(s) or a single hadron in lepton-hadron collisions was introduced to describe events at HERA with large momentum transfer, 
where the virtuality of the exchanged photon $Q^2$ remains very small. The photon virtuality $Q^2$
is controlled by imposing an experimental cut on the transverse momentum of the scattered lepton $\ell'_T$ to limit the value of $Q^2$ to be less than a pre-chosen number (or a parameter).  

The terms ``Direct'' and ``Resolved'' photon contributions were also introduced to describe the photoproduction in lepton-hadron collisions at the HERA~\cite{Chyla:1993xj},
\begin{equation}
    d\sigma_{eh\to e+Jet+X}
    \approx f_{\gamma/e}\otimes 
    \sum_b
    \left[ d\hat{\sigma}_{\gamma b\to Jet+X}^\mathrm{Direct} + 
    \sum_a f_{a/\gamma} \otimes d\hat{\sigma}_{a b\to Jet+X}^\mathrm{Resolved}\right] \otimes f_{b/h}\, ,
    \label{eq:resolved}
\end{equation}
where $\otimes$ represents the convolution of a momentum fraction in the same way as in Eq.~\eqref{eq:lh-fac}, and the photon distribution of an unpolarized electron $f_{\gamma/e}$ was approximated by various WW distributions~\cite{Hinderer:2015hra,Kniehl:1996we}. To ensure the dominance of photoproduction (at very small $Q^2$), an experimental cut was applied to the scattered final-state electron $e$. The so-called ``Resolved" photon contribution on the right-hand side of Eq.~\eqref{eq:resolved} should be small 
when the $Q^2$ is large, such that the exchanged photon has less time to fluctuate into a hadronic state. The resolved photon contribution should be more important when $Q^2$ is small. However, the fluctuation for the photon to a partonic state, $f_{a/\gamma}$ in Eq.~\eqref{eq:resolved}, could be nonperturbative.
On the other hand, the resolved photon contribution is naturally taken care of in the joint QCD+QED factorization approach as a part of the $\sum_i$ with $i=q,\bar{q},g$ of factorization formula in Eq.~\eqref{eq:lh-fac} without further approximating the $f_{i/e}$ to $f_{\gamma/e}\otimes f_{i/\gamma}$. That is, the joint QCD+QED factorization approach to single inclusive high-$P_T$ hadron or jet production in lepton-hadron collisions naturally contains all contributions, including the so-called photoproduction, without the need to introduce an experimental cut (or parameter) to help select the ``photon'' contribution, which could introduce additional uncertainty from potentially large logarithms associated with this extra parameter.  Furthermore, the joint QCD+QED factorization approach does not require introducing so-called resolved photon contributions and photon structure. Those are naturally taken care of by the factorization formalism without the need to make additional approximations to isolate these contributions.

\subsection{Kinematics}

\begin{figure}[t]
\begin{center}
\includegraphics[width=\textwidth]{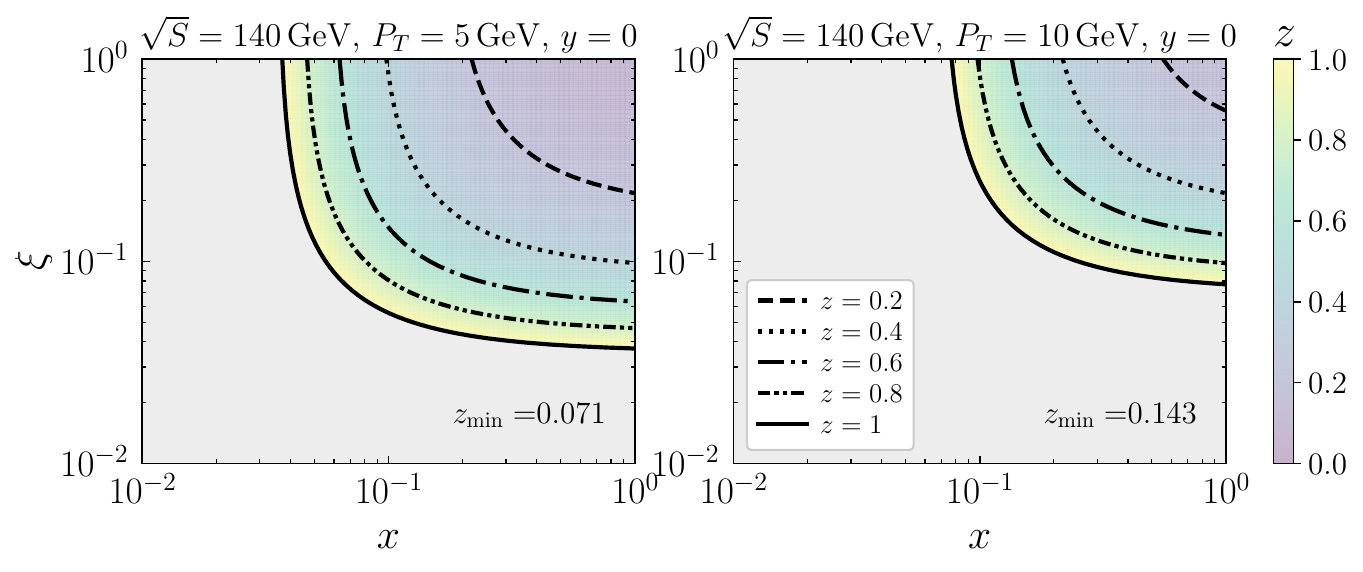}
\includegraphics[width=\textwidth]{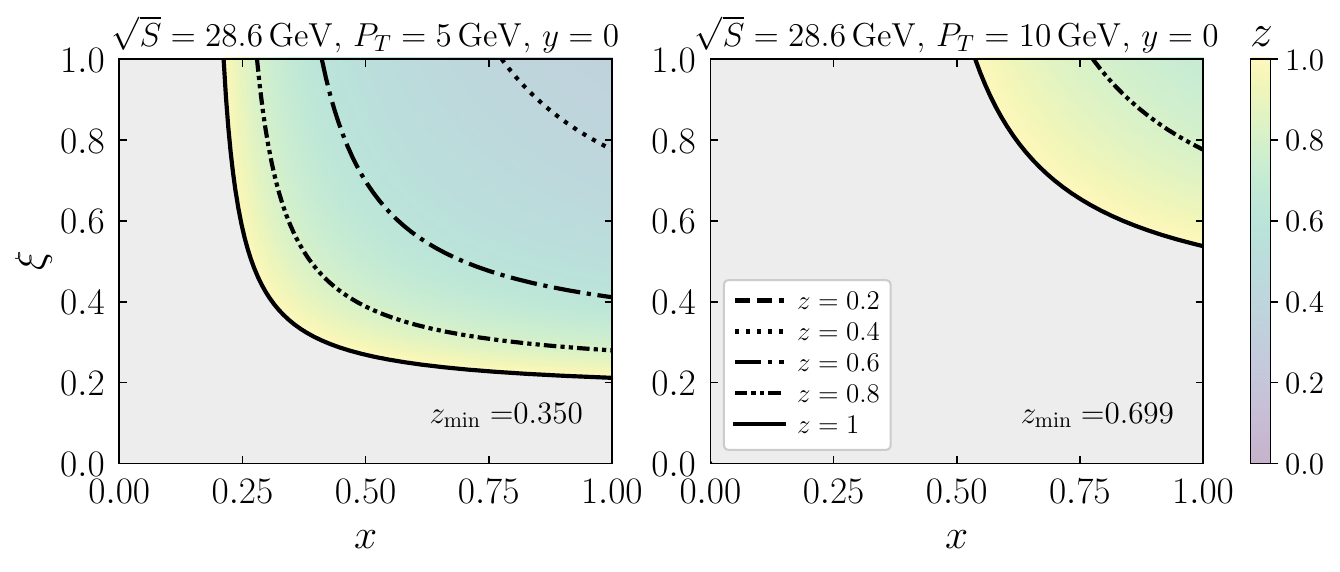}
\caption{Allowed kinematic regions in the $(\xi, x)$ plane for LO partonic scattering at the EIC at $\sqrt{S}=140\,\mathrm{GeV}$ \textbf{(top)} and $\sqrt{S}=28.6\,\mathrm{GeV}$ \textbf{(bottom)}. 
}
\label{fig:kinematics}
\end{center}
\end{figure}

In the joint QCD+QED factorization approach, both leptons and photons participate in lepton-parton scattering processes in lepton-hadron collisions, similar to parton-parton scattering processes in hadron-hadron collisions. It would be worthwhile to clarify the allowable kinematic regions for such partonic scatterings. For the time being, we shall restrict our discussion to the LO kinematics. The momentum conservation at LO, $\hat{s}+\hat{t}+\hat{u}=0$ 
with $\hat{t}=(x/z)T$ and $\hat{u}=(\xi/z)U$,
leads to
\begin{align}\label{eq:kinematics}
\xi=\frac{xP_{T}\,e^{+y}}{zx\sqrt{S}-P_{T}\,e^{-y}},
\end{align}
with transverse momentum \pT\ and rapidity $y$ of the observed hadron. 
Since momentum fractions $\xi \leq 1$, $z \leq 1$, and $x\leq 1$, kinematically, one finds the lower limit of the momentum fractions $z$ and $x$ on the right-hand side of Eq.~\eqref{eq:kinematics},
\begin{equation}\label{eq:limits-lo}
    z_\mathrm{min} = \frac{2P_{T}}{\sqrt{S}}\, \cosh(y)
    \quad\quad \text{and}\quad\quad 
    x_\mathrm{min}=\frac{P_{T}\,e^{-y}}{z\sqrt{S}-P_{T}\, e^{+y}}\, ,
\end{equation}
which are consistent with the general limits in Eq.~\eqref{eq:limits} beyond the LO kinematics.
In Fig.~\ref{fig:kinematics} we show kinematic range of incoming lepton/parton momentum fractions ($\xi,x$) for given values of $z$, which are kinematically limited by Eqs.~\eqref{eq:kinematics} and \eqref{eq:limits-lo} at various values of $P_{T}$, $y$ and collision energy $\sqrt{S}$ at the EIC.  With $z$ as the momentum fraction of the final-state parton produced, carried by the observed hadron, large values of $z$ allow us to access smaller values of $\xi$ and $x$, which are clearly evident in Fig.~\ref{fig:kinematics}. 
With sufficiently high $P_{T}$ to ensure factorization, unlike inclusive DIS, hadron production at EIC energies will not be able to access the small-$x$ regime, as shown in Eq.~\eqref{eq:limits-lo} or in Fig.~\ref{fig:kinematics}. As discussed in Appendix~\ref{app:high_z}, like collisions between hadrons, the observed single hadron production cross sections are also dominated by the phase space where $z$ is large while $x$ is relatively small due to the unique features of hadron PDFs: steep growth when $x\to 0$ and fast fall when $x\to 1$.

As illustrated in Fig.~\ref{fig:kinematics}, the interval $10^{-2} < x < 1$ constitutes the primary kinematic domain for single hadron production at mid-rapidity for EIC energies. Within this region, the $u$-quark density in the proton dominates over other quark flavors across the entire range, whereas the gluon density overwhelms the quarks at $x \lesssim 0.1$. This behavior is consistently observed across several PDF sets, including JAM20~\cite{Moffat:2021dji} and NNPDF3.1~\cite{NNPDF:2017mvq}, the latter of which was used to extract the MAP1.0 FF set~\cite{Khalek:2021gxf}.

\subsection{Numerical results}

\begin{figure}[t]
\begin{center}
\includegraphics[width=\textwidth]{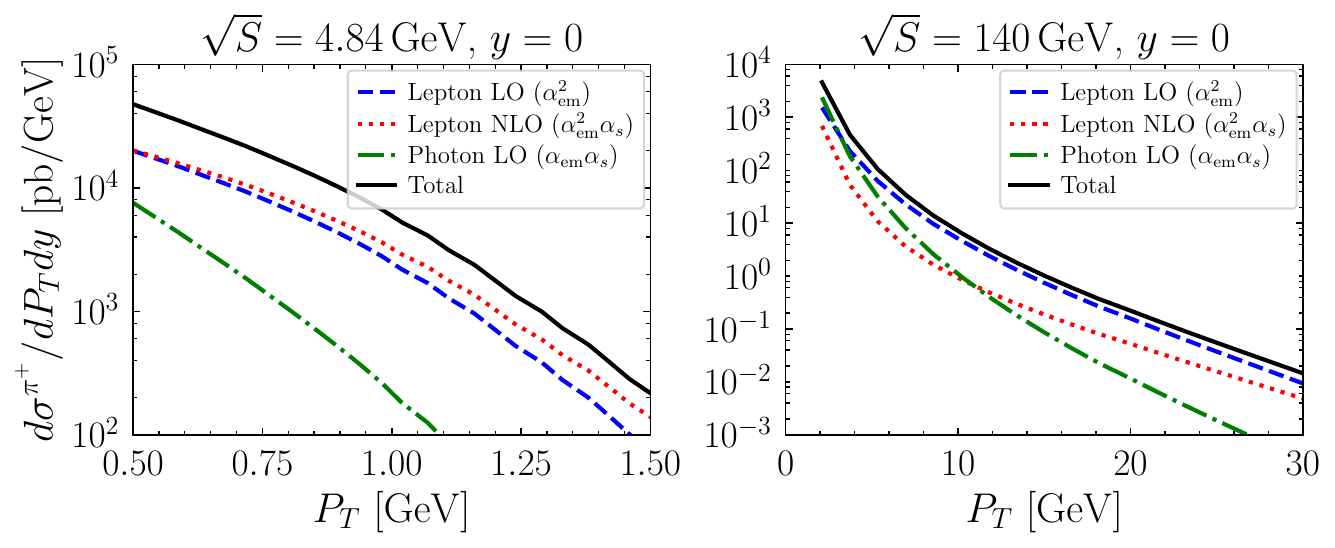}
\caption{Differential cross section for inclusive $\pi^+$ production as a function of $P_T$ in electron-proton scatterings at JLab \textbf{(left)} and EIC \textbf{(right)}.
When $P_T < m_c$, the scale was fixed at $\mu=m_c$, freezing the evolution of the universal functions.
}
\label{fig:pip-production-pT}
\end{center}
\end{figure}
\begin{figure}[t]
\begin{center}
\includegraphics[width=\textwidth]{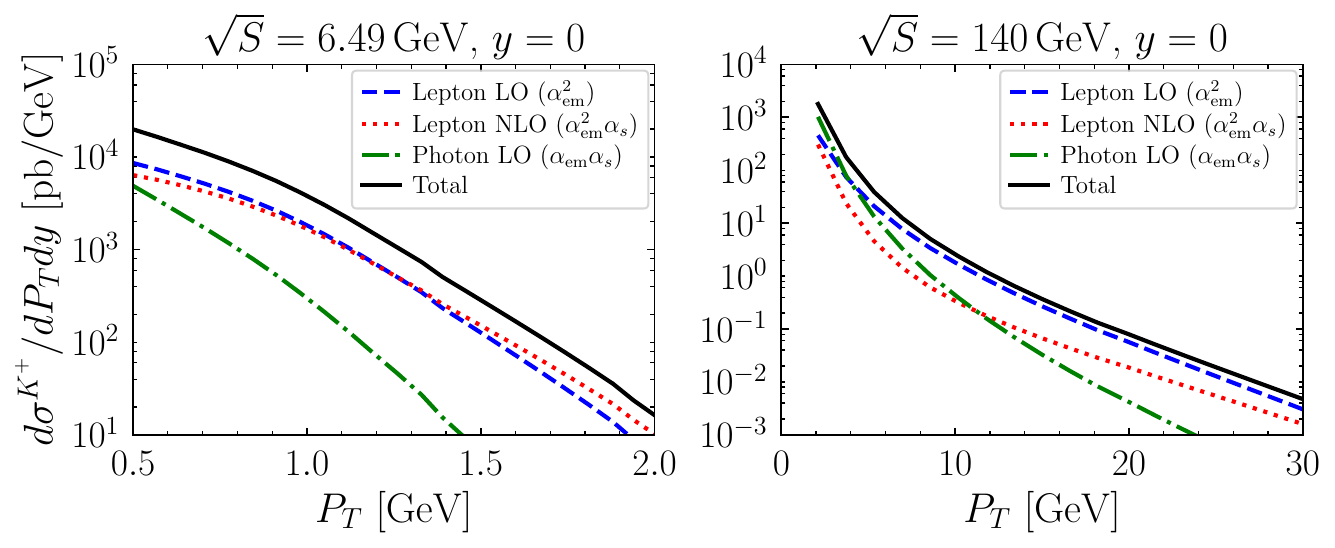}
\caption{Same as Fig.~\ref{fig:pip-production-pT}, but for inclusive $K^+$ production. When $P_T < m_c$, the scale was fixed at $\mu=m_c$, freezing the evolution of the universal functions. For JLab, predictions for the possible $22\,\mathrm{GeV}$ upgrade plan are shown.}
\label{fig:kp-production-pT}
\end{center}
\end{figure}

We use the factorized differential cross-section in Eq.~\eqref{eq:lh-fac} to compute single inclusive high-$P_T$ hadron ($H=\pi^+, K^+$) production in electron-proton ($ep$) collisions without tagging a scattered electron. With different parton-to-hadron FFs, we could also compute production rates for inclusive $\pi^-$, $K^-$, or proton production in $ep$ collisions.
The partonic hard parts $\widehat{H}$ of the LO processes were first studied in Ref.~\cite{Kang:2011jw}. The calculations at NLO and NNLO were then performed in Refs.~\cite{Hinderer:2015hra,Abelof:2016pby}, respectively.
See also \cite{Hinderer:2017ntk} for NLO calculations of polarized cross sections.
In this paper, we kept the partonic hard parts $\widehat{H}_{ib\to c(p_c)X}$ defined in Eq.~\eqref{eq:H} at the NLO accuracy and used the results consistent with those derived in Ref.~\cite{Hinderer:2015hra}, 
where all perturbative collinear divergences in NLO partonic scattering cross sections are factorized into the non-perturbative universal PDFs, FFs, and LDFs.
This choice is also consistent with the accuracy of PDFs and FFs that we used.

In Fig.~\ref{fig:pip-production-pT}, we present numerical results for the single inclusive $\pi^+$ production in $ep$ collision at the CEBAF energy at JLab (left panel) and one of future EIC energies (right panel), which are calculated by using the central set of JAM20 proton PDFs and pion FFs~\cite{Moffat:2021dji}, along with our default LDFs. 
We set the renormalization, factorization, and fragmentation scales,
$\mu_r^2=\mu_e^2=\mu_h^2=\mu_H^2=P_{T}^2$ with $P_{T}$ being the transverse momentum of the produced hadron. 
For the inclusive hadron production of $P_T$ less than the charm quark mass $m_c$, we imposed $\mu$ to be $m_c$ ($>P_T$), so that the evolution of the universal functions is frozen in our numerical calculations. This prescription also holds for any numerical results shown below.
We will discuss the scale dependence of our calculated results later in this section.

Keeping the partonic hard parts at the LO in QED, the leading-power collision-induced and process-independent QED radiative corrections are represented by the LDFs. As described in the previous section, our default LDFs were derived by solving the evolution equations with both QCD and QED evolution kernels, including mixing evolution between leptons and partons (i.e., between QCD and QED).
For perturbatively calculated partonic hard parts, subprocesses such as $e q\to e q$ with different quark flavors are referred to as ``LO $(\alpha_{em}^2)$'', and $eq\to e q+g$ 
with either the produced quark or gluon fragmenting to the observed $\pi^+$
as ``NLO $(\alpha_{em}^2\alpha_s)$''. Those are also used to label the plotted curves generated with the corresponding hard subprocesses in Fig.~\ref{fig:pip-production-pT}. In addition, subprocesses involving the real photon from the unpolarized electron, such as $\gamma q\to qg$ with either the produced quark or gluon fragmenting to $\pi^+$, are referred to as the LO short-distance contribution from photoproduction, labeled as ``Photon $(\alpha_{em}\alpha_s)$" in the plots. We stress that the photoproduction of $\pi^+$ is a natural part of the $ep$ collisions, and the corresponding subprocesses are not only allowed but also treated on the same footing as the lepton production in the joint QCD+QED factorization formalism.

\begin{figure}[t]
\begin{center}
\includegraphics[width=\textwidth]{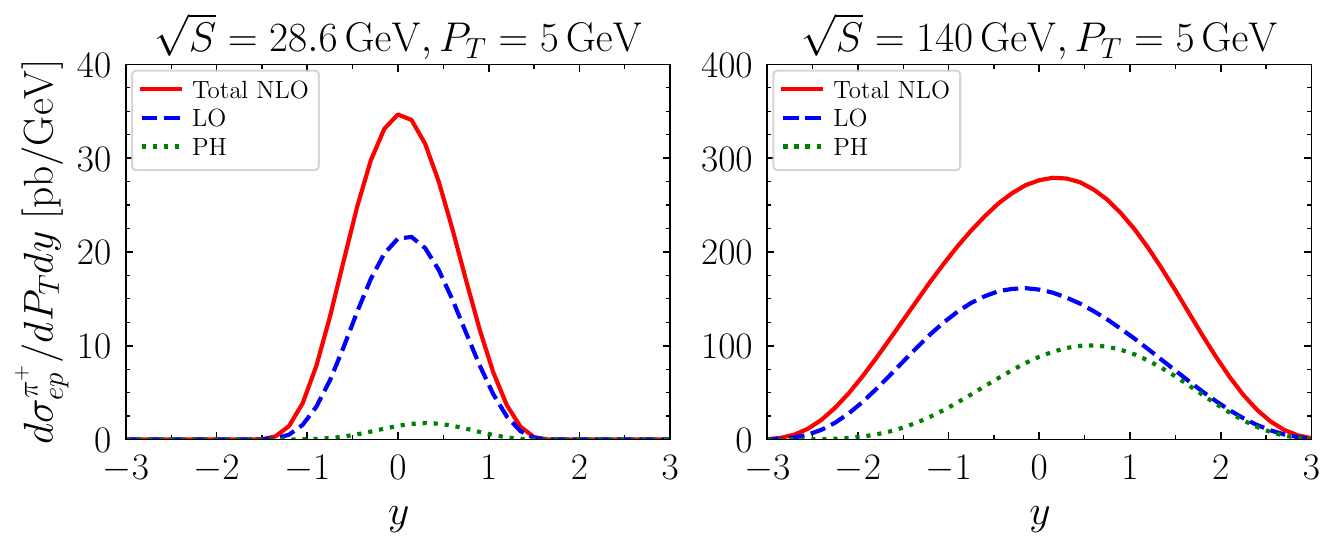}
\includegraphics[width=\textwidth]{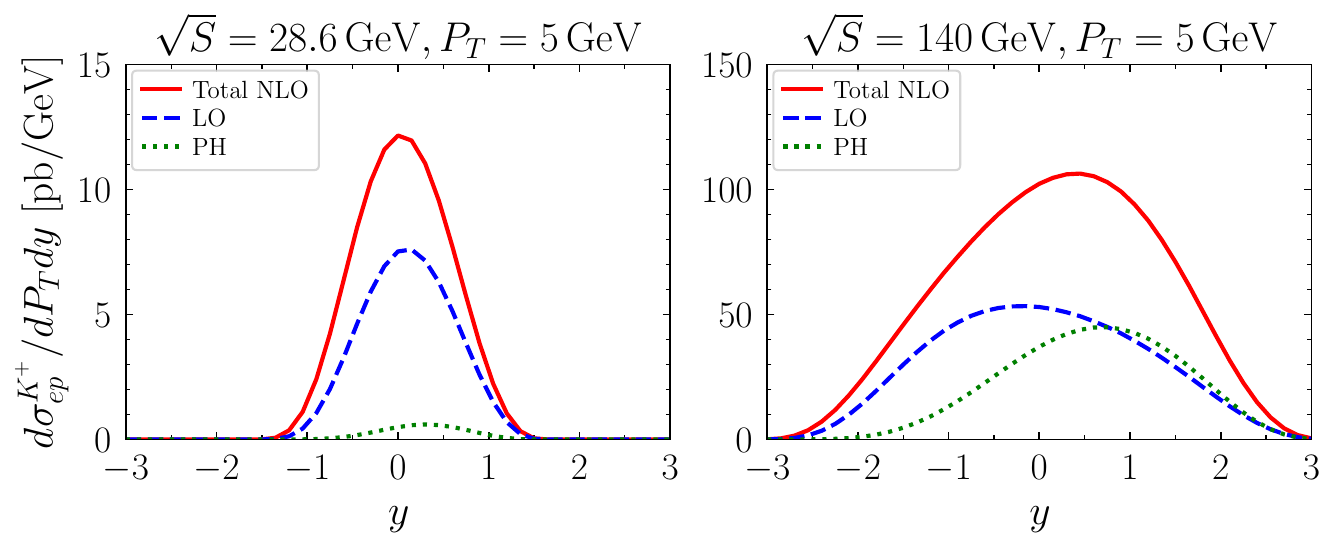}
\caption{Rapidity distributions of the inclusive $\pi^+$ (\textbf{top}) and $K^+$ (\textbf{bottom}) production cross sections at fixed $P_{T}$ in electron-proton collisions at two different EIC energies.}
\label{fig:hadron-production-eta}
\end{center}
\end{figure}

The $\pi^+$ leptoproduction at LO and NLO is the main production process at high $P_{T}$, while the photoproduction is predominant over the leptoproduction at low $P_{T}$. For the first time, the leptoproduction and photoproduction of a hadron in $ep$ collisions are described naturally in the unified factorization formalism without introducing any kinematical and experimental cuts to separate them. One could apply the QCD+QED joint factorization formalism to other phenomenology without introducing additional constraints and parameters to separate the so-called ``Direct" and ``Resolved" photon contributions at HERA energies, as discussed in Sec.~\ref{subsec:photoproduction}. In addition, as described in the last section and shown in Fig.~\ref{fig:WWmodel}, the photon distribution of our default LDFs, which is evolved with both QCD and QED evolution kernels, differs from the various WW distributions used in photoproduction calculations at HERA.

Similarly, Fig.~\ref{fig:kp-production-pT} shows our numerical results for single inclusive $K^+$ production in $ep$ collision at CEBAF and EIC energies comparable to those in Fig.~\ref{fig:pip-production-pT}
by using the JAM20 hadron FFs for producing a $K^+$~\cite{Moffat:2021dji}. We can see a similar pattern of contributions from various orders or partonic scattering channels.

Figure~\ref{fig:hadron-production-eta} shows the rapidity distribution of single inclusive $\pi^+$ and $K^+$ production at $P_T=5\,\mathrm{GeV}$ in $ep$ collisions at two different EIC energies, $\sqrt{S}=28.6\,\mathrm{GeV}$ and 140\,GeV, covering the expected range of collision energies. 
At $P_T=5\,\mathrm{GeV}$, the contribution from photoproduction could be as important as that from leptoproduction at the high end of the EIC energy.

\begin{figure}[t]
\begin{center}
\includegraphics[width=\textwidth]{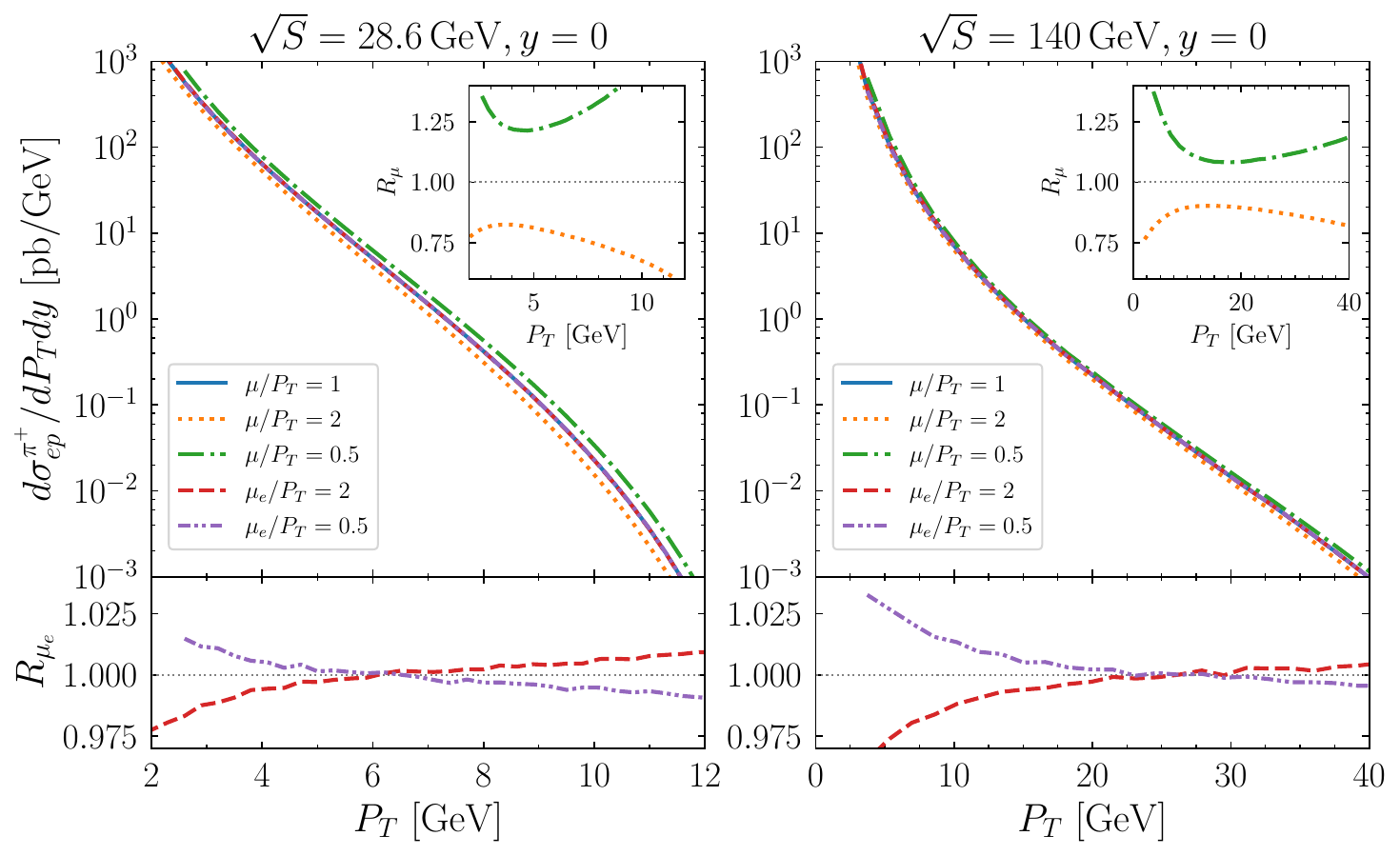}
\caption{Scale dependence of the $\pi^+$ production $P_T$ spectrum at $\sqrt{S}=28.6\,\mathrm{GeV}$ \textbf{(left)} and $140\,\mathrm{GeV}$ \textbf{(right)}. 
See text for details.}
\label{fig:mu-dep}
\end{center}
\end{figure}

We show in Fig.~\ref{fig:mu-dep} the scale dependence of the inclusive $\pi^+$ production cross section in the mid-rapidity region at EIC energies, where 
the $\pi^+$ $P_T$ spectra in solid, dotted, and dot-dashed lines were obtained by setting $\mu=\mu_r=\mu_e=\mu_h=\mu_H=\{1, 2,0.5\} P_T$, respectively.
The insets show the ratio of spectra calculated with $\mu\neq P_T$ to that with $\mu=P_T$, denoted by $R_\mu$. At $\sqrt{S}=28.6\,\mathrm{GeV}$, the uncertainty associated with $\mu$ is about 25\% at low $P_T$, but increases with $P_T$. Meanwhile, at $\sqrt{S}=140\,\mathrm{GeV}$, the uncertainty becomes smaller, especially at high $P_T$ (about 20\% at $P_T=40\,\mathrm{GeV}$). The bottom windows show $R_{\mu_e}$, the ratio of spectra calculated with only $\mu_e\neq P_T$ to that with $\mu=P_T$. It is clear from the figure that the uncertainty associated with the choice of $\mu_e$ for LDFs is of the order of a few \% but smaller than that for PDFs and FFs because the QCD evolution is much faster than the QED evolution.

\subsection{Leading Power QED Radiative Corrections}

\begin{figure}[t]
\begin{center}
\includegraphics[width=\textwidth]{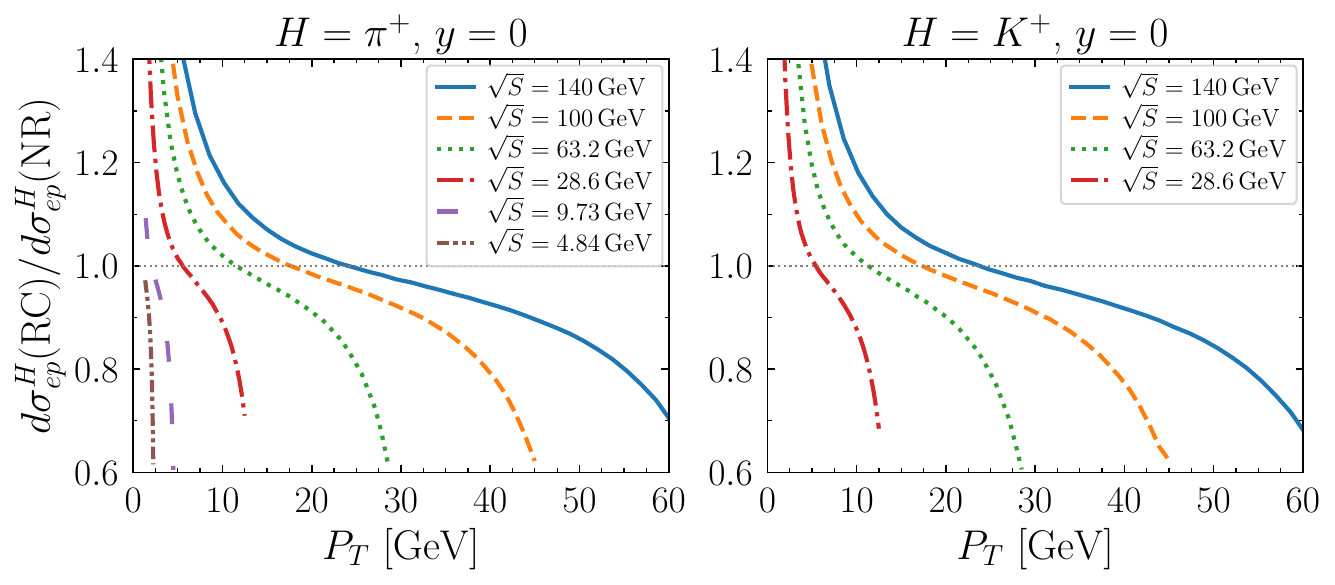}
\caption{Ratio of the inclusive hadron $H$ production $P_{T}$ spectrum with QED radiative corrections (RC) to that without corrections (NR) for various collision energies from CEBAF to EIC. \textbf{Left:} $H=\pi^+$. \textbf{Right:} $H=K^+$.}
\label{fig:RCs-hadron-allrs}
\end{center}
\end{figure}

In the joint QCD+QED factorization approach to single inclusive hadron production in lepton-hadron collisions, 
the collision-induced QED contributions fall into three distinct types.
The first is leading-power, process-independent, or universal contributions in the $1/P_T$ power expansion of the inclusive cross sections. 
They are systematically included in the universal LDFs, as well as in the PDFs and FFs, if we include QED evolution kernels to evaluate the factorization scale dependence of those universal functions.
The second is leading-power process-dependent, but perturbatively calculable contributions to the factorized short-distance hard coefficients in powers of $\alpha_{em}$. 
The third is power-suppressed contributions of order $1/P_T^n$ with $n$ depending on whether polarized or unpolarized cross sections. Such power corrections may not be factorizable, which represent the uncertainty of the joint QCD+QED factorization formalism.
If we keep the LO QED contributions to the factorized hard coefficients and the scale dependence of PDFs and hadron FFs, as an approximation, 
since $\alpha_{em} \ll \alpha_s$ for $\mu\sim P_T$, significant leading-power QED contributions to the single inclusive hadron production in lepton-hadron collisions are mainly from the universal and process-independent LDFs.
It is the universality of the LDFs that provides the predictive power for the QED contributions to the single inclusive hadron production in lepton-hadron collisions, or the so-called QED radiative corrections, traditionally at the CEBAF of JLab and the future EIC.

To study the impact of such QED radiative corrections or QED contributions, 
Fig.~\ref{fig:RCs-hadron-allrs} displays the ratio of cross sections with $d\sigma^H_{ep}(\mathrm{RC})$, evaluated with our default LDFs that include the leading-power collision-induced QED radiation, and $d\sigma^H_{ep}(\mathrm{NR})$,
evaluated by setting electron LDF $f_{e/e}(\xi) = \delta(1-\xi)$ and all other flavor LDFs to be zero, which effectively removes all collision-induced QED radiative corrections. As a general feature, collision-induced QED radiation reduces the probability of finding an electron as $\xi\to 1$. 
Consequently, it reduces the production cross section at high $P_{T}$, while generating a non-vanishing probability of finding an electron and a radiated photon at lower $\xi$ in the parent beam electron, leading to the enhancement of the cross sections at lower $P_{T}$. Figure~\ref{fig:RCs-hadron-allrs} shows clearly that the collision-induced QED radiation significantly changes the $P_T$ distribution of the hadron production cross section.
The change in the hadron $P_T$ spectrum can be very large, as large as 50\% or even more, depending on the value of $P_T$, and sensitive to the kinematics and the phase space available for the collision-induced QED radiation. 
The notable advantage of the joint QCD+QED factorization approach is that such a large impact on theoretical calculations of single inclusive hadron production cross sections is precisely predictable, once we determine the universal LDFs.

\subsection{Uncertainty from various sets of FFs}
\label{subsec:ffs}

\begin{figure}[t]
\begin{center}
\includegraphics[width=\textwidth]{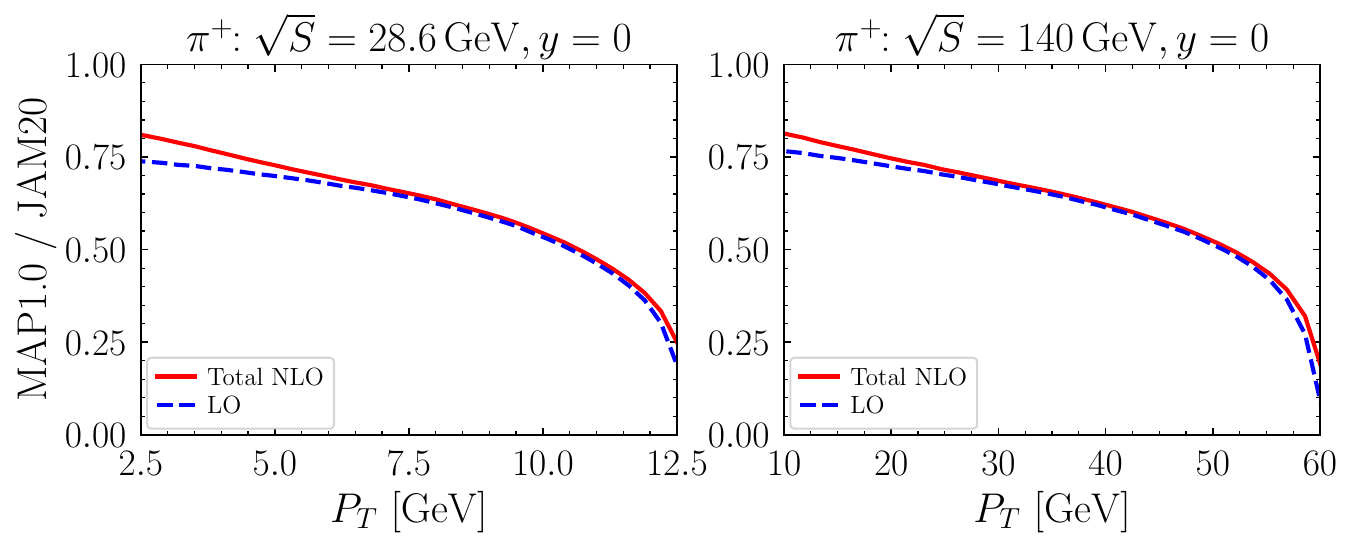}
\caption{Ratio of the $\pi^+$ $P_T$ spectrum at $y=0$ calculated with the MAP1.0 FF set~\cite{Khalek:2021gxf} to that with the JAM20 FF set~\cite{Moffat:2021dji} at $\sqrt{S}=28.6\,\mathrm{GeV}$ \textbf{(left)} and $140\,\mathrm{GeV}$ \textbf{(right)}. Dashed-blue curves: LO contribution. Solid-red curves: Total NLO contribution, including photoproduction. }
\label{fig:pion-JAM-MAP-pT}
\end{center}
\end{figure}
\begin{figure}[t]
\begin{center}
\includegraphics[width=\textwidth]{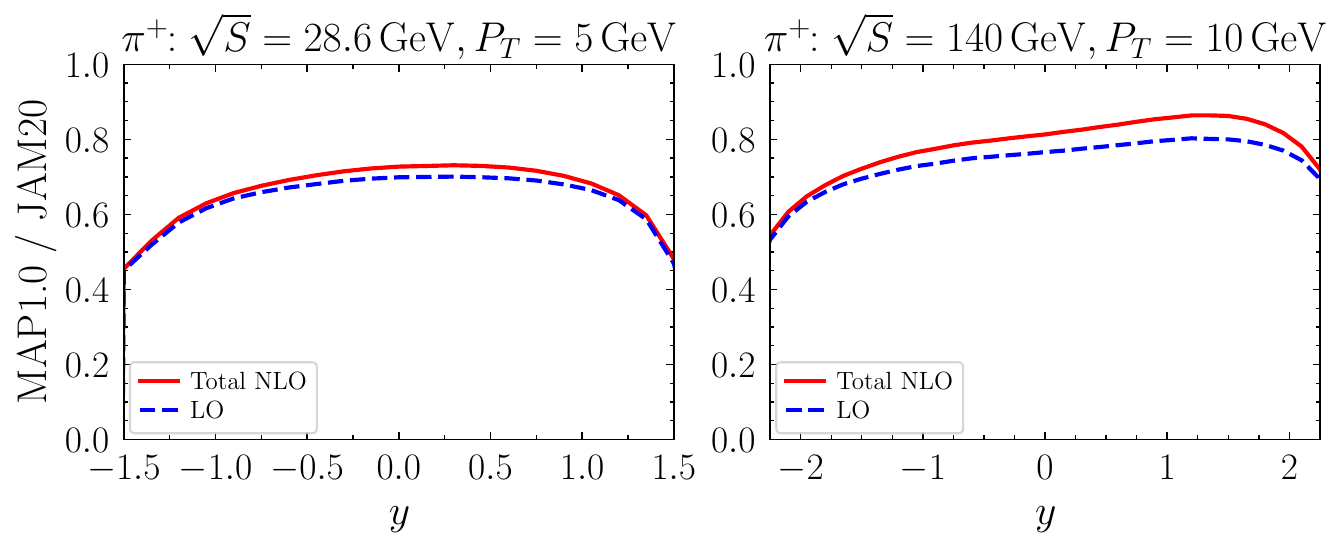}
\caption{Same as Fig.~\ref{fig:pion-JAM-MAP-pT}, but for the rapidity dependence of the $\pi^+$ production cross section at the EIC for different collision energies. \textbf{Left:} $\sqrt{S}=28.6\,\mathrm{GeV}$ and $P_T=5\,\mathrm{GeV}$. \textbf{Right:} $\sqrt{S}=140\,\mathrm{GeV}$ and $P_T=10\,\mathrm{GeV}$. }
\label{fig:pion-JAM-MAP-eta}
\end{center}
\end{figure}
\begin{figure}[t]
\begin{center}
\includegraphics[width=\textwidth]{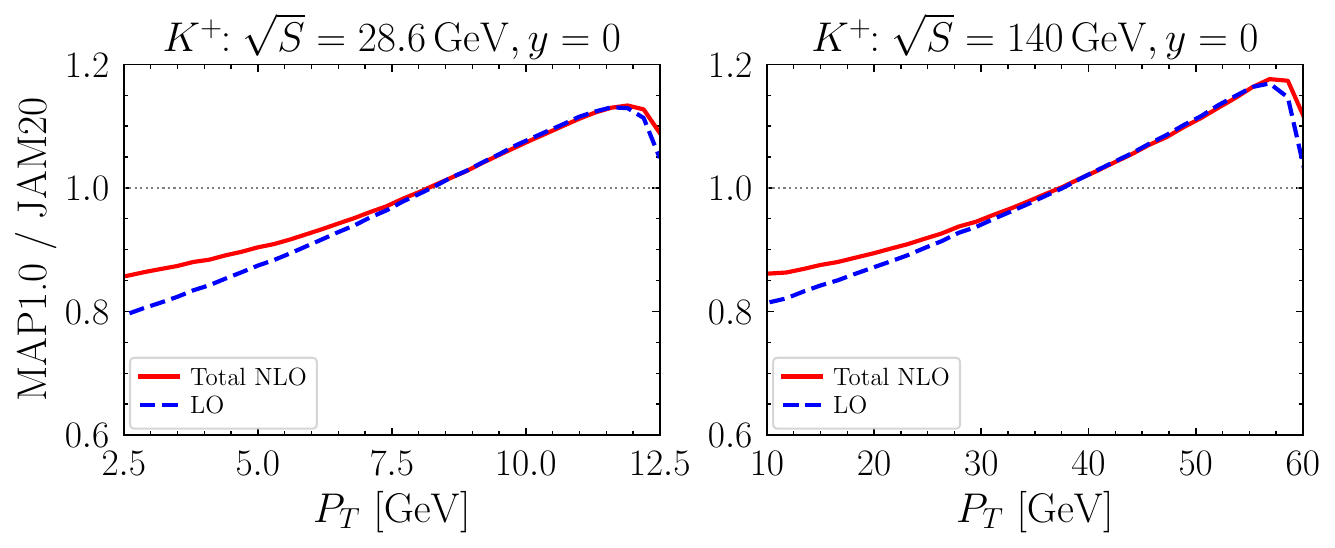}
\caption{Same as Fig.~\ref{fig:pion-JAM-MAP-pT}, but for the inclusive $K^+$ production at EIC for different collision energies.}
\label{fig:kaon-JAM-MAP-pT}
\end{center}
\end{figure}
\begin{figure}[t]
\begin{center}
\includegraphics[width=\textwidth]{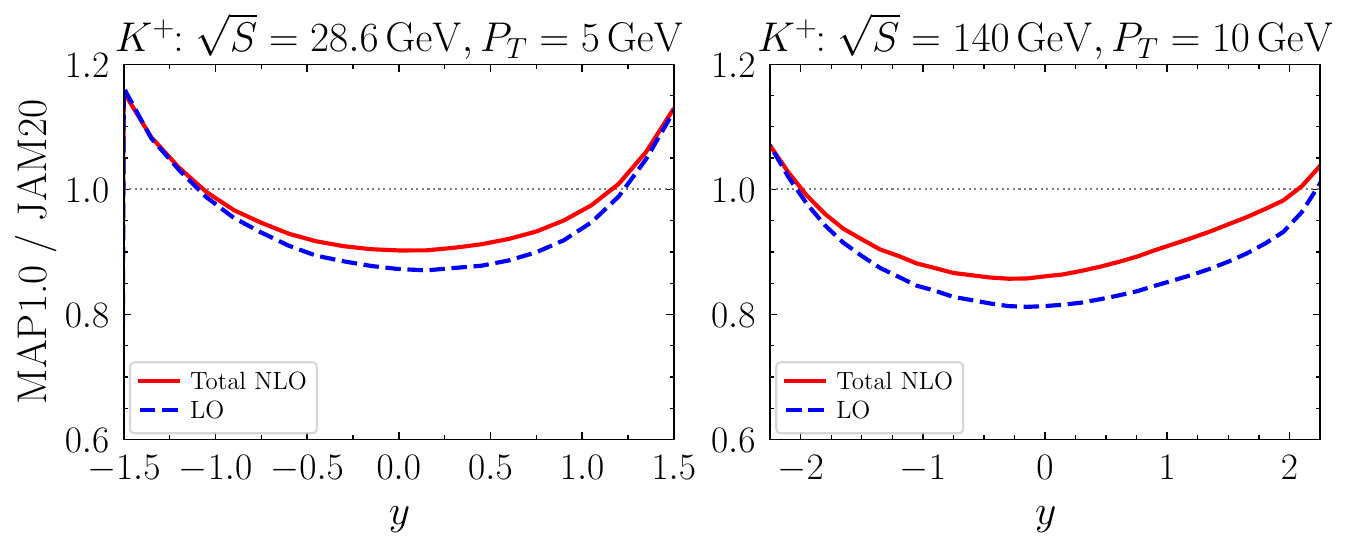}
\caption{Same as Fig.~\ref{fig:pion-JAM-MAP-eta}, but for the inclusive $K^+$ production at the EIC for different collision energies.}
\label{fig:kaon-JAM-MAP-eta}
\end{center}
\end{figure}

In addition to the uncertainty associated with the choice of renormalization and factorization scales, there is a systematic uncertainty arising from different sets of the universal functions, such as PDFs and FFs extracted from global fits. Several groups have performed QCD global analyses of existing data to extract PDFs available in the literature; their results are broadly consistent within uncertainties, especially at high $\mu^2$. 
We have confirmed that the dominant $u$-quark density from the JAM20 proton PDF central set agrees within 10\% with the NNPDF3.1 central set~\cite{NNPDF:2017mvq}, which was used to extract the MAP1.0 FF set, in the dominant kinematic region $10^{-2} < x \lesssim 0.5$ at $\mu=20\,\mathrm{GeV}$. Meanwhile, a small discrepancy of a few \% between these two sets is observed for the gluon density at $10^{-2} < x < 0.1$ at the same scale, where the gluon density dominates over the quark distributions.
In contrast, existing sets of FFs extracted for light-hadron production still carry large uncertainties, as detailed in Appendix~\ref{app:ffs}.

To demonstrate the uncertainty of high-$P_T$ single hadron production in lepton-hadron collisions, caused by the difference between various FFs, we show in Fig.~\ref{fig:pion-JAM-MAP-pT} the ratio of the $\pi^+$ $P_T$ spectrum calculated with the MAP1.0 FF central set~\cite{Khalek:2021gxf} to that with the JAM20 FF central set~\cite{Moffat:2021dji} at $y=0$ at $\sqrt{S}=28.6\,\mathrm{GeV}$ and $140\,\mathrm{GeV}$, where the JAM20 PDF central set is used for both figures. Figure~\ref{fig:pion-JAM-MAP-pT} shows clearly that the ratio is far from unity, indicating a major impact of the uncertainty associated with the choice of hadron FF sets.
It is also noted that the deviation from unity increases with $P_T$.  
As shown in Fig.~\ref{fig:zmax} of Appendix~\ref{app:high_z}, the $\pi^+$ cross section at high $P_T$ is dominated by the large-$z$ behavior of the FFs.
The fact that the MAP1.0 FF central set for $u\to \pi^+$ is slightly smaller than the JAM20 central set at large $z$, as shown in Fig.~\ref{fig:FF-JAM-MAP} of Appendix~\ref{app:ffs}, is responsible for the ratio of the $\pi^+$ cross section in Fig.~\ref{fig:pion-JAM-MAP-pT} less than unity, in particular, in the large $P_T$ region.

In Fig.~\ref{fig:pion-JAM-MAP-eta}, we show the same ratio as a function of rapidity $y$ at fixed $P_T$ ($=5, 10\,\mathrm{GeV}$ for $\sqrt{S}=28.6, 140\,\mathrm{GeV}$,
respectively), which is calculated with the two different FF sets. 
The ratio shows relatively weaker $y$-dependence at given $P_T$ values.
As shown in Figs.~\ref{fig:pion-JAM-MAP-pT} and \ref{fig:pion-JAM-MAP-eta}, we find that the uncertainty in the $\pi^+$ cross section obtained by using different FF sets for $\pi^+$ is much greater than that associated with the choice of factorization and renormalization scales. Therefore, 
future data on single inclusive $\pi^+$ production in lepton-hadron scattering could be a valuable resource for determining the $\pi^+$ FFs further.

Similarly, Figs.~\ref{fig:kaon-JAM-MAP-pT} and \ref{fig:kaon-JAM-MAP-eta} compare the high-$P_T$ single $K^+$ production cross section calculated with the JAM20 FF set to that calculated with the MAP1.0 FF set. In this particular production channel, one can also see a significant difference in the shape of the calculated $P_T$-spectrum generated by using the two sets of FFs, which could be a result of quite different $z$-distribution of the JAM20 FFs and the MAP1.0 FFs for $K^+$, as shown in Appendix~\ref{app:ffs}. 
Thus, studying high-$P_T$ single inclusive pion and kaon production in lepton-hadron collisions could help determine the $\pi^+$, $K^+$, and other hadron FFs.

\begin{figure}[t]
\centering
\includegraphics[width=\textwidth]{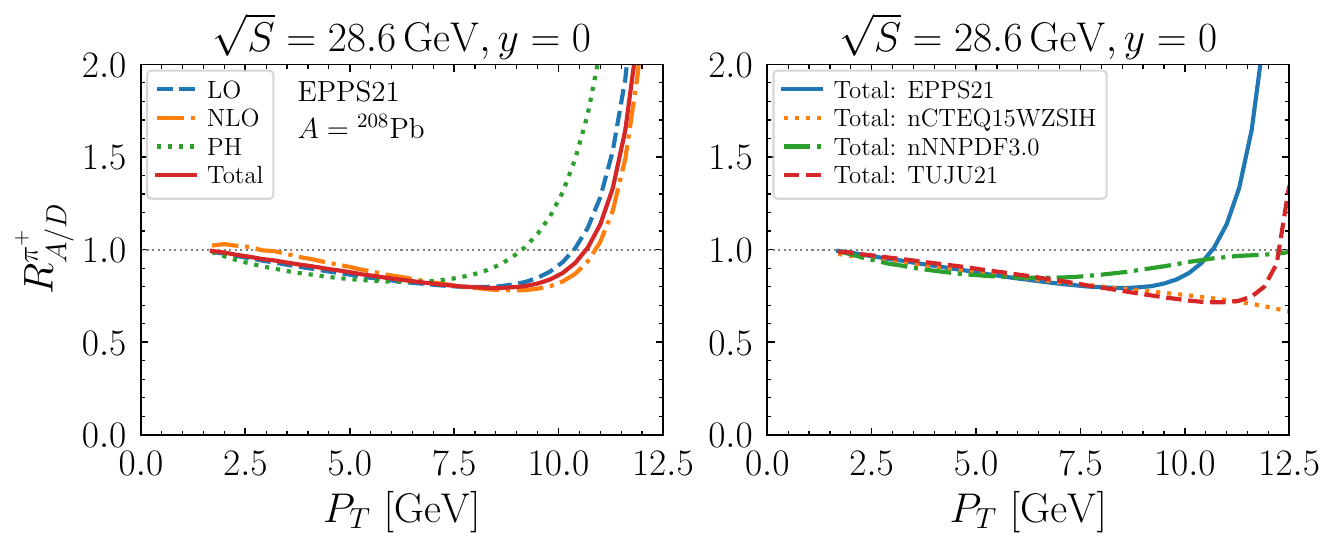}
\caption{\textbf{Left:} Nuclear modification factor for the $\pi^+$ $P_T$ spectrum in $e$-Pb collisions at the EIC, calculated using the EPPS21 nPDF central set~\cite{Eskola:2021nhw} and the CT18ANLO proton PDF central set~\cite{Hou:2019efy}. \textbf{Right:} Comparison of the same observable calculated with different nPDF central sets. Solid curves: EPPS21 (with CT18ANLO). Dotted curves: nCTEQ15WZSIH~\cite{Kusina:2020lyz}. Dash-dotted curves: nNNPDF3.0~\cite{AbdulKhalek:2022fyi}. Dashed curves: TUJU21~\cite{Helenius:2021tof}.}
\label{fig:nPDFset-dep}
\end{figure}
\begin{figure}[t]
\centering
\includegraphics[width=\textwidth]{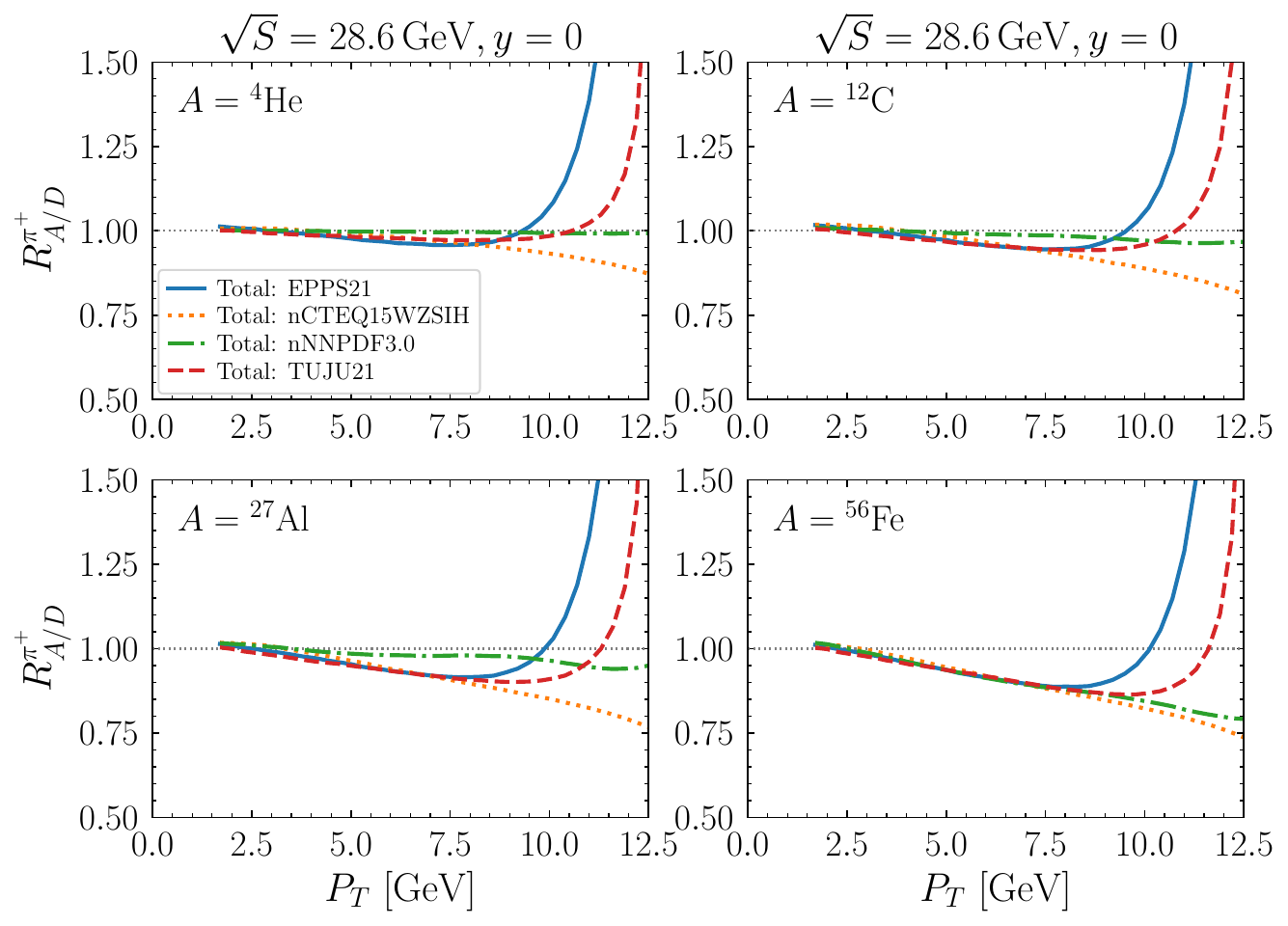}
\caption{Same as Fig.~\ref{fig:nPDFset-dep} but for light and medium nuclear targets. \textbf{Top left:} ${^4\mathrm{He}}$. \textbf{Top right:} ${^{12}\mathrm{C}}$. \textbf{Bottom left:} ${^{27}\mathrm{Al}}$. \textbf{Bottom right:} ${^{56}\mathrm{Fe}}$.}
\label{fig:nPDFset-dep-lightmedium}
\end{figure}
\begin{figure}[t]
\centering
\includegraphics[width=\textwidth]{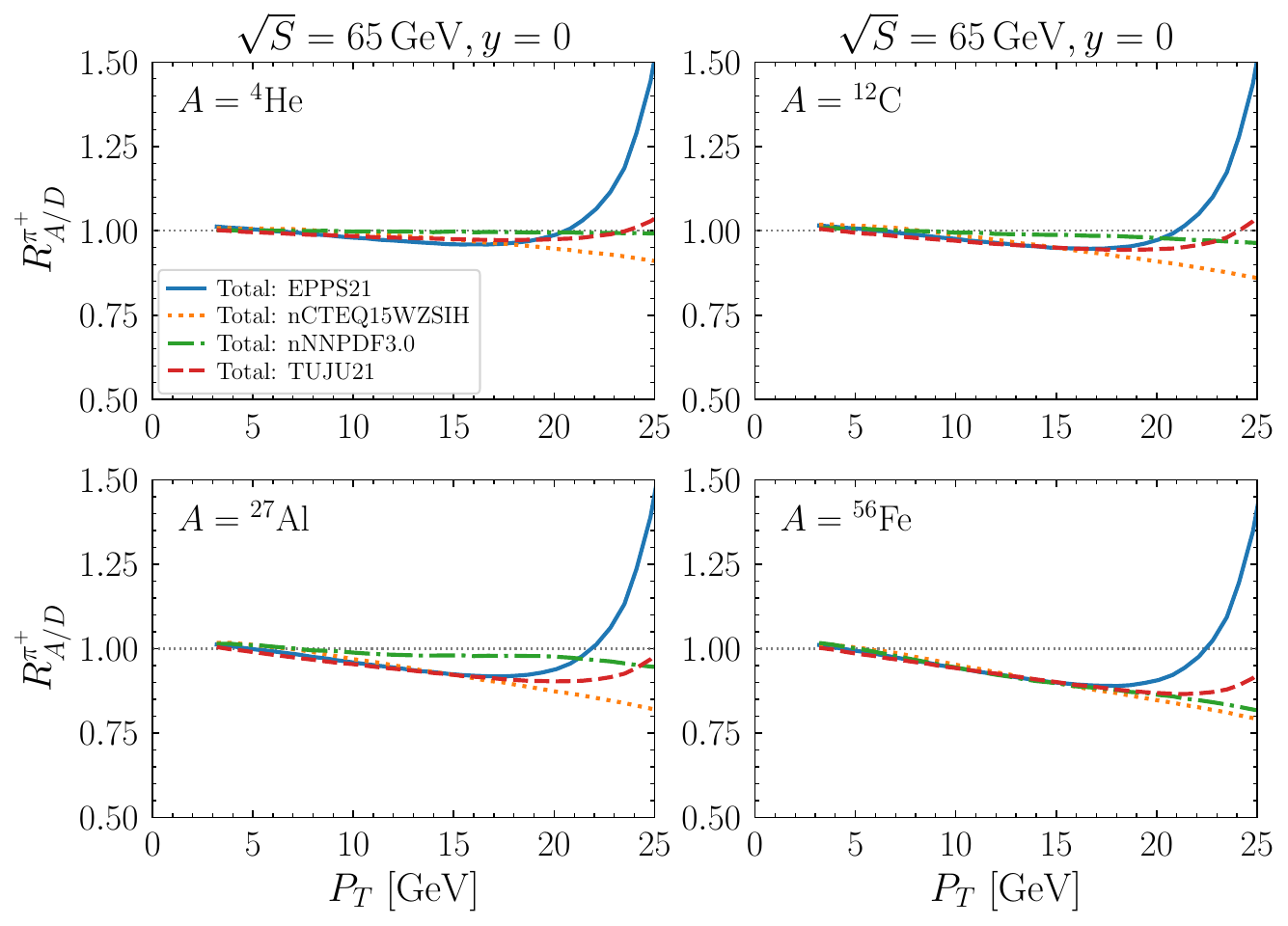}
\caption{Same as Fig.~\ref{fig:nPDFset-dep-lightmedium}, but for $\sqrt{S}=65\,\mathrm{GeV}$.}
\label{fig:nPDFset-dep-lightmedium-rs65GeV}
\end{figure}

\subsection{Nuclear Dependence}

\begin{figure}[t]
\centering
\includegraphics[width=\textwidth]{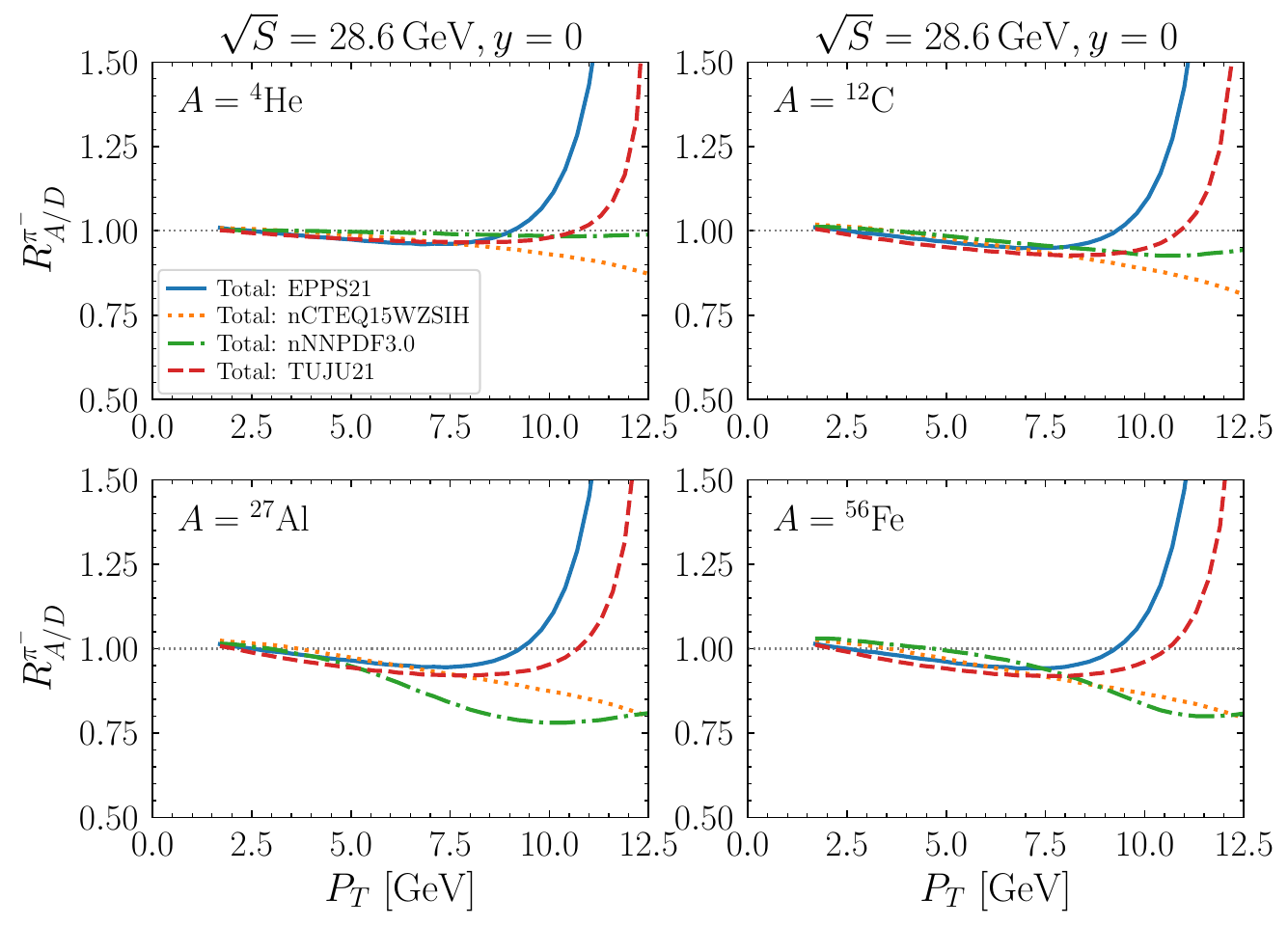}
\caption{Same as Fig.~\ref{fig:nPDFset-dep-lightmedium}, but for $\pi^-$ production.}
\label{fig:pim-nPDFset-dep-lightmedium}
\end{figure}

Lepton-nucleus collisions with large momentum transfer provide excellent probes for studying strong-interaction dynamics between quarks and gluons in the nuclear environment.
Due to collision-induced radiation from large momentum transfer of the collisions, it is also necessary to study the inclusive high-$P_T$ hadron production in lepton-nucleus collisions in terms of the joint QCD+QED factorization approach.

The justification of the joint QCD+QED factorization for inclusive high-$P_T$ hadron production in lepton-hadron collisions at the leading power does not depend on the details of the target hadron state. 
Therefore, we expect the factorization formalism in Eq.~\eqref{eq:lh-fac} to hold for the production in lepton-nucleus collisions, so long as the observed hadron's $P_T$ is sufficiently large. The size of power corrections is suppressed by powers of $1/P_T$, but could be enhanced by nuclear size in lepton-nucleus collisions~\cite{Botts:1990uy,Qiu:2001hj}.
When the joint factorization formalism in Eq.~\eqref{eq:lh-fac} is applied to lepton-nucleus collisions, we only need to replace the proton's PDFs with nuclear PDFs (nPDFs) to take into account the leading power, process-independent nuclear effects.

To quantify the dependence of single high-$P_T$ hadron production in lepton-nucleus collisions on the details of nPDFs, we evaluate the ratio of the hadron production cross section in lepton-nucleus collisions, $d\sigma_{eA\to H+X}/dP_T dy$ normalized by the atomic weight of the nucleus $A$, over that in lepton-deuteron collisions: 
\begin{align}
R_{A/D}^H=\frac{2}{A}\frac{d\sigma_{eA\to H+X}/dP_T dy}{d\sigma_{eD\to H+X}/dP_T dy}.
\end{align}
Figure~\ref{fig:nPDFset-dep} (left panel) shows numerical results of $R_{A/D}^H$ for $\pi^+$ production in the central rapidity region at $\sqrt{S}=28.6\,\mathrm{GeV}$, obtained by using the EPPS21 central set for nPDFs with $A={^{208}\mathrm{Pb}}$, along with contributions from different subprocesses. For the deuteron target, we assumed isospin symmetry and used the CT18ANLO central set to evaluate both the free proton and free neutron PDFs without additional nuclear effects.
Deuteron scattering data have been accounted for in the global fit of the CT18 set by neglecting small nuclear corrections~\cite{Hou:2019efy}. Ref.~\cite{Eskola:2021nhw} used the CT18ANLO set as baseline to extract universal nuclear effects at leading power encoded in nPDFs, so that we adopt the same treatment here.
Clearly, the nuclear dependence of the $\pi^+$ production in lepton-nucleus collisions is a nontrivial function of $P_T$, with suppression below $P_T=10\,\mathrm{GeV}$ and enhancement at large $P_T>10\,\mathrm{GeV}$. We find that these nuclear suppression and enhancement can be attributed to the so-called EMC effect and the Fermi-motion effect of the nPDFs, respectively.  For collisions with light nuclei, no significant difference in the feature of suppression and enhancement is observed as long as the central set of nPDFs is used.

\begin{figure}[t]
\centering
\includegraphics[width=\textwidth]{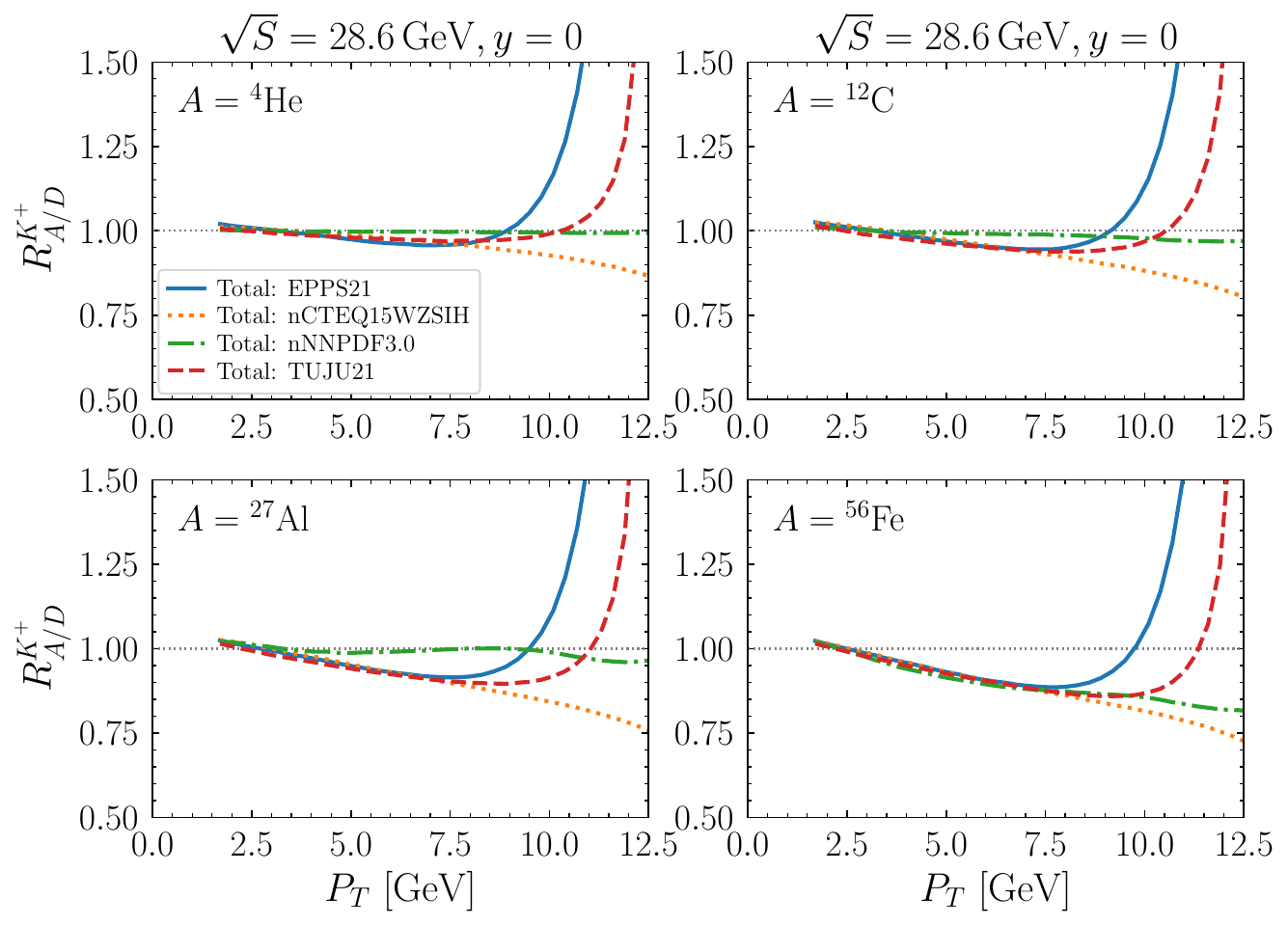}
\caption{Same as Fig.~\ref{fig:nPDFset-dep-lightmedium}, but for $K^+$ production.}
\label{fig:Kp-nPDFset-dep-lightmedium}
\end{figure}
\begin{figure}[t]
\centering
\includegraphics[width=\textwidth]{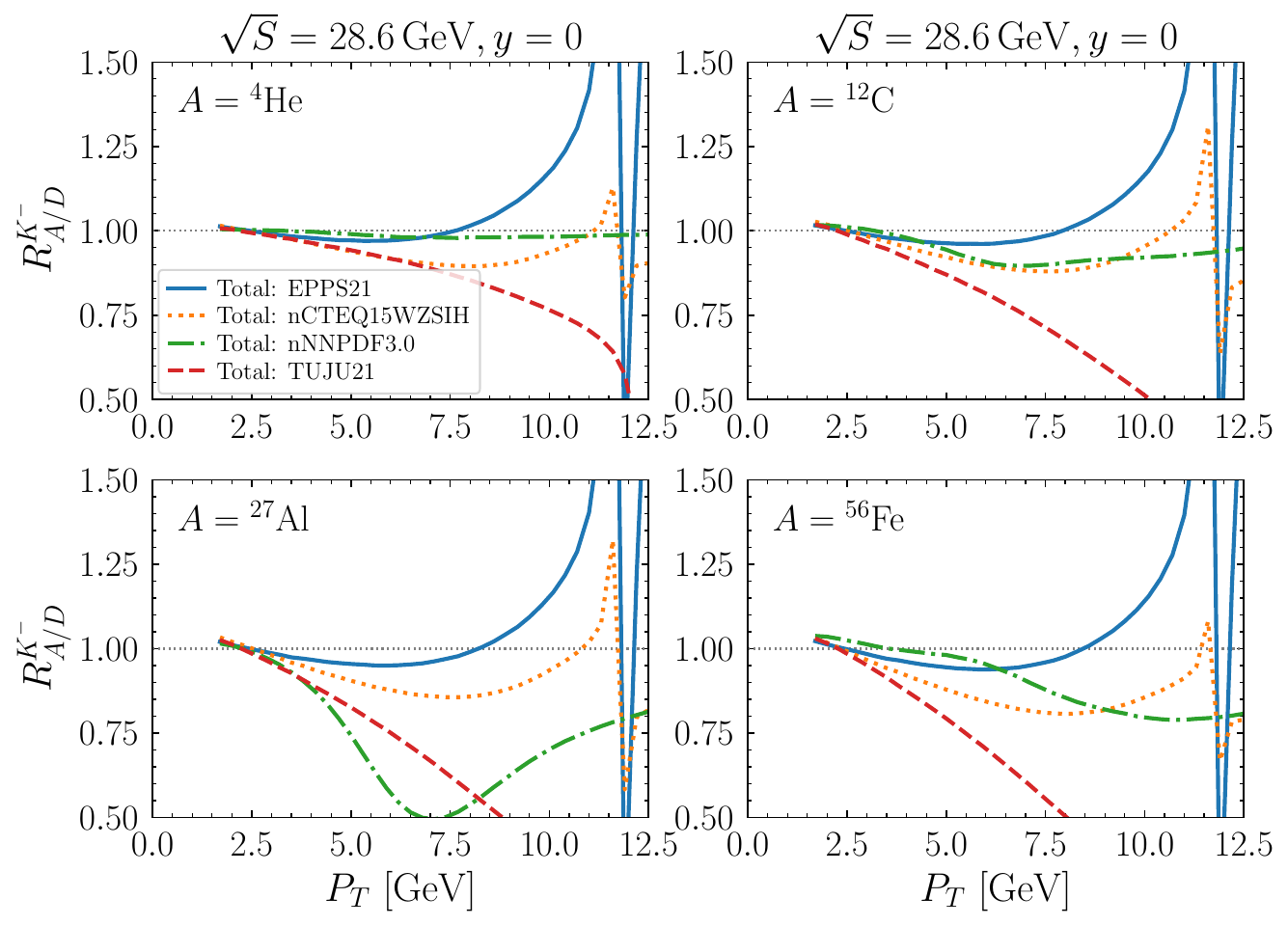}
\caption{Same as Fig.~\ref{fig:nPDFset-dep-lightmedium}, but for $K^-$ production.}
\label{fig:Km-nPDFset-dep-lightmedium}
\end{figure}

To test the sensitivities of the nuclear modification for the high-$P_T$ hadron production on nPDFs, we have also performed the same calculation with other sets of nPDFs, including nCTEQ15WZSIH~\cite{Kusina:2020lyz}, nNNPDF3.0~\cite{AbdulKhalek:2022fyi}, and TUJU21~\cite{Helenius:2021tof}, as shown in Fig.~\ref{fig:nPDFset-dep} (right panel).
We also plot the same ratios for light nuclear targets in Fig.~\ref{fig:nPDFset-dep-lightmedium}. By comparing the results calculated with the EPPS21 set to those with the other sets at $\sqrt{S}=28.6\,\mathrm{GeV}$, one can find that the $\pi^+$ nuclear modification pattern at high $P_T>7.5\,\mathrm{GeV}$ depends on the choice of nPDF sets. Similarly, Fig.~\ref{fig:nPDFset-dep-lightmedium-rs65GeV} presents our results of the $\pi^+$ production in lepton-nucleus collisions at $\sqrt{S} = 65\,\mathrm{GeV}$. 
Figs.~\ref{fig:nPDFset-dep-lightmedium} and \ref{fig:nPDFset-dep-lightmedium-rs65GeV} clearly indicate that single high-$P_T$ hadron production in lepton-nucleus collisions is an excellent probe of nuclear effects encoded in nPDFs in the large-$x$ region.

In Figs.~\ref{fig:pim-nPDFset-dep-lightmedium}-\ref{fig:Km-nPDFset-dep-lightmedium}, we present our numerical predictions for the nuclear modification factors $R_{A/D}^H$ of the inclusive $\pi^-$ and $K^\pm$ productions, respectively. All variables and notations are the same as those in Fig.~\ref{fig:nPDFset-dep-lightmedium}. 
While the calculated nuclear modifications are similar across single high-$P_T$ hadron production channels, $K^-$ production exhibits a much larger variation in Fig.~\ref{fig:Km-nPDFset-dep-lightmedium}.
The strength of nuclear modification to the $\pi$ and $K$ production cross section depends on the convolution of nPDFs and hadron FFs for each active flavor. The relatively large variation of the nuclear modification for $K^-$ production in Fig.~\ref{fig:Km-nPDFset-dep-lightmedium} can be attributed to the functional form of the central set of $K^-$ FFs.  
In conclusion, high-$P_T$ single inclusive hadron production in lepton-nucleus collisions at the EIC is an excellent probe for constraining nPDFs in the EMC regime and their flavor dependence. In addition, single hadron production in lepton-nucleus collisions is complementary to inclusive lepton-nucleus DIS, providing an opportunity to study hadron FFs with possible additional final-state nuclear effects~\cite{Doradau:2024wli}.

\section{Single inclusive jet production}
\label{sec:single-jet}

High-$P_T$ jet production in lepton-hadron collisions is an important complement to single hadron production. Unlike single hadron production, jet production does not require non-perturbative FFs.
To describe single inclusive jet production, two major approaches are commonly used: Monte Carlo (MC) event generators, such as \textsc{Pythia}~\cite{Sjostrand:2014zea,Bierlich:2022pfr}, \textsc{Sherpa}~\cite{Sherpa:2019gpd}, and \textsc{Herwig}~\cite{Bellm:2015jjp}, or the factorized QCD formalism in Eq.~\eqref{eq:hh-fac} for hadron-hadron collisions (or in Eq.~\eqref{eq:lh-fac} for lepton-hadron or something similar for lepton-lepton collisions) with the single parton-to-hadron FFs replaced by single parton-to-jet Jet Functions (JFs). Both approaches implement jet clustering algorithms. In this paper, 
we employ the JFs derived in Refs.~\cite{Jager:2004jh,Mukherjee:2012uz,Kaufmann:2015hma,Kang:2016mcy}. The QCD factorization formalism with the JFs for single inclusive jet production is justified when all particles inside the jet are collimated along the jet axis, and it reproduces MC results well, especially for small jet cone sizes: $R\sim 0.4-0.7$~\cite{Jager:2004jh}.
The JFs can be expanded as
\begin{align}\label{eq:jet_function}
{\mathcal J}_i(z, \mu^2) = \sum_{n=0}\left(\dfrac{\alpha_s(\mu^2)}{2\pi}\right)^n\hat{\mathcal{J}}^{(n)}_i(z, \mu^2)={\mathcal J}_i^{(0)}(z, \mu^2) + {\mathcal J}_i^{(1)}(z, \mu^2)+\cdots,
\end{align}
where $i=q,g$. In this work, we will implement the JFs up to NLO, $\mathcal{O}(\alpha_s)$. Note that the JF is the same for all quarks and anti-quarks: ${\mathcal J}_{q} = {\mathcal J}_{\bar{q}}$. The variable $z = \omega_J/\omega$ represents the fraction of the large light-cone momentum $\omega$ of the initiating parton that is carried by the produced jet of light-cone momentum $\omega_J$. The variable $\mu$ represents the jet production scale, analogous to the fragmentation scale for hadron production.

The JFs at LO simply read ${\mathcal J}_i^{(0)}(z, \omega_J, \mu^2)=\delta(1-z)$, which is the same for $i=q,g$. The JFs at NLO for $i=q,g$ are given by~\cite{Kang:2016mcy}
\begin{align}
{\mathcal J}_q^{(1)}(z, \omega_J, \mu^2)
=&\,L P^{(0,1)}_{qq}(z,\mu^2)+\left[L - 2\ln(1-z)\right] P^{(0,1)}_{gq}(z,\mu^2)\non
&-\frac{
\alpha_s(\mu^2)}{2\pi}\bigg\{C_F\left[2(1+z^2)\left(\frac{\ln(1-z)}{1-z}\right)_+ +1\right]
-\delta(1-z)d_J^{q,\mathrm{alg}} \bigg\},
\label{eq:jet_func_q}
\\
{\mathcal J}_g^{(1)}(z, \omega_J, \mu^2)
=&\,L P^{(0,1)}_{gg}(z,\mu^2)+2n_f \left[L-2\ln(1-z)\right] P^{(0,1)}_{qg}(z,\mu^2)\non
&-\frac{\alpha_s(\mu^2)}{2\pi}\bigg\{\frac{4C_A(1-z+z^2)^2}{z}\left(\frac{\ln(1-z)}{1-z}\right)_+ 
-\delta(1-z)d_J^{g,\mathrm{alg}}\non
&+ 4n_f T_R z(1-z) \bigg\},
\label{eq:jet_func_g}
\end{align}
with $T_R=1/2$. The variable $L$ represents the logarithmic term defined as
\begin{align}
L=\ln\frac{\mu^2}{\omega_J^2\tan^2({\mathcal R}/2)},  \quad 
{\mathcal R}\equiv \frac{R}{\cosh\eta}
\end{align}
where $\omega_J=2P_T\cosh\eta$, with $P_T$ and $\eta$ being the transverse momentum and rapidity of the jet, respectively, and $R$ is the jet cone size (parameter)\footnote{For a small jet cone size, the logarithmic term becomes $L \overset{R\ll 1}{\approx} \ln (\mu^2/ (P_TR)^2)$, making the $\ln R$ dependence explicit.}.
The QCD splitting functions $P^{(0,1)}_{qq}, P^{(0,1)}_{gg}, P^{(0,1)}_{qg}, P^{(0,1)}_{gq}$ are given in Appendix~\ref{app:kernels}.
In the NLO expressions above, the constant terms $d_J^{i,\mathrm{alg}}$ with $i=q,g$ depend on the jet clustering algorithm~\cite{Cacciari:2008gp}. For the anti-$k_T$ algorithm used in this work, they are given by
\begin{align}
d_J^{q,\mathrm{anti\text{-}k_T}}=C_F\left(\frac{13}{2}-\frac{2\pi^2}{3}\right),
\quad
d_J^{g,\mathrm{anti\text{-}k_T}}=C_A\left(\frac{67}{9}-\frac{2\pi^2}{3}\right)-T_Rn_f\left(\frac{23}{9}\right).
\end{align}
The corresponding expressions for other algorithms, such as $k_T$ and Cambridge/Aachen, can be found in Ref.~\cite{Kang:2016mcy}.

For small jet sizes ($R<1$), the JFs involve a large single logarithm $\ln R$, which can be resummed using the standard time-like DGLAP evolution equation. This small-$R$ resummation is typically essential for inclusive jet production at the LHC. On the other hand, 
resummation effects for inclusive jet production may not be sizable at the EIC, where the jet cone size is expected to be $R\sim 1$, although measuring jets with smaller cone sizes should be pursued as far as detector technology allows.
Therefore, in this paper, we will not explicitly resum $(\alpha_s\ln R)^n$, and instead directly apply the fixed-order JFs at NLO to inclusive jet production in lepton-hadron collisions. 
Investigating both the impact of quantum evolution on the JFs in inclusive jet production and their higher-order perturbative corrections at NNLO and beyond presents a significant theoretical challenge~\cite{deFlorian:2013qia,Dasgupta:2016bnd,Lee:2024icn,Generet:2025vth} that lies beyond the scope of this paper.
Therefore, we leave such precision studies for future work.

Explicitly considering the $\alpha_s$ power counting of the perturbative JFs, the inclusive jet production cross section up to $\mathcal{O}(\alpha_s)$ can be written schematically as
\begin{align}\label{eq:jet_xsection_fixed_order}
d\sigma_{J}=&\,\sum_{i=e,\bar{e}}\sum_{j=q,\bar{q}}\sum_{k=q,\bar{q}}f_{i/e}(\mu_e)\otimes_{\xi} f_{j/p}(\mu_h)\otimes_{x} {\mathcal J}_k^{(0)}(\mu_J)\otimes_z H_{ij\to k}^{(0)}(\mu_r)\non
&+\sum_{i=e,\bar{e},\gamma}\sum_{j=q,\bar{q},g}\sum_{k=q,\bar{q},g}f_{i/e}(\mu_e)\otimes_{\xi} f_{j/p}(\mu_h) \otimes_{x} {\mathcal J}_k^{(0)}(\mu_J)\otimes_{z} H_{ij\to k}^{(1)}(\mu_r)\non
&+\sum_{i=e,\bar{e}}\sum_{j=q,\bar{q}}\sum_{k=q,\bar{q}}f_{i/e}(\mu_e)\otimes_{\xi} f_{j/p}(\mu_h) \otimes_{x} {\mathcal J}_k^{(1)}(\mu_J)\otimes_{z} H_{ij\to k}^{(0)}(\mu_r)
\end{align}
where the superscript $(0)$ and $(1)$ indicate the corresponding order in $\alpha_s$. In the expression above, we have omitted higher-order contributions of $\mathcal{O}(\alpha_s^2)$, including the $g\to J$ channel with ${\mathcal J}_g^{(1)}$, for perturbative consistency.
Regarding the choice of scales, for simplicity, we define our baseline of fixed-order calculations by setting $\mu_r=\mu_e=\mu_h=\mu_J=P_T$, where $\mu_J$ represents the JF scale. As discussed in Ref.~\cite{Kang:2016mcy}, for narrow jets ($R\ll 1$), one can choose the natural jet scale to be $\mu_J\sim P_T R$ and evolve the JFs up to a higher hard scale $\mu\sim P_T>\mu_J$, thereby resumming the $\ln R$-enhanced corrections. However, because our primary interest is inclusive jet production at the EIC across a wide range of jet sizes ($R\lesssim 1$), we will perform our numerical calculations by setting $\mu_J\sim P_T R$ for various values of $R$ to estimate the associated scale uncertainties.

\begin{figure}[t]
\centering
\includegraphics[width=\textwidth]{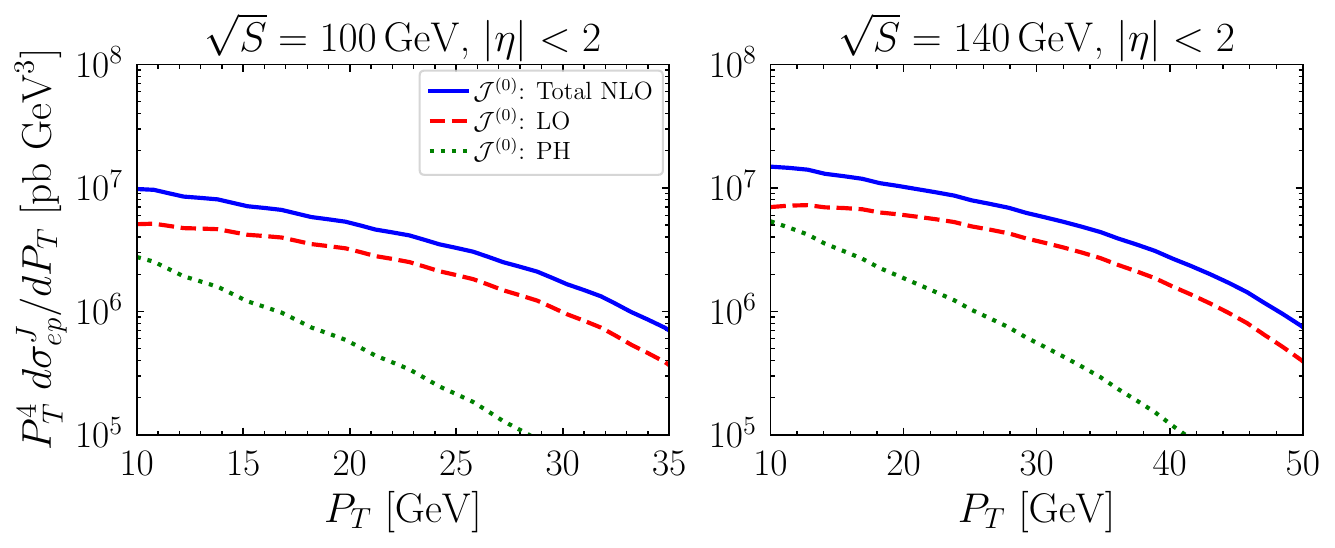}
\caption{Inclusive jet production cross section as a function of $P_T$ at mid-rapidity $|\eta|<2$ in lepton-hadron collisions at $\sqrt{S}=100\,\mathrm{GeV}$ \textbf{(left)} and $140\,\mathrm{GeV}$ \textbf{(right)}. Dashed curves: Leptoproduction at LO in $\alpha_s$. Dotted curves: Photoproduction contribution, included in the second term of Eq.~\eqref{eq:jet_xsection_fixed_order}. Solid curves: Total NLO partonic cross section, including photoproduction, calculated with the LO JF.}
\label{fig:jetfunction_LO}
\end{figure}
\begin{figure}[t]
\centering
\includegraphics[width=\textwidth]{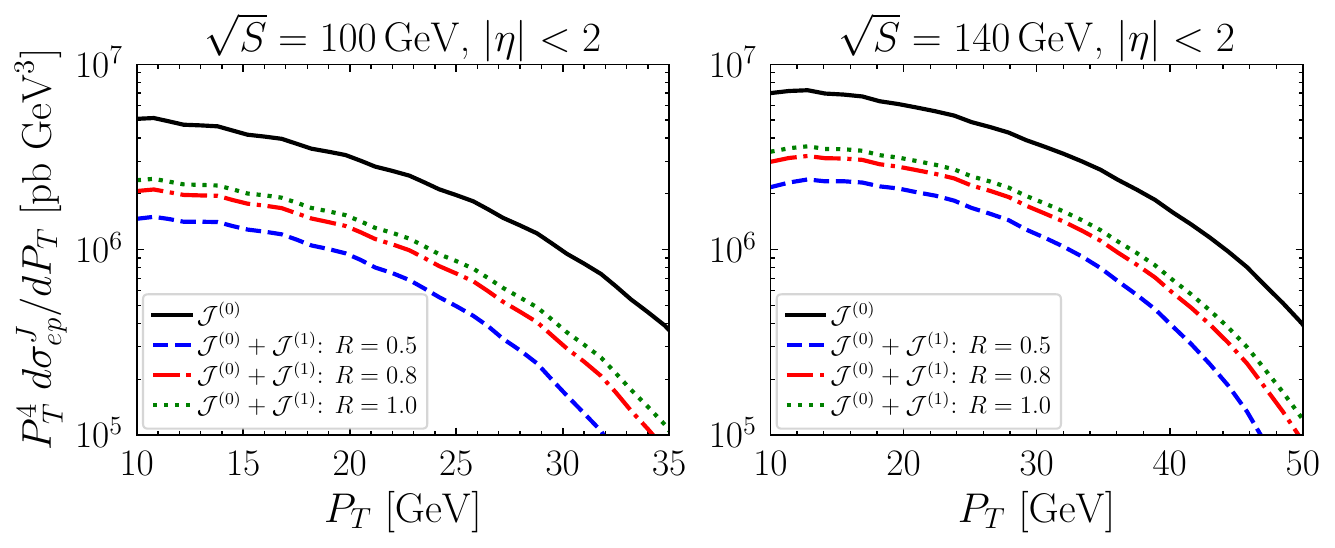}
\caption{Inclusive jet production cross section as a function of \pT\ at mid-rapidity $|\eta|<2$ in lepton-hadron collisions, calculated by convoluting the NLO JF with LO hard parts ($H^{(0)}$) using the anti-$k_T$ algorithm. Dashed blue curves: $\mathcal{J}^{(0)}+\mathcal{J}^{(1)}$ with $R=0.5$. Dash-dotted red curves: $\mathcal{J}^{(0)}+\mathcal{J}^{(1)}$ with $R=0.8$. Dotted green curves: $\mathcal{J}^{(0)}+\mathcal{J}^{(1)}$ with $R=1$. Solid black curves: $\mathcal{J}^{(0)}$ (LO result). }
\label{fig:jetfunction_NLO}
\end{figure}
\begin{figure}[t]
\centering
\includegraphics[width=\textwidth]{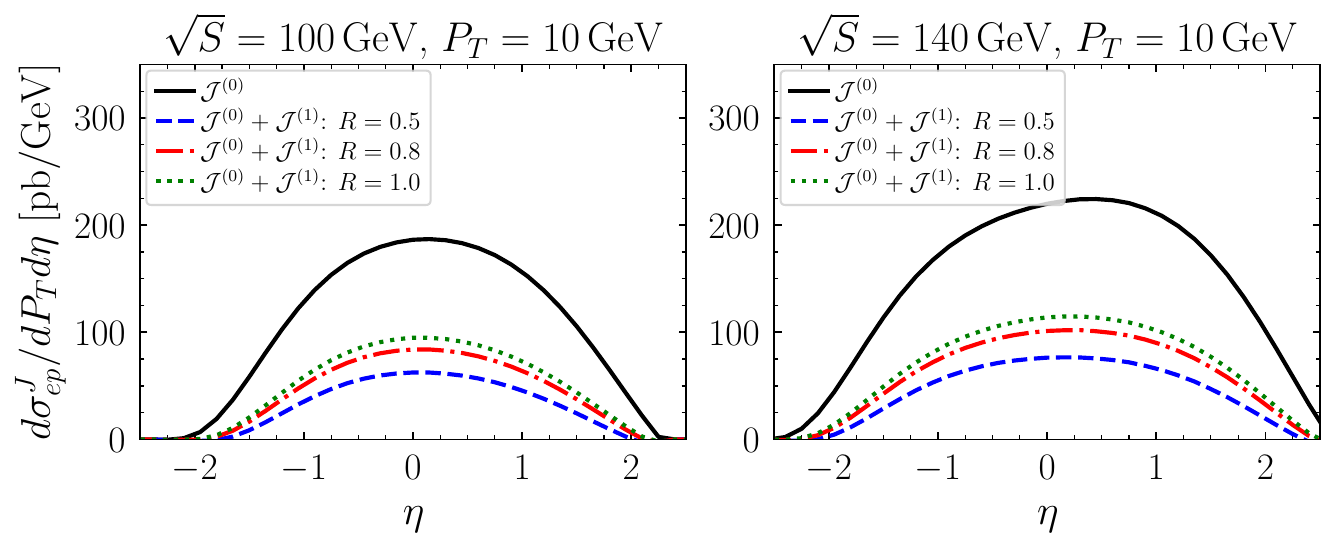}
\caption{Same as Fig~\ref{fig:jetfunction_NLO}, but for the rapidity dependence of the jet production cross section.}
\label{fig:jetfunction_NLO_eta}
\end{figure}

Figure~\ref{fig:jetfunction_LO} shows numerical results for the inclusive jet production cross section as a function of \pT\ at mid-rapidity ($|\eta|<2$) in lepton-hadron collisions at the EIC. These results are evaluated using the LO JF $\mathcal{J}^{(0)}$, corresponding to the first and second lines of Eq.~\eqref{eq:jet_xsection_fixed_order}. In this particular case, the jet is initiated by a single produced quark or gluon. The figure compares the results at center-of-mass energies of $\sqrt{S}=100\,\mathrm{GeV}$ and $140\,\mathrm{GeV}$. The \pT\ spectrum is scaled by $P_T^4$ to better highlight the difference between jet leptoproduction and photoproduction.

In Fig.~\ref{fig:jetfunction_NLO} we show the full results for inclusive jet production, incorporating both the LO and NLO JFs for various jet cone sizes.
We examine the quantitative behavior of the NLO JF contributions using the anti-$k_T$ algorithm, while employing the LO short-distance hard parts for parton production. The NLO corrections from the JF reduce the overall jet production cross section because they are negative. For a typical jet cone size of $R=1$ at the EIC, the high-\pT\ jet production cross section is reduced by approximately half. Figure~\ref{fig:jetfunction_NLO_eta} shows the rapidity dependence of the inclusive jet cross section at $P_T=10\,\mathrm{GeV}$ at the EIC, which is obtained by using the NLO JF.

\begin{figure}[t]
\centering
\includegraphics[width=\textwidth]{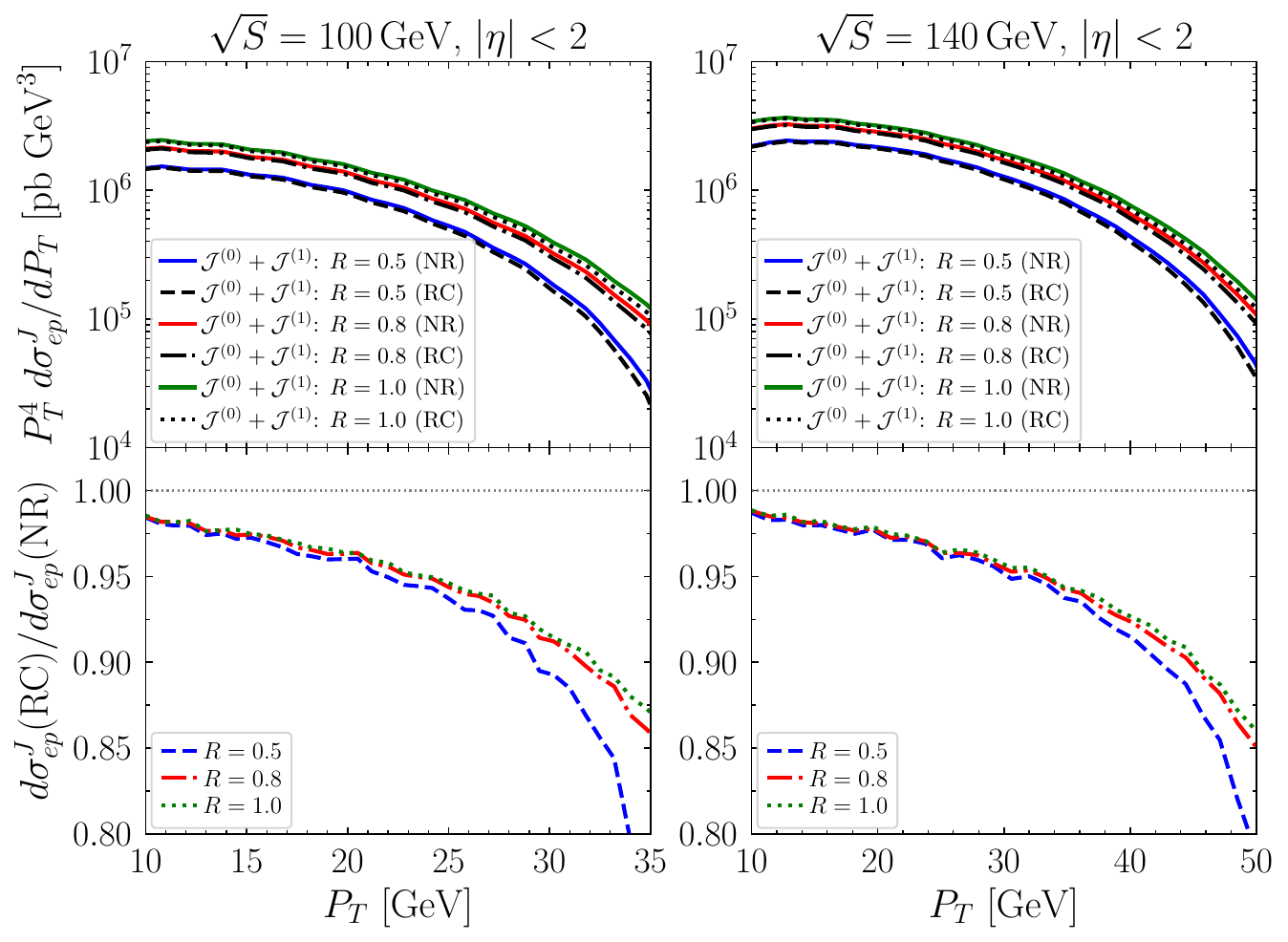}
\caption{
\textbf{Top:} Same as Fig.~\ref{fig:jetfunction_NLO}, but comparing the jet $P_T$ spectra calculated using the NLO JF with QED radiative corrections (RC) and without corrections (NR) at different collision energies. \textbf{Bottom:} Corresponding RC-to-NR ratios.
}
\label{fig:jet_RC_NR_ratio}
\end{figure}

Figure~\ref{fig:jet_RC_NR_ratio} illustrates the quantitative impact of QED radiative corrections on the inclusive jet production cross section, calculated using the LO hard parts for parton production. The QED radiative corrections suppress the electron LDF at large $\xi$ and enhance it at small $\xi$. This modification of the electron LDF leads to a corresponding depletion in the differential jet production cross section at high $P_T$, resulting in a significant reduction of approximately $10\%$ to $20\%$.

Before concluding this section, we give 
some remarks on profound quantum corrections at large $R$ in inclusive jet production.
At large jet cone sizes, $R\simeq1$, the so-called narrow jet approximation can receive substantial corrections, including perturbative power corrections of $\mathcal{O}(R^2)$~\cite{Ellis:1992qq,Kang:2016mcy,Qiu:2019sfj}. Furthermore, comparisons to MC event generators at large $R$ are largely influenced by non-perturbative features of the background, where the underlying event activity scales with the jet area, i.e., $R^2$~\cite{Dasgupta:2007wa}. Comprehensive studies of inclusive jet production in lepton-hadron collisions within the joint QCD+QED factorization formalism will be pursued in future papers.

Finally, similar to single high-$P_T$ hadron production, single inclusive jet production in lepton-nucleus collisions can serve as an important probe of nPDFs, analogous to the approach discussed in Refs.~\cite{Kang:2012zr,Kang:2013wca,Kang:2013lga,Cao:2024ota}.
In these works, it was shown that the so-called $N$-Jettiness~\cite{Stewart:2010tn}, denoted by $\tau_N$, provides a useful framework for studying nPDFs, especially through the 1-Jettiness distribution~\cite{Jouttenus:2011wh,Kang:2013nha}. See also recent progress on 1-Jettiness in inclusive DIS in Ref.~\cite{Dotson:2026ttc}. Further study of nPDFs and nuclear-enhanced power corrections using 1-Jettiness with QED radiative corrections is left for future work.

\section{Summary and outlook}
\label{sec:summary}

This paper has presented the first numerical calculation of single inclusive hadron and jet production at high \pT\ in lepton-hadron collisions within a joint QCD+QED collinear factorization framework. By treating collision-induced QED radiation on the same footing as QCD radiative corrections, we can factorize the hadron production cross section into short-distance hard parts convolved with universal LDFs, along with standard PDFs and FFs. In this paper, we demonstrated that the LDFs obey DGLAP-type evolution equations with mixed QCD and QED splitting kernels. Since all lepton-hadron scattering data had some kind of radiative correction applied to them, which cannot be used to extract the LDFs, we have developed a default set of non-perturbative LDFs at the input scale $\mu_0=m_c$, from which we derived the first set of LDFs at $\mu \geq \mu_0$ by solving the joint QCD+QED evolution equations.

By applying the joint QCD+QED factorization formalism, we evaluated the inclusive production of light hadrons ($\pi$ and $K$) at JLab and future EIC energies, and quantified the impact of the leading power QED radiative corrections from the universal LDFs on single high-$P_T$ light hadron production in lepton-hadron collisions. 
A major advantage of the QCD+QED joint factorization approach is that it allows the systematic description of hadron leptoproduction and photoproduction in a unified framework, without introducing kinematic cuts on the scattered lepton to keep the exchanged photon close to real. Our numerical studies indicate that theoretical uncertainties are currently dominated by the choice of FFs rather than by the scale dependence. 

We also demonstrated the predictive power of the joint QCD+QED factorization approach for the so-called radiative corrections to single hadron production in lepton-hadron collisions. If we keep the factorized hard parts to the lowest order in QED, all leading power contributions from collision-induced QED radiation are included in the LDFs (and LFFs if the scattered electron is measured). Since LDFs (and LFFs) are universal, once determined, the size of leading power contributions from collision-induced QED radiation in lepton-hadron collisions is uniquely fixed and predicted, just as shown in Fig.~\ref{fig:RCs-hadron-allrs} for single high-$P_T$ hadron production.

We extended our analysis to lepton-nucleus collisions at future EIC energies to investigate cold nuclear matter effects by using the joint QCD+QED collinear factorization approach. Replacing free-proton PDFs with collinear nPDFs, we computed the nuclear modification factors of single inclusive high-$P_T$ hadron production for various nuclei. The numerical results exhibit distinctive medium modifications, such as nuclear suppression at moderate \pT\ attributed to the EMC effect and nuclear enhancement at high \pT\ attributed to the Fermi motion in the nuclear medium, although the pattern of nuclear modifications depends entirely on the choice of nPDF sets. Nevertheless, the results emphasize the feasibility of single inclusive hadron \pT\ spectra as a valuable probe for studying dynamics in the nuclear medium, and our ability to distinguish between different parameterizations of nPDFs, especially for the EMC regime.

In addition to studying inclusive hadron production, we investigated single inclusive jet production using perturbatively calculated JFs with the anti-$k_T$ clustering algorithm. The leading-power QED radiative corrections, encoded in the LDFs, result in a significant depletion of the high \pT\ jet production cross section by approximately 10\% to 20\%. The modification of the jet production cross section can be further studied quantitatively in the future by considering the quantum evolution of the JFs as the jet cone size changes. We argued that single inclusive high-$P_T$ jet in lepton-nucleus collisions at the EIC can be a very good probe of nPDFs, complementary to the 1-Jettiness distribution based on event shape.

We also showed that high-\pT\ hadron production in lepton-hadron collisions is particularly sensitive to the large-$z$ region of hadron FFs, offering a complementary probe for extracting hadron FFs from lepton-lepton scattering data, such as $e^+e^-$ annihilation, which are more sensitive to hadron FFs at lower $z$. Combining hadron-hadron, lepton-hadron, and lepton-lepton scattering data enables us to precisely constrain hadron FFs across a wider phase space in the future. Meanwhile, lepton-lepton annihilation data from Belle and lepton-hadron scattering data from JLab and the future EIC offer an important opportunity to systematically extract universal hadron FFs along with LDFs. 

Future work will focus on extending this joint factorization formalism to polarized DIS, parity-violating DIS~\cite{Whitehill:2026swa,CQWZ:pvdis}, and SIDIS, as well as to heavy quarkonium and open heavy flavor production, and BSM physics~\cite{Cammarota:2026ylh} in lepton-hadron collisions. All of these are critically important for the science program at the future EIC.
Such extensive studies of lepton-hadron scattering across a wide range of kinematic regimes within the joint factorization approach are required to pave the way for future global data analysis of unknown but universal LDFs, FFs, and nPDFs. We will address such extensive studies in future papers.

\bigskip
\noindent{\bf Code Availability.} The numerical code developed for this work is maintained in a private GitHub repository. Researchers interested in reproducing or extending these results may obtain access from the corresponding author for academic research purposes upon reasonable request.

\section*{Acknowledgments}
The authors are grateful to J.~Cammarota, J.-Y.~Zhang, Y.~Mehtar-Tani, R.~Venugopalan, H.~Enyo, T.~Liu, N.~Sato, W.~Melnitchouk, R.~Seidl, and E.~C.~Aschenauer for fruitful discussion and insightful comments.
K.W. thanks the BNL EIC Theory Institute and JLab Theory Center for their hospitality and financial support, during which part of this work took place.
The work of J.Q. is supported by the U.S. Department of Energy, Office of Science, Office of Nuclear Physics under Contract No.~89243126CSC000213.
The work of K.W. is supported by JSPS KAKENHI Grant No.~JP25K07286.

\clearpage
\appendix

\section{QCD and QED splitting functions for evolution of LDFs}
\label{app:kernels}

In this appendix, we summarize the leading-order QCD and QED splitting functions for the evolution of LDFs, as defined in Eq.~(\ref{eq:kernel_def}), 
\begin{align}
&P^{(1,0)}_{ee}(\xi,\mu^2)=\frac{\alpha_{em}(\mu^2)}{2\pi}\left[\frac{1+\xi^2}{[1-\xi]_+}+\frac{3}{2}\delta(1-\xi)\right],\\
&P^{(1,0)}_{e\gamma}(\xi,\mu^2)=\frac{\alpha_{em}(\mu^2)}{2\pi}\left[(1-\xi)^2+\xi^2\right],\\
&P^{(1,0)}_{\gamma e}(\xi,\mu^2)=\frac{\alpha_{em}(\mu^2)}{2\pi}\left[\frac{(1-\xi)^2+1}{\xi}\right],\\
&P^{(1,0)}_{\gamma\gamma}(\xi,\mu^2)=\frac{\alpha_{em}(\mu^2)}{2\pi}\left[-\frac{2}{3}n_l\,\delta(1-\xi)\right], \label{eq:evo_gg}\\
&P^{(1,0)}_{q\gamma}(\xi,\mu^2)=e_q^2 N_c \frac{\alpha_{em}(\mu^2)}{2\pi}\left[(1-\xi)^2+\xi^2\right]=e_q^2 N_c P^{(1,0)}_{e\gamma}(\xi,\mu^2),\\
&P^{(1,0)}_{\gamma q}(\xi,\mu^2)=e_q^2\frac{\alpha_{em}(\mu^2)}{2\pi}\left[\frac{(1-\xi)^2+1}{\xi}\right]=e_q^2P^{(1,0)}_{\gamma e}(\xi,\mu^2),\\
&P^{(0,1)}_{qq}(\xi,\mu^2)=\frac{\alpha_{s}(\mu^2)}{2\pi}C_F\left[\frac{1+\xi^2}{[1-\xi]_+}+\frac{3}{2}\delta(1-\xi)\right],\\
&P^{(0,1)}_{qg}(\xi,\mu^2)=\frac{\alpha_{s}(\mu^2)}{2\pi}T_R\left[(1-\xi)^2+\xi^2\right],\\
&P^{(0,1)}_{gq}(\xi,\mu^2)=\frac{\alpha_{s}(\mu^2)}{2\pi}C_F\left[\frac{(1-\xi)^2+1}{\xi}\right],\\
&P^{(0,1)}_{gg}(\xi,\mu^2)=\frac{\alpha_{s}(\mu^2)}{2\pi} \left\{2N_c\left[\frac{\xi}{[1-\xi]_+}+\frac{1-\xi}{\xi}+\xi(1-\xi)\right]+\left[\frac{11}{6}N_c-\frac{n_f}{3}\right]\delta(1-\xi)\right\}\, ,
\end{align}
where $e_q$ is the electric charge fraction of a quark $q$, color factor $T_R=1/2$ and $C_F=(N_c^2-1)/(2N_c)$ with the number of color $N_c=3$, and the ``+''-description is defined as
\begin{equation}
\int_0^1 dx\frac{f(x)}{(1-x)_+} 
=\int_0^1 dx \frac{f(x) - f(1)}{1-x} \, ,
\label{eq:plus-description} 
\end{equation}
for a test function $f(x)$.

\section{Dominance of high-$z$ region of FFs in high-$P_T$ hadron production}
\label{app:high_z}

\begin{figure}[t]
\begin{center}
\includegraphics[width=\textwidth]{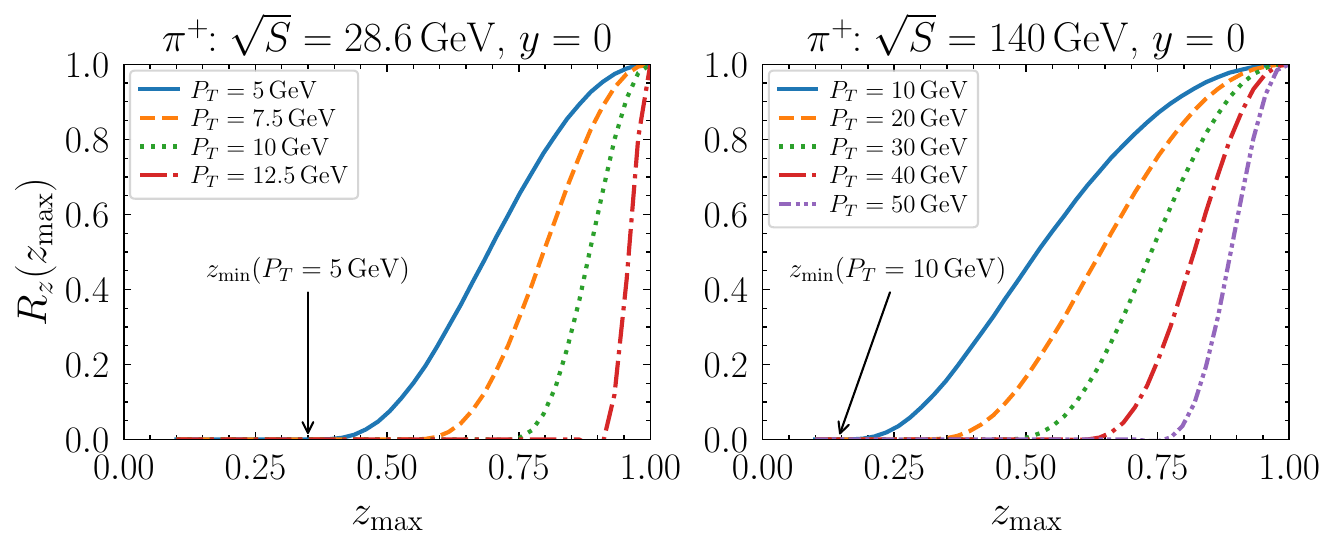}
\caption{Ratio of the $\pi^+$ \pT\ spectrum with $z_\mathrm{max}<1$ to that with $z_\mathrm{max}=1$ at $\sqrt{S}=28.6\,\mathrm{GeV}$ \textbf{(left)} and $140\,\mathrm{GeV}$ \textbf{(right)}. The different curves correspond to various values of \pT. $z_\mathrm{min}$ for the lowest \pT\ case is indicated on each panel. }
\label{fig:zmax}
\end{center}
\end{figure}

High-\pT\ hadron production in lepton-hadron scatterings is sensitive to the FFs in the large-$z$ region, in contrast to production in lepton-lepton scatterings~\cite{Belle:2013lfg,Belle:2020pvy,BaBar:2013yrg}, where the cross section is largely dominated by the small-$z$ region of FFs. This is because the observed hadron momentum is proportional to $(\xi x z)\sqrt{S}$
in lepton-hadron collisions, and the fact that PDF $f(x)$ is a much steeper falling function of momentum fraction than FF $D(z)$, so that a combination of small $x$ and large $z$, while $\xi \sim 1$, dominates the production rate.
We note that this feature is more pronounced in inclusive hadron production in hadron-hadron collisions due to the fact that the produced hadron momentum is proportional to $(x_1 x_2 z)\sqrt{S}$, where $x_{1,2}$ the momentum fractions of the incoming partons, and the production cross section involves two initial state PDFs and one final state hadron FF~\cite{Berger:2001wr}.

This feature can be verified quantitatively by considering the normalized cross section, defined by~\cite{Berger:2001wr,Lee:2021oqr}
\begin{align}
R_z(z_\mathrm{max}) = \frac{d\sigma^H_{ep}(z_\mathrm{max})/d^2P_T dy}{d\sigma^H_{ep}(z_\mathrm{max}=1)/d^2P_T dy},
\quad
\frac{d\sigma^H_{ep}(z_\mathrm{max})}{d^2P_T dy} \equiv \sum_c \int_{z_\mathrm{min}}^{z_\mathrm{max}} \frac{dz}{z^2} \, D_{c\to H}(z) \, \frac{d\hat{\sigma}^c_{ep}}{d^2p_{cT} dy},
\end{align}
where $z_\mathrm{min}$ is fully determined by the kinematics of single inclusive hadron production in lepton-hadron collisions, while $z_\mathrm{max}$ in $R_z$ is varied from $z_\mathrm{min}$ to 1. 

In Fig.~\ref{fig:zmax}, we plot the ratio $R_z$ as a function of $z_\mathrm{max}$ for $\pi^+$ production at EIC energies, $\sqrt{S}=28.6\,\mathrm{GeV}$ and $140\,\mathrm{GeV}$, which is obtained using the JAM20 central set for PDFs and FFs together with our default LDFs. One can see, for example, that about 50\% of the cross section originates from the region $z \ge 0.75$ at $P_T=7.5\,\mathrm{GeV}$ and $\sqrt{S}=28.6\,\mathrm{GeV}$. For fixed $\sqrt{S}$, the larger the hadron \pT, the more sensitive the cross section becomes to the large-$z$ region. Comparing Fig.~\ref{fig:zmax} (left panel) and (right panel), one finds that a larger $\sqrt{S}$ opens a broader phase space for hadron production, causing the cross section to receive larger contributions from the lower-$z$ region. The above discussion is also expected to hold for inclusive jet production with JFs. Thus, understanding hadron FFs in the large-$z$ region is essential for single inclusive hadron production in lepton-hadron collisions, and consequently for studying the internal structure of hadrons via hadron production.

\section{Uncertainty of hadron FFs: MAP1.0 vs. JAM20}
\label{app:ffs}

\begin{figure}[t]
\begin{center}
\includegraphics[width=\textwidth]{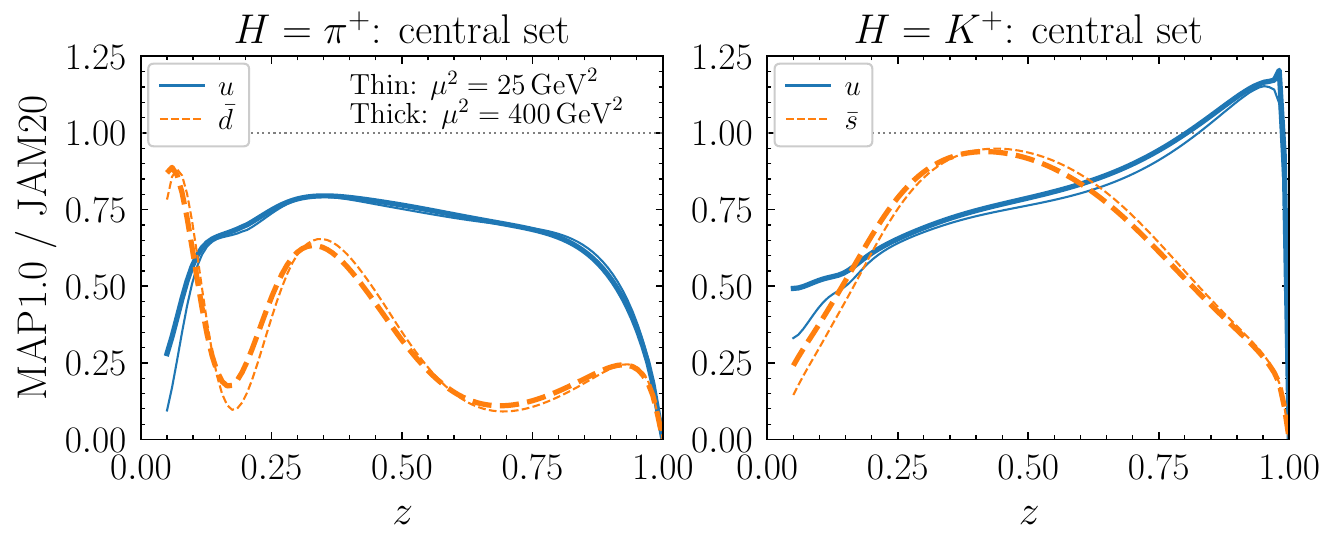}
\caption{Ratio of the MAP1.0 FF central set~\cite{Khalek:2021gxf} to the JAM20 FF central set~\cite{Moffat:2021dji} for $\pi^+$ production \textbf{(left)} and $K^+$ production \textbf{(right)}. Solid lines: $u\to \pi^+,K^+$. Dashed lines: $\bar{d}\to \pi^+$ and $\bar{s}\to K^+$.}
\label{fig:FF-JAM-MAP}
\end{center}
\end{figure}

Like all QCD factorization approaches to describe physical observables, the predictive power of the joint QCD+QED factorization approach to single high-$P_T$ hadron production in lepton-hadron collisions relies on our knowledge of those non-perturbative but universal LDFs, PDFs, and especially FFs as discussed in Sec.~\ref{subsec:ffs}. In this appendix, we explicitly show the differences and similarities of the two sets of hadron FFs used in our calculations: MAP1.0~\cite{Khalek:2021gxf} and JAM20~\cite{Moffat:2021dji}. In Fig.~\ref{fig:FF-JAM-MAP}, we plot the ratio of parton-to-hadron FFs between the central sets of MAP1.0 and JAM20 for $\pi^+$ production (left panel) and $K^+$ production (right panel) at fixed values of the fragmentation scale $\mu$. 
We restrict our consideration to the dominant $u$ and $\bar{d}$ ($\bar{s}$) quark channels for $\pi^+$ ($K^+$) production, as those channels feature the largest FF values. Nevertheless, the final hadron production rate in lepton-hadron collisions exhibits the highest sensitivity to the $u$-quark channel because of its dominance in the proton PDF and the larger electric charge fraction of $u$-type quarks compared to $d$-type quarks.
It is clear that these two sets of commonly used hadron FFs are very different as a function of momentum fraction $z$. 
Substantial differences are also observed when comparing them to other alternative sets available in the literature~\cite{deFlorian:2014xna,deFlorian:2017lwf,Bertone:2017tyb}. Therefore, the single high-$P_T$ hadron production in lepton-hadron collisions can be an excellent resource for studying and constraining hadron FFs.

\clearpage

\bibliographystyle{ptephy}
\bibliography{reference}

@article{Bloom:1969kc,
    author = "Bloom, Elliott D. and others",
    title = "{High-Energy Inelastic e p Scattering at 6-Degrees and 10-Degrees}",
    reportNumber = "SLAC-PUB-0642",
    doi = "10.1103/PhysRevLett.23.930",
    journal = "Phys. Rev. Lett.",
    volume = "23",
    pages = "930--934",
    year = "1969"
}

@article{Brambilla:2014jmp,
    author = "Brambilla, N. and others",
    title = "{QCD and Strongly Coupled Gauge Theories: Challenges and Perspectives}",
    eprint = "1404.3723",
    archivePrefix = "arXiv",
    primaryClass = "hep-ph",
    reportNumber = "CCQCN-2014-24, CCTP-2014-5, CERN-PH-TH-2014-033, DF-1-2014, HIP-2014-03-TH, ITEP-LAT-2014-1, JLAB-THY-14-1865, MITP-14-016, NT@UW-14-04, RUB-TPII-01-2014, TUM-EFT-46-14, FERMILAB-PUB-14-024-T, LLNL-JRNL-651216, UWTHPH-2014-006",
    doi = "10.1140/epjc/s10052-014-2981-5",
    journal = "Eur. Phys. J. C",
    volume = "74",
    number = "10",
    pages = "2981",
    year = "2014"
}

@article{Gross:2022hyw,
    author = "Gross, Franz and others",
    title = "{50 Years of Quantum Chromodynamics}",
    eprint = "2212.11107",
    archivePrefix = "arXiv",
    primaryClass = "hep-ph",
    doi = "10.1140/epjc/s10052-023-11949-2",
    journal = "Eur. Phys. J. C",
    volume = "83",
    pages = "1125",
    year = "2023"
}

@article{Accardi:2012qut,
    author = "Accardi, A. and others",
    editor = "Deshpande, A. and Meziani, Z. E. and Qiu, J. W.",
    title = "{Electron Ion Collider: The Next QCD Frontier}: {Understanding the glue that binds us all}",
    eprint = "1212.1701",
    archivePrefix = "arXiv",
    primaryClass = "nucl-ex",
    reportNumber = "BNL-98815-2012-JA, JLAB-PHY-12-1652",
    doi = "10.1140/epja/i2016-16268-9",
    journal = "Eur. Phys. J. A",
    volume = "52",
    number = "9",
    pages = "268",
    year = "2016"
}

@inproceedings{Proceedings:2020eah,
    author = "Hatta, Yoshitaka and others",
    title = "{Probing Nucleons and Nuclei in High Energy Collisions: Dedicated to the Physics of the Electron Ion Collider}: {Seattle (WA), United States, October 1 - November 16, 2018}",
    eprint = "2002.12333",
    archivePrefix = "arXiv",
    primaryClass = "hep-ph",
    doi = "10.1142/11684",
    publisher = "WSP",
    month = "2",
    year = "2020"
}

@article{AbdulKhalek:2021gbh,
    author = "Abdul Khalek, R. and others",
    title = "{Science Requirements and Detector Concepts for the Electron-Ion Collider}: {EIC Yellow Report}",
    eprint = "2103.05419",
    archivePrefix = "arXiv",
    primaryClass = "physics.ins-det",
    reportNumber = "BNL-220990-2021-FORE, JLAB-PHY-21-3198, LA-UR-21-20953",
    doi = "10.1016/j.nuclphysa.2022.122447",
    journal = "Nucl. Phys. A",
    volume = "1026",
    pages = "122447",
    year = "2022"
}

@article{Boer:2024ylx,
    author = {Boer, Dani{\"e}l and others},
    title = "{Physics case for quarkonium studies at the Electron Ion Collider}",
    eprint = "2409.03691",
    archivePrefix = "arXiv",
    primaryClass = "hep-ph",
    doi = "10.1016/j.ppnp.2025.104162",
    journal = "Prog. Part. Nucl. Phys.",
    volume = "142",
    pages = "104162",
    year = "2025"
}

@inproceedings{Proceedings:2026xrb,
    author = "Alexandrou, C. and others",
    title = "{Precision QCD with the Electron-Ion Collider}",
    eprint = "2604.04765",
    archivePrefix = "arXiv",
    primaryClass = "hep-ph",
    reportNumber = "INT-PUB-26-011",
    month = "4",
    year = "2026"
}

@book{Roberts:1990ww,
    author = "Roberts, R. G.",
    title = "{The Structure of the proton: Deep inelastic scattering}",
    doi = "10.1017/CBO9780511564062",
    isbn = "978-0-521-44944-1, 978-1-139-24244-8",
    publisher = "Cambridge University Press",
    series = "Cambridge Monographs on Mathematical Physics",
    month = "2",
    year = "1994"
}

@book{Tung:2001cv,
    author = "Tung, W. K.",
    title = "{Perturbative QCD and the parton structure of the nucleon}",
    publisher     = "World Scientific",
    year = "2001",
    booktitle = {At The Frontier of Particle Physics},
    chapter = {},
    pages = {887-971},
    doi = {10.1142/9789812810458_0024},
    URL = {https://www.worldscientific.com/doi/abs/10.1142/9789812810458_0024},
}

@article{Cammarota:2025jyr,
    author = "Cammarota, Justin and Qiu, Jian-Wei and Watanabe, Kazuhiro and Zhang, Jia-Yue",
    title = "{Factorized QED and QCD contribution to deeply inelastic scattering}",
    eprint = "2505.23487",
    archivePrefix = "arXiv",
    primaryClass = "hep-ph",
    reportNumber = "JLAB-THY-25-4341",
    doi = "10.1103/2d8y-ljwx",
    journal = "Phys. Rev. D",
    volume = "112",
    number = "5",
    pages = "056007",
    year = "2025"
}

@article{Charchula:1994kf,
    author = "Charchula, K. and Schuler, G. A. and Spiesberger, H.",
    title = "{Combined QED and QCD radiative effects in deep inelastic lepton - proton scattering: The Monte Carlo generator DJANGO6}",
    reportNumber = "CERN-TH-7133-94",
    doi = "10.1016/0010-4655(94)90086-8",
    journal = "Comput. Phys. Commun.",
    volume = "81",
    pages = "381--402",
    year = "1994"
}

@article{Aschenauer:2013iia,
    author = "Aschenauer, Elke C. and Burton, Thomas and Martini, Till and Spiesberger, Hubert and Stratmann, Marco",
    title = "{Prospects for Charged Current Deep-Inelastic Scattering off Polarized Nucleons at a Future Electron-Ion Collider}",
    eprint = "1309.5327",
    archivePrefix = "arXiv",
    primaryClass = "hep-ph",
    reportNumber = "HU-EP-13-35, MITP-13-055",
    doi = "10.1103/PhysRevD.88.114025",
    journal = "Phys. Rev. D",
    volume = "88",
    pages = "114025",
    year = "2013"
}

@article{Liu:2021jfp,
    author = "Liu, Tianbo and Melnitchouk, W. and Qiu, Jian-Wei and Sato, N.",
    title = "{A new approach to semi-inclusive deep-inelastic scattering with QED and QCD factorization}",
    eprint = "2108.13371",
    archivePrefix = "arXiv",
    primaryClass = "hep-ph",
    reportNumber = "JLAB-THY-21-3489",
    doi = "10.1007/JHEP11(2021)157",
    journal = "JHEP",
    volume = "11",
    pages = "157",
    year = "2021"
}

@article{Liu:2020rvc,
    author = "Liu, Tianbo and Melnitchouk, W. and Qiu, Jian-Wei and Sato, N.",
    title = "{Factorized approach to radiative corrections for inelastic lepton-hadron collisions}",
    eprint = "2008.02895",
    archivePrefix = "arXiv",
    primaryClass = "hep-ph",
    reportNumber = "JLAB-THY-20-3233",
    doi = "10.1103/PhysRevD.104.094033",
    journal = "Phys. Rev. D",
    volume = "104",
    number = "9",
    pages = "094033",
    year = "2021"
}

@inproceedings{Cammarota:2024vxj,
    author = "Cammarota, Justin and Qiu, Jian-Wei and Watanabe, Kazuhiro and Zhang, Jia-Yue",
    title = "{Factorized QED Contribution to Lepton-Hadron DIS}",
    booktitle = "{31st International Workshop on Deep-Inelastic Scattering and Related Subjects}",
    eprint = "2408.08377",
    archivePrefix = "arXiv",
    primaryClass = "hep-ph",
    reportNumber = "JLAB-THY-24-4144",
    month = "8",
    year = "2024"
}

@article{Mo:1968cg,
    author = "Mo, Luke W. and Tsai, Yung-Su",
    title = "{Radiative Corrections to Elastic and Inelastic e p and mu p Scattering}",
    reportNumber = "SLAC-PUB-0380",
    doi = "10.1103/RevModPhys.41.205",
    journal = "Rev. Mod. Phys.",
    volume = "41",
    pages = "205--235",
    year = "1969"
}

@article{Bardin:1989vz,
    author = "Bardin, D. Yu. and Burdik, K. C. and Khristova, P. Kh. and Riemann, T.",
    title = "{ELECTROWEAK RADIATIVE CORRECTIONS TO DEEP INELASTIC SCATTERING AT HERA! CHARGED CURRENT SCATTERING}",
    reportNumber = "JINR-E2-89-145",
    doi = "10.1007/BF01548593",
    journal = "Z. Phys. C",
    volume = "44",
    pages = "149",
    year = "1989"
}

@article{Badelek:1994uq,
    author = "Badelek, B. and Bardin, Dmitri Yu. and Kurek, K. and Scholz, C.",
    title = "{Radiative correction schemes in deep inelastic muon scattering}",
    eprint = "hep-ph/9403238",
    archivePrefix = "arXiv",
    reportNumber = "TSL-ISV-94-0092",
    doi = "10.1007/BF01579633",
    journal = "Z. Phys. C",
    volume = "66",
    pages = "591--600",
    year = "1995"
}

@article{Kripfganz:1990vm,
    author = "Kripfganz, J. and Mohring, H. J. and Spiesberger, H.",
    title = "{Higher order leading logarithmic QED corrections to deep inelastic e p scattering at very high-energies}",
    reportNumber = "DESY-90-096",
    doi = "10.1007/BF01549704",
    journal = "Z. Phys. C",
    volume = "49",
    pages = "501--510",
    year = "1991"
}

@article{Spiesberger:1994dm,
    author = "Spiesberger, H.",
    title = "{QED radiative corrections for parton distributions}",
    eprint = "hep-ph/9412286",
    archivePrefix = "arXiv",
    primaryClass = "hep-ph",
    reportNumber = "BI-TP-94-60, BI-TP.94-60",
    doi = "10.1103/PhysRevD.52.4936",
    journal = "Phys. Rev. D",
    volume = "52",
    pages = "4936--4940",
    year = "1995"
}

@article{Blumlein:2002fy,
    author = "Blumlein, Johannes and Kawamura, Hiroyuki",
    title = "{O(alpha**2 L) radiative corrections to deep inelastic ep scattering}",
    eprint = "hep-ph/0211191",
    archivePrefix = "arXiv",
    reportNumber = "DESY-02-193",
    doi = "10.1016/S0370-2693(02)03194-5",
    journal = "Phys. Lett. B",
    volume = "553",
    pages = "242--250",
    year = "2003"
}

@article{Kang:2011jw,
    author = "Kang, Zhong-Bo and Metz, Andreas and Qiu, Jian-Wei and Zhou, Jian",
    title = "{Exploring the structure of the proton through polarization observables in l p {\textbackslash}to jet X}",
    eprint = "1106.3514",
    archivePrefix = "arXiv",
    primaryClass = "hep-ph",
    doi = "10.1103/PhysRevD.84.034046",
    journal = "Phys. Rev. D",
    volume = "84",
    pages = "034046",
    year = "2011"
}

@article{Hinderer:2015hra,
    author = "Hinderer, Patriz and Schlegel, Marc and Vogelsang, Werner",
    title = "{Single-Inclusive Production of Hadrons and Jets in Lepton-Nucleon Scattering at NLO}",
    eprint = "1505.06415",
    archivePrefix = "arXiv",
    primaryClass = "hep-ph",
    doi = "10.1103/PhysRevD.92.014001",
    journal = "Phys. Rev. D",
    volume = "92",
    number = "1",
    pages = "014001",
    year = "2015",
    note = "[Erratum: Phys.Rev.D 93, 119903 (2016)]"
}

@article{Abelof:2016pby,
    author = "Abelof, Gabriel and Boughezal, Radja and Liu, Xiaohui and Petriello, Frank",
    title = "{Single-inclusive jet production in electron{\textendash}nucleon collisions through next-to-next-to-leading order in perturbative QCD}",
    eprint = "1607.04921",
    archivePrefix = "arXiv",
    primaryClass = "hep-ph",
    reportNumber = "NSF-KITP-16-054",
    doi = "10.1016/j.physletb.2016.10.022",
    journal = "Phys. Lett. B",
    volume = "763",
    pages = "52--59",
    year = "2016"
}

@article{Qiu:2020xum,
    author = "Qiu, Jian-Wei and Wang, Xiang-Peng and Xing, Hongxi",
    title = "{Exploring $J/\psi$ Production Mechanism at the Future Electron-Ion Collider}",
    eprint = "2005.10832",
    archivePrefix = "arXiv",
    primaryClass = "hep-ph",
    doi = "10.1088/0256-307X/38/4/041201",
    journal = "Chin. Phys. Lett.",
    volume = "38",
    number = "4",
    pages = "041201",
    year = "2021"
}

@article{Boussarie:2023izj,
    author = "Boussarie, Renaud and others",
    title = "{TMD Handbook}",
    eprint = "2304.03302",
    archivePrefix = "arXiv",
    primaryClass = "hep-ph",
    reportNumber = "JLAB-THY-23-3780, LA-UR-21-20798, MIT-CTP/5386",
    month = "4",
    year = "2023"
}

@article{ZEUS:2023zie,
    author = "Abt, I. and others",
    collaboration = "ZEUS",
    title = "{Measurement of jet production in deep inelastic scattering and NNLO determination of the strong coupling at ZEUS}",
    eprint = "2309.02889",
    archivePrefix = "arXiv",
    primaryClass = "hep-ex",
    reportNumber = "DESY-23-129",
    doi = "10.1140/epjc/s10052-023-12180-9",
    journal = "Eur. Phys. J. C",
    volume = "83",
    number = "11",
    pages = "1082",
    year = "2023"
}

@article{Cridge:2021pxm,
    author = "Cridge, T. and Harland-Lang, L. A. and Martin, A. D. and Thorne, R. S.",
    title = "{QED parton distribution functions in the MSHT20 fit}",
    eprint = "2111.05357",
    archivePrefix = "arXiv",
    primaryClass = "hep-ph",
    reportNumber = "IPPP/21/44",
    doi = "10.1140/epjc/s10052-022-10028-2",
    journal = "Eur. Phys. J. C",
    volume = "82",
    number = "1",
    pages = "90",
    year = "2022"
}

@article{Xie:2021equ,
    author = "Xie, Keping and Hobbs, T. J. and Hou, Tie-Jiun and Schmidt, Carl and Yan, Mengshi and Yuan, C. -P.",
    collaboration = "CTEQ-TEA",
    title = "{Photon PDF within the CT18 global analysis}",
    eprint = "2106.10299",
    archivePrefix = "arXiv",
    primaryClass = "hep-ph",
    reportNumber = "MSUHEP-21-013, PITT-PACC-2112, FERMILAB-PUB-21-370-QIS-SCD-T, SMU-HEP-21-06, SMU-HEP-21-06, FERMILAB-PUB-21-370-QIS-SCD-T",
    doi = "10.1103/PhysRevD.105.054006",
    journal = "Phys. Rev. D",
    volume = "105",
    number = "5",
    pages = "054006",
    year = "2022"
}

@article{NNPDF:2024djq,
    author = "Ball, Richard D. and others",
    collaboration = "NNPDF",
    title = "{Photons in the proton: implications for the LHC}",
    eprint = "2401.08749",
    archivePrefix = "arXiv",
    primaryClass = "hep-ph",
    reportNumber = "TIF-UNIMI-2023-17, Edinburgh 2023/19, CERN-TH-2023-159",
    doi = "10.1140/epjc/s10052-024-12731-8",
    journal = "Eur. Phys. J. C",
    volume = "84",
    number = "5",
    pages = "540",
    year = "2024"
}

@article{Khalek:2021gxf,
    author = "Khalek, Rabah Abdul and Bertone, Valerio and Nocera, Emanuele R.",
    collaboration = "MAP (Multi-dimensional Analyses of Partonic distributions)",
    title = "{Determination of unpolarized pion fragmentation functions using semi-inclusive deep-inelastic-scattering data}",
    eprint = "2105.08725",
    archivePrefix = "arXiv",
    primaryClass = "hep-ph",
    doi = "10.1103/PhysRevD.104.034007",
    journal = "Phys. Rev. D",
    volume = "104",
    number = "3",
    pages = "034007",
    year = "2021"
}

@inproceedings{Manohar:2024ndc,
    author = "Manohar, Aneesh and Nason, Paolo and Salam, Gavin and Zanderighi, Giulia",
    title = "{The photon parton distribution function: updates and applications}",
    booktitle = "{International Congress of Basic Science}",
    eprint = "2408.12719",
    archivePrefix = "arXiv",
    primaryClass = "hep-ph",
    month = "8",
    year = "2024"
}

@article{Eskola:1993mb,
    author = "Eskola, K. J. and Qiu, Jian-wei and Wang, Xin-Nian",
    title = "{Perturbative gluon shadowing in heavy nuclei}",
    eprint = "nucl-th/9307025",
    archivePrefix = "arXiv",
    reportNumber = "LBL-34156, LBL-34163",
    doi = "10.1103/PhysRevLett.72.36",
    journal = "Phys. Rev. Lett.",
    volume = "72",
    pages = "36--39",
    year = "1994"
}

@article{Eskola:2002yc,
    author = "Eskola, K. J. and Honkanen, H. and Kolhinen, V. J. and Qiu, Jian-wei and Salgado, C. A.",
    title = "{Nonlinear corrections to the DGLAP equations in view of the hera data}",
    eprint = "hep-ph/0211239",
    archivePrefix = "arXiv",
    reportNumber = "HIP-2002-52-TH, CERN-TH-2002-322",
    doi = "10.1016/S0550-3213(03)00257-8",
    journal = "Nucl. Phys. B",
    volume = "660",
    pages = "211--224",
    year = "2003"
}

@article{Eskola:2021nhw,
    author = "Eskola, Kari J. and Paakkinen, Petja and Paukkunen, Hannu and Salgado, Carlos A.",
    title = "{EPPS21: a global QCD analysis of nuclear PDFs}",
    eprint = "2112.12462",
    archivePrefix = "arXiv",
    primaryClass = "hep-ph",
    doi = "10.1140/epjc/s10052-022-10359-0",
    journal = "Eur. Phys. J. C",
    volume = "82",
    number = "5",
    pages = "413",
    year = "2022"
}

@article{Kusina:2020lyz,
    author = "Kusina, A. and others",
    title = "{Impact of LHC vector boson production in heavy ion collisions on strange PDFs}",
    eprint = "2007.09100",
    archivePrefix = "arXiv",
    primaryClass = "hep-ph",
    reportNumber = "IFJPAN-IV-2020-4, KA-TP-07-2020, MS-TP-20-27, P3H-20-035, SMU-HEP-20-04",
    doi = "10.1140/epjc/s10052-020-08532-4",
    journal = "Eur. Phys. J. C",
    volume = "80",
    number = "10",
    pages = "968",
    year = "2020"
}

@article{AbdulKhalek:2022fyi,
    author = "Abdul Khalek, Rabah and Gauld, Rhorry and Giani, Tommaso and Nocera, Emanuele R. and Rabemananjara, Tanjona R. and Rojo, Juan",
    title = "{nNNPDF3.0: evidence for a modified partonic structure in heavy nuclei}",
    eprint = "2201.12363",
    archivePrefix = "arXiv",
    primaryClass = "hep-ph",
    reportNumber = "Nikhef-2021-028, BONN-TH-2021-14",
    doi = "10.1140/epjc/s10052-022-10417-7",
    journal = "Eur. Phys. J. C",
    volume = "82",
    number = "6",
    pages = "507",
    year = "2022"
}

@article{Helenius:2021tof,
    author = "Helenius, Ilkka and Walt, Marina and Vogelsang, Werner",
    title = "{NNLO nuclear parton distribution functions with electroweak-boson production data from the LHC}",
    eprint = "2112.11904",
    archivePrefix = "arXiv",
    primaryClass = "hep-ph",
    doi = "10.1103/PhysRevD.105.094031",
    journal = "Phys. Rev. D",
    volume = "105",
    number = "9",
    pages = "094031",
    year = "2022"
}

@article{Nayak:2005rt,
    author = "Nayak, Gouranga C. and Qiu, Jian-Wei and Sterman, George F.",
    title = "{Fragmentation, NRQCD and NNLO factorization analysis in heavy quarkonium production}",
    eprint = "hep-ph/0509021",
    archivePrefix = "arXiv",
    reportNumber = "YITP-SB-05-26",
    doi = "10.1103/PhysRevD.72.114012",
    journal = "Phys. Rev. D",
    volume = "72",
    pages = "114012",
    year = "2005"
}

@article{QSY:2026,
    author = "Qiu, Jian-Wei and Sterman, George and Yu, Zhite",
    title = "{Factorization of inclusive production of single hadron at large $P_{h_T}$ in hadronic collisions}",
    journal = "in preparation",
    year = "2026"
}

@article{Dokshitzer:1977sg,
    author = "Dokshitzer, Yuri L.",
    title = "{Calculation of the Structure Functions for Deep Inelastic Scattering and e+ e- Annihilation by Perturbation Theory in Quantum Chromodynamics.}",
    journal = "Sov. Phys. JETP",
    volume = "46",
    pages = "641--653",
    year = "1977"
}

@article{Gribov:1972ri,
    author = "Gribov, V. N. and Lipatov, L. N.",
    title = "{Deep inelastic e p scattering in perturbation theory}",
    reportNumber = "IPTI-381-71",
    journal = "Sov. J. Nucl. Phys.",
    volume = "15",
    pages = "438--450",
    year = "1972"
}

@article{Lipatov:1974qm,
    author = "Lipatov, L. N.",
    title = "{The parton model and perturbation theory}",
    journal = "Sov. J. Nucl. Phys.",
    volume = "20",
    pages = "94--102",
    year = "1975"
}

@article{Altarelli:1977zs,
    author = "Altarelli, Guido and Parisi, G.",
    title = "{Asymptotic Freedom in Parton Language}",
    reportNumber = "LPTENS-77-6",
    doi = "10.1016/0550-3213(77)90384-4",
    journal = "Nucl. Phys. B",
    volume = "126",
    pages = "298--318",
    year = "1977"
}

@article{Kniehl:1996we,
    author = "Kniehl, Bernd A. and Kramer, G. and Spira, M.",
    title = "{Large p(T) photoproduction of D*+- mesons in e p collisions}",
    eprint = "hep-ph/9610267",
    archivePrefix = "arXiv",
    reportNumber = "DESY-96-210, CERN-TH-96-274, MPI-PHT-96-103",
    doi = "10.1007/s002880050591",
    journal = "Z. Phys. C",
    volume = "76",
    pages = "689--700",
    year = "1997"
}

@article{Buckley:2014ana,
    author = {Buckley, Andy and Ferrando, James and Lloyd, Stephen and Nordstr{\"o}m, Karl and Page, Ben and R{\"u}fenacht, Martin and Sch{\"o}nherr, Marek and Watt, Graeme},
    title = "{LHAPDF6: parton density access in the LHC precision era}",
    eprint = "1412.7420",
    archivePrefix = "arXiv",
    primaryClass = "hep-ph",
    reportNumber = "GLAS-PPE-2014-05, MCNET-14-29, IPPP-14-111, DCPT-14-222",
    doi = "10.1140/epjc/s10052-015-3318-8",
    journal = "Eur. Phys. J. C",
    volume = "75",
    pages = "132",
    year = "2015"
}

@article{Martin:2004dh,
    author = "Martin, A. D. and Roberts, R. G. and Stirling, W. J. and Thorne, R. S.",
    title = "{Parton distributions incorporating QED contributions}",
    eprint = "hep-ph/0411040",
    archivePrefix = "arXiv",
    reportNumber = "IPPP-04-62, DCPT-04-124, CAVENDISH-HEP-2004-28",
    doi = "10.1140/epjc/s2004-02088-7",
    journal = "Eur. Phys. J. C",
    volume = "39",
    pages = "155--161",
    year = "2005"
}

@article{Roth:2004ti,
    author = "Roth, Markus and Weinzierl, Stefan",
    title = "{QED corrections to the evolution of parton distributions}",
    eprint = "hep-ph/0403200",
    archivePrefix = "arXiv",
    reportNumber = "MPP-2004-32",
    doi = "10.1016/j.physletb.2004.04.009",
    journal = "Phys. Lett. B",
    volume = "590",
    pages = "190--198",
    year = "2004"
}

@article{Manohar:2016nzj,
    author = "Manohar, Aneesh and Nason, Paolo and Salam, Gavin P. and Zanderighi, Giulia",
    title = "{How bright is the proton? A precise determination of the photon parton distribution function}",
    eprint = "1607.04266",
    archivePrefix = "arXiv",
    primaryClass = "hep-ph",
    reportNumber = "CERN-TH-2016-155",
    doi = "10.1103/PhysRevLett.117.242002",
    journal = "Phys. Rev. Lett.",
    volume = "117",
    number = "24",
    pages = "242002",
    year = "2016"
}

@article{Moffat:2021dji,
    author = "Moffat, Eric and Melnitchouk, Wally and Rogers, T. C. and Sato, Nobuo",
    collaboration = "Jefferson Lab Angular Momentum (JAM)",
    title = "{Simultaneous Monte~Carlo analysis of parton densities and fragmentation functions}",
    eprint = "2101.04664",
    archivePrefix = "arXiv",
    primaryClass = "hep-ph",
    reportNumber = "JLAB-THY-21-3304",
    doi = "10.1103/PhysRevD.104.016015",
    journal = "Phys. Rev. D",
    volume = "104",
    number = "1",
    pages = "016015",
    year = "2021"
}

@article{Hou:2019efy,
    author = "Hou, Tie-Jiun and others",
    title = "{New CTEQ global analysis of quantum chromodynamics with high-precision data from the LHC}",
    eprint = "1912.10053",
    archivePrefix = "arXiv",
    primaryClass = "hep-ph",
    reportNumber = "MSUHEP-19-025, PITT-PACC-1911, SMU-HEP-19-03",
    doi = "10.1103/PhysRevD.103.014013",
    journal = "Phys. Rev. D",
    volume = "103",
    number = "1",
    pages = "014013",
    year = "2021"
}

@article{Chyla:1993xj,
    author = "Chyla, Jiri",
    title = "{Direct versus resolved photon: An Exercise in factorization}",
    eprint = "hep-ph/9306287",
    archivePrefix = "arXiv",
    reportNumber = "PRA-HEP-93-07",
    doi = "10.1016/0370-2693(94)90844-3",
    journal = "Phys. Lett. B",
    volume = "320",
    pages = "186--192",
    year = "1994"
}

@article{Hinderer:2017ntk,
    author = "Hinderer, Patriz and Schlegel, Marc and Vogelsang, Werner",
    title = "{Double-Longitudinal Spin Asymmetry in Single-Inclusive Lepton Scattering at NLO}",
    eprint = "1703.10872",
    archivePrefix = "arXiv",
    primaryClass = "hep-ph",
    doi = "10.1103/PhysRevD.96.014002",
    journal = "Phys. Rev. D",
    volume = "96",
    number = "1",
    pages = "014002",
    year = "2017"
}

@article{NNPDF:2017mvq,
    author = "Ball, Richard D. and others",
    collaboration = "NNPDF",
    title = "{Parton distributions from high-precision collider data}",
    eprint = "1706.00428",
    archivePrefix = "arXiv",
    primaryClass = "hep-ph",
    reportNumber = "EDINBURGH-2017-08, NIKHEF-2017-006, OUTP-17-04P, TIF-UNIMI-2017-3, CAVENDISH-HEP-17-06, CERN-TH-2017-077, Edinburgh 2017/08, Nikhef/2017-006, OUTP-17-04P,TIF-UNIMI-2017-3",
    doi = "10.1140/epjc/s10052-017-5199-5",
    journal = "Eur. Phys. J. C",
    volume = "77",
    number = "10",
    pages = "663",
    year = "2017"
}

@article{Botts:1990uy,
    author = "Botts, James and Qiu, Jian-wei and Sterman, George F.",
    editor = "Matthews, J. L. and Donnelly, T. W. and Farhi, E. H. and Osborne, L. S.",
    title = "{Elastic amplitudes and power corrections in QCD}",
    reportNumber = "ITP-SB-90-68",
    doi = "10.1016/0375-9474(91)90159-4",
    journal = "Nucl. Phys. A",
    volume = "527",
    pages = "577--580",
    year = "1991"
}

@article{Qiu:2001hj,
    author = "Qiu, Jian-wei and Sterman, George F.",
    title = "{QCD and rescattering in nuclear targets}",
    eprint = "hep-ph/0111002",
    archivePrefix = "arXiv",
    reportNumber = "YITP-SB-01-65",
    doi = "10.1142/S0218301303001235",
    journal = "Int. J. Mod. Phys. E",
    volume = "12",
    pages = "149",
    year = "2003"
}

@article{Doradau:2024wli,
    author = "Doradau, Matias and Martinez, Ramiro Tomas and Sassot, Rodolfo and Stratmann, Marco",
    title = "{Pion nuclear fragmentation functions revisited}",
    eprint = "2411.08222",
    archivePrefix = "arXiv",
    primaryClass = "hep-ph",
    doi = "10.1103/PhysRevD.111.034045",
    journal = "Phys. Rev. D",
    volume = "111",
    number = "3",
    pages = "034045",
    year = "2025"
}

@article{Sjostrand:2014zea,
    author = {Sj{\"o}strand, Torbj{\"o}rn and Ask, Stefan and Christiansen, Jesper R. and Corke, Richard and Desai, Nishita and Ilten, Philip and Mrenna, Stephen and Prestel, Stefan and Rasmussen, Christine O. and Skands, Peter Z.},
    title = "{An introduction to PYTHIA 8.2}",
    eprint = "1410.3012",
    archivePrefix = "arXiv",
    primaryClass = "hep-ph",
    reportNumber = "LU-TP-14-36, MCNET-14-22, CERN-PH-TH-2014-190, FERMILAB-PUB-14-316-CD, DESY-14-178, SLAC-PUB-16122",
    doi = "10.1016/j.cpc.2015.01.024",
    journal = "Comput. Phys. Commun.",
    volume = "191",
    pages = "159--177",
    year = "2015"
}

@article{Bierlich:2022pfr,
    author = "Bierlich, Christian and others",
    title = "{A comprehensive guide to the physics and usage of PYTHIA 8.3}",
    eprint = "2203.11601",
    archivePrefix = "arXiv",
    primaryClass = "hep-ph",
    reportNumber = "LU-TP 22-16, MCNET-22-04, FERMILAB-PUB-22-227-SCD",
    doi = "10.21468/SciPostPhysCodeb.8",
    journal = "SciPost Phys. Codeb.",
    volume = "2022",
    pages = "8",
    year = "2022"
}

@article{Sherpa:2019gpd,
    author = "Bothmann, Enrico and others",
    collaboration = "Sherpa",
    title = "{Event Generation with Sherpa 2.2}",
    eprint = "1905.09127",
    archivePrefix = "arXiv",
    primaryClass = "hep-ph",
    reportNumber = "FERMILAB-PUB-19-218-T, SLAC-PUB-17433, IPPP/19/42, MCNET-19-11",
    doi = "10.21468/SciPostPhys.7.3.034",
    journal = "SciPost Phys.",
    volume = "7",
    number = "3",
    pages = "034",
    year = "1919"
}

@article{Bellm:2015jjp,
    author = "Bellm, Johannes and others",
    title = "{Herwig 7.0/Herwig++ 3.0 release note}",
    eprint = "1512.01178",
    archivePrefix = "arXiv",
    primaryClass = "hep-ph",
    reportNumber = "CERN-PH-TH-2015-289, MAN-HEP-2015-15, IFJPAN-IV-2015-13, KA-TP-18-2015, DCPT-15-142, MCNET-15-28, IPPP-15-71, HERWIG-2015-01",
    doi = "10.1140/epjc/s10052-016-4018-8",
    journal = "Eur. Phys. J. C",
    volume = "76",
    number = "4",
    pages = "196",
    year = "2016"
}

@article{Jager:2004jh,
    author = "Jager, B. and Stratmann, M. and Vogelsang, W.",
    title = "{Single inclusive jet production in polarized $p p$ collisions at $O(alpha^3_s)$}",
    eprint = "hep-ph/0404057",
    archivePrefix = "arXiv",
    reportNumber = "BNL-NT-04-11, RBRC-412",
    doi = "10.1103/PhysRevD.70.034010",
    journal = "Phys. Rev. D",
    volume = "70",
    pages = "034010",
    year = "2004"
}

@article{Mukherjee:2012uz,
    author = "Mukherjee, Asmita and Vogelsang, Werner",
    title = "{Jet production in (un)polarized pp collisions: dependence on jet algorithm}",
    eprint = "1209.1785",
    archivePrefix = "arXiv",
    primaryClass = "hep-ph",
    doi = "10.1103/PhysRevD.86.094009",
    journal = "Phys. Rev. D",
    volume = "86",
    pages = "094009",
    year = "2012",
    note = "[Erratum: Phys.Rev.D 107, 119901 (2023)]"
}

@article{Kaufmann:2015hma,
    author = "Kaufmann, Tom Alignment and Mukherjee, Asmita and Vogelsang, Werner",
    title = "{Hadron Fragmentation Inside Jets in Hadronic Collisions}",
    eprint = "1506.01415",
    archivePrefix = "arXiv",
    primaryClass = "hep-ph",
    doi = "10.1103/PhysRevD.92.054015",
    journal = "Phys. Rev. D",
    volume = "92",
    number = "5",
    pages = "054015",
    year = "2015",
    note = "[Erratum: Phys.Rev.D 101, 079901 (2020)]"
}

@article{Kang:2016mcy,
    author = "Kang, Zhong-Bo and Ringer, Felix and Vitev, Ivan",
    title = "{The semi-inclusive jet function in SCET and small radius resummation for inclusive jet production}",
    eprint = "1606.06732",
    archivePrefix = "arXiv",
    primaryClass = "hep-ph",
    doi = "10.1007/JHEP10(2016)125",
    journal = "JHEP",
    volume = "10",
    pages = "125",
    year = "2016"
}

@article{Cacciari:2008gp,
    author = "Cacciari, Matteo and Salam, Gavin P. and Soyez, Gregory",
    title = "{The anti-$k_t$ jet clustering algorithm}",
    eprint = "0802.1189",
    archivePrefix = "arXiv",
    primaryClass = "hep-ph",
    reportNumber = "LPTHE-07-03",
    doi = "10.1088/1126-6708/2008/04/063",
    journal = "JHEP",
    volume = "04",
    pages = "063",
    year = "2008"
}

@article{deFlorian:2013qia,
    author = "de Florian, Daniel and Hinderer, Patriz and Mukherjee, Asmita and Ringer, Felix and Vogelsang, Werner",
    title = "{Approximate next-to-next-to-leading order corrections to hadronic jet production}",
    eprint = "1310.7192",
    archivePrefix = "arXiv",
    primaryClass = "hep-ph",
    doi = "10.1103/PhysRevLett.112.082001",
    journal = "Phys. Rev. Lett.",
    volume = "112",
    pages = "082001",
    year = "2014"
}

@article{Dasgupta:2016bnd,
    author = "Dasgupta, Mrinal and Dreyer, Fr{\'e}d{\'e}ric A. and Salam, Gavin P. and Soyez, Gregory",
    title = "{Inclusive jet spectrum for small-radius jets}",
    eprint = "1602.01110",
    archivePrefix = "arXiv",
    primaryClass = "hep-ph",
    reportNumber = "CERN-TH-2016-020",
    doi = "10.1007/JHEP06(2016)057",
    journal = "JHEP",
    volume = "06",
    pages = "057",
    year = "2016"
}

@article{Lee:2024icn,
    author = "Lee, Kyle and Moult, Ian and Zhang, Xiaoyuan",
    title = "{Revisiting single inclusive jet production: timelike factorization and reciprocity}",
    eprint = "2409.19045",
    archivePrefix = "arXiv",
    primaryClass = "hep-ph",
    reportNumber = "MIT-CTP 5766",
    doi = "10.1007/JHEP05(2025)129",
    journal = "JHEP",
    volume = "05",
    pages = "129",
    year = "2025"
}

@article{Generet:2025vth,
    author = "Generet, Terry and Lee, Kyle and Moult, Ian and Poncelet, Rene and Zhang, Xiaoyuan",
    title = "{Small radius inclusive jet production at the LHC through NNLO+NNLL}",
    eprint = "2503.21866",
    archivePrefix = "arXiv",
    primaryClass = "hep-ph",
    reportNumber = "Cavendish-HEP-25/02, MIT-CTP 5857, IFJPAN-IV-2025-7",
    doi = "10.1007/JHEP08(2025)015",
    journal = "JHEP",
    volume = "08",
    pages = "015",
    year = "2025"
}

@article{Ellis:1992qq,
    author = "Ellis, Stephen D. and Kunszt, Zoltan and Soper, Davison E.",
    title = "{Jets at hadron colliders at order $\alpha-s^{3:}$ A Look inside}",
    eprint = "hep-ph/9208249",
    archivePrefix = "arXiv",
    reportNumber = "UW-PT-92-01, DOE-ER-40614-16",
    doi = "10.1103/PhysRevLett.69.3615",
    journal = "Phys. Rev. Lett.",
    volume = "69",
    pages = "3615--3618",
    year = "1992"
}

@article{Qiu:2019sfj,
    author = "Qiu, Jian-Wei and Ringer, Felix and Sato, Nobuo and Zurita, Pia",
    title = "{Factorization of jet cross sections in heavy-ion collisions}",
    eprint = "1903.01993",
    archivePrefix = "arXiv",
    primaryClass = "hep-ph",
    reportNumber = "JLAB-THY-19-2896",
    doi = "10.1103/PhysRevLett.122.252301",
    journal = "Phys. Rev. Lett.",
    volume = "122",
    number = "25",
    pages = "252301",
    year = "2019"
}

@article{Dasgupta:2007wa,
    author = "Dasgupta, Mrinal and Magnea, Lorenzo and Salam, Gavin P.",
    title = "{Non-perturbative QCD effects in jets at hadron colliders}",
    eprint = "0712.3014",
    archivePrefix = "arXiv",
    primaryClass = "hep-ph",
    reportNumber = "DFTT-27-2007, MAN-HEP-2007-41",
    doi = "10.1088/1126-6708/2008/02/055",
    journal = "JHEP",
    volume = "02",
    pages = "055",
    year = "2008"
}

@article{Kang:2012zr,
    author = "Kang, Zhong-Bo and Mantry, Sonny and Qiu, Jian-Wei",
    title = "{N-Jettiness as a Probe of Nuclear Dynamics}",
    eprint = "1204.5469",
    archivePrefix = "arXiv",
    primaryClass = "hep-ph",
    doi = "10.1103/PhysRevD.86.114011",
    journal = "Phys. Rev. D",
    volume = "86",
    pages = "114011",
    year = "2012"
}

@article{Kang:2013wca,
    author = "Kang, Zhong-Bo and Liu, Xiaohui and Mantry, Sonny and Qiu, Jian-Wei",
    title = "{Probing nuclear dynamics in jet production with a global event shape}",
    eprint = "1303.3063",
    archivePrefix = "arXiv",
    primaryClass = "hep-ph",
    doi = "10.1103/PhysRevD.88.074020",
    journal = "Phys. Rev. D",
    volume = "88",
    pages = "074020",
    year = "2013"
}

@article{Kang:2013lga,
    author = "Kang, Zhong-Bo and Liu, Xiaohui and Mantry, Sonny",
    title = "{1-jettiness DIS event shape: NNLL+NLO results}",
    eprint = "1312.0301",
    archivePrefix = "arXiv",
    primaryClass = "hep-ph",
    doi = "10.1103/PhysRevD.90.014041",
    journal = "Phys. Rev. D",
    volume = "90",
    number = "1",
    pages = "014041",
    year = "2014"
}

@article{Cao:2024ota,
    author = "Cao, Haotian and Kang, Zhong-Bo and Liu, Xiaohui and Mantry, Sonny",
    title = "{One-jettiness DIS event shape at N3LL+O({\ensuremath{\alpha}}s2)}",
    eprint = "2401.01941",
    archivePrefix = "arXiv",
    primaryClass = "hep-ph",
    doi = "10.1103/PhysRevD.110.014045",
    journal = "Phys. Rev. D",
    volume = "110",
    number = "1",
    pages = "014045",
    year = "2024"
}

@article{Stewart:2010tn,
    author = "Stewart, Iain W. and Tackmann, Frank J. and Waalewijn, Wouter J.",
    title = "{N-Jettiness: An Inclusive Event Shape to Veto Jets}",
    eprint = "1004.2489",
    archivePrefix = "arXiv",
    primaryClass = "hep-ph",
    reportNumber = "MIT-CTP-4139",
    doi = "10.1103/PhysRevLett.105.092002",
    journal = "Phys. Rev. Lett.",
    volume = "105",
    pages = "092002",
    year = "2010"
}

@article{Jouttenus:2011wh,
    author = "Jouttenus, Teppo T. and Stewart, Iain W. and Tackmann, Frank J. and Waalewijn, Wouter J.",
    title = "{The Soft Function for Exclusive N-Jet Production at Hadron Colliders}",
    eprint = "1102.4344",
    archivePrefix = "arXiv",
    primaryClass = "hep-ph",
    doi = "10.1103/PhysRevD.83.114030",
    journal = "Phys. Rev. D",
    volume = "83",
    pages = "114030",
    year = "2011"
}

@article{Kang:2013nha,
    author = "Kang, Daekyoung and Lee, Christopher and Stewart, Iain W.",
    title = "{Using 1-Jettiness to Measure 2 Jets in DIS 3 Ways}",
    eprint = "1303.6952",
    archivePrefix = "arXiv",
    primaryClass = "hep-ph",
    reportNumber = "MIT-CTP-4375, LA-UR-13-20960",
    doi = "10.1103/PhysRevD.88.054004",
    journal = "Phys. Rev. D",
    volume = "88",
    pages = "054004",
    year = "2013"
}

@article{Dotson:2026ttc,
    author = "Dotson, Andrew and Ee, June-Haak and Lee, Christopher and Makris, Yiannis and Terry, John",
    title = "{Centauric 1-Jettiness in DIS and Universal Power Corrections}",
    eprint = "2606.20825",
    archivePrefix = "arXiv",
    primaryClass = "hep-ph",
    reportNumber = "LA-UR-26-21204, MIT-CTP/5994",
    month = "6",
    year = "2026"
}

@article{Whitehill:2026swa,
    author = "Whitehill, R. M. and Dalton, M. M. and Liu, T. and Melnitchouk, W. and Sato, N.",
    title = "{Impact of parity-violating deep-inelastic scattering on the weak mixing angle and high-$x$ parton distributions}",
    eprint = "2606.26056",
    archivePrefix = "arXiv",
    primaryClass = "hep-ph",
    reportNumber = "JLAB-THY-26-4810",
    month = "6",
    year = "2026"
}

@article{CQWZ:pvdis,
    author = "Cammarota, Justin and Qiu, Jian-Wei and Watanabe, Kazuhiro and Zhang, Jia-Yue",
    title = "{Factorized QED QCD Contribution to Parity-Violating Deep Inelastic Scattering}",
    journal = "in preparation",
    year = "2026"
}

@article{Cammarota:2026ylh,
    author = "Cammarota, Justin and Carlton, John and Gardner, Susan",
    title = "{New Energy-Loss Constraints on Dark Sectors from Deeply Inelastic Scattering with Initial State Radiation}",
    eprint = "2606.28500",
    archivePrefix = "arXiv",
    primaryClass = "hep-ph",
    reportNumber = "DESY-26-086",
    month = "6",
    year = "2026"
}

@article{Belle:2013lfg,
    author = "Leitgab, M. and others",
    collaboration = "Belle",
    title = "{Precision Measurement of Charged Pion and Kaon Differential Cross Sections in e+e- Annihilation at s=10.52  GeV}",
    eprint = "1301.6183",
    archivePrefix = "arXiv",
    primaryClass = "hep-ex",
    reportNumber = "BELLE-PREPRINT-2013-2, KEK-PREPRINT-2012-38",
    doi = "10.1103/PhysRevLett.111.062002",
    journal = "Phys. Rev. Lett.",
    volume = "111",
    pages = "062002",
    year = "2013"
}

@article{Belle:2020pvy,
    author = "Seidl, R. and others",
    collaboration = "Belle",
    title = "{Update of inclusive cross sections of single and pairs of identified light charged hadrons}",
    eprint = "2001.10194",
    archivePrefix = "arXiv",
    primaryClass = "hep-ex",
    reportNumber = "Belle Preprint 2020-01, KEK Preprint 2019-56",
    doi = "10.1103/PhysRevD.101.092004",
    journal = "Phys. Rev. D",
    volume = "101",
    number = "9",
    pages = "092004",
    year = "2020"
}

@article{BaBar:2013yrg,
    author = "Lees, J. P. and others",
    collaboration = "BaBar",
    title = "{Production of charged pions, kaons, and protons in $e^+e^-$ annihilations into hadrons at $\sqrt{s}$=10.54  GeV}",
    eprint = "1306.2895",
    archivePrefix = "arXiv",
    primaryClass = "hep-ex",
    reportNumber = "BABAR-PUB-13-003, SLAC-PUB-15524",
    doi = "10.1103/PhysRevD.88.032011",
    journal = "Phys. Rev. D",
    volume = "88",
    pages = "032011",
    year = "2013"
}

@article{Berger:2001wr,
    author = "Berger, Edmond L. and Qiu, Jian-wei and Zhang, Xiao-fei",
    title = "{QCD factorized Drell-Yan cross-section at large transverse momentum}",
    eprint = "hep-ph/0107309",
    archivePrefix = "arXiv",
    reportNumber = "ANL-HEP-01-057",
    doi = "10.1103/PhysRevD.65.034006",
    journal = "Phys. Rev. D",
    volume = "65",
    pages = "034006",
    year = "2002"
}

@article{Lee:2021oqr,
    author = "Lee, Kyle and Qiu, Jian-Wei and Sterman, George and Watanabe, Kazuhiro",
    title = "{QCD factorization for hadronic quarkonium production at high $p_T$}",
    eprint = "2108.00305",
    archivePrefix = "arXiv",
    primaryClass = "hep-ph",
    reportNumber = "JLAB-THY-21-3477",
    doi = "10.21468/SciPostPhysProc.8.143",
    journal = "SciPost Phys. Proc.",
    volume = "8",
    pages = "143",
    year = "2022"
}

@article{deFlorian:2014xna,
    author = "de Florian, Daniel and Sassot, R. and Epele, Manuel and Hern{\'a}ndez-Pinto, Roger J. and Stratmann, Marco",
    title = "{Parton-to-Pion Fragmentation Reloaded}",
    eprint = "1410.6027",
    archivePrefix = "arXiv",
    primaryClass = "hep-ph",
    doi = "10.1103/PhysRevD.91.014035",
    journal = "Phys. Rev. D",
    volume = "91",
    number = "1",
    pages = "014035",
    year = "2015"
}

@article{deFlorian:2017lwf,
    author = "de Florian, D. and Epele, M. and Hernandez-Pinto, R. J. and Sassot, R. and Stratmann, M.",
    title = "{Parton-to-Kaon Fragmentation Revisited}",
    eprint = "1702.06353",
    archivePrefix = "arXiv",
    primaryClass = "hep-ph",
    doi = "10.1103/PhysRevD.95.094019",
    journal = "Phys. Rev. D",
    volume = "95",
    number = "9",
    pages = "094019",
    year = "2017"
}

@article{Bertone:2017tyb,
    author = "Bertone, Valerio and Carrazza, Stefano and Hartland, Nathan P. and Nocera, Emanuele R. and Rojo, Juan",
    collaboration = "NNPDF",
    title = "{A determination of the fragmentation functions of pions, kaons, and protons with faithful uncertainties}",
    eprint = "1706.07049",
    archivePrefix = "arXiv",
    primaryClass = "hep-ph",
    reportNumber = "CERN-TH-2017-122, OUTP-16-15P, NIKHEF-2016-047",
    doi = "10.1140/epjc/s10052-017-5088-y",
    journal = "Eur. Phys. J. C",
    volume = "77",
    number = "8",
    pages = "516",
    year = "2017"
}

\end{document}